\documentclass[12pt]{article}

\usepackage[nodisplayskipstretch]{setspace} % must come before hyperref
\usepackage[margin=1.25in]{geometry}

\usepackage[dvipsnames,svgnames]{xcolor}
\usepackage{amsmath, bbm, amssymb, amsfonts, graphicx, pdflscape, mathtools, amsthm, upgreek,bm,pgfplots,csvsimple,multirow, multicol, booktabs, import, placeins, subcaption,etoolbox}

\let\originalleft\left
\let\originalright\right
\renewcommand{\left}{\mathopen{}\mathclose\bgroup\originalleft}
\renewcommand{\right}{\aftergroup\egroup\originalright}

\usepackage{tikz}
\usetikzlibrary{shapes.geometric, arrows, fit, positioning}
\usepackage[multiple]{footmisc}
\usepackage{verbatim}
\usepackage[notlof,notlot,nottoc]{tocbibind}
\usepackage{microtype}

\usepackage{amsfonts}
\usepackage{amssymb}
\usepackage[T1]{fontenc}
\usepackage[utf8]{inputenc}
\usepackage[backend=biber,sorting=ynt,style=authoryear-comp,giveninits=true,maxbibnames=9,citetracker=true,maxcitenames=2,uniquename=false,uniquelist=false,natbib,isbn=false,doi=false,url=false,eprint=false]{biblatex}
\DeclareFieldFormat{pages}{#1} %Removes pp. and p. from bibliography
\renewbibmacro{in:}{} %Removes "In:" from bibliography

\usepackage{titlesec}

\titleformat{\subsubsection}[runin]
  {\normalfont\normalsize\bfseries}
  {\thesubsubsection}{1em}{}
  []
\titlespacing{\subsubsection}
  {0pt}% left
  {1.5ex plus 1ex minus .2ex}% before
  {1em}% after

\titleformat{\section}
  {\bfseries\Large}{%
    {\bfseries\Large\thesection}
  }{1em}{}
  \titleformat{\subsection}
  {\bfseries\large}{%
    {\bfseries\large\thesubsection}
  }{1em}{}

\titlespacing{\subsection}
  {0pt}% left
  {1.75ex plus 1ex minus .2ex}% before
  {1ex}% after

\titlespacing{\section}
  {0pt}% left
  {1.75ex plus 1ex minus .2ex}% before
  {1ex}% after

\usepackage{verbatim}
\usepackage{accents}
\newcommand{\ubar}[1]{\underaccent{\bar}{#1}}
\usepackage{needspace}

\usepackage{comment}
\usepackage{enumitem}
    \setlist[itemize]{noitemsep,nolistsep}
    \setlist[enumerate,1]{noitemsep,nolistsep,label=(\arabic*)}
    \setlist[enumerate,2]{noitemsep,nolistsep,label=(\alph*)}
    \setlist[enumerate,3]{noitemsep,nolistsep,label=(\roman*)}
\usepackage{tocvsec2}
\usepackage{threeparttable}
\usepackage[page,title,toc]{appendix}
\usetikzlibrary{calc, cd}
\setlist{noitemsep, topsep=0pt}
\usepackage[skip = 10pt]{caption}
\allowdisplaybreaks[1]
\allowdisplaybreaks[1]

\makeatletter
\renewcommand{\paragraph}{%
	\@startsection{paragraph}{4}%
	{\z@}{1.5ex \@plus 1ex \@minus .2ex}{-0.7em}%
	{\normalfont\normalsize\bfseries}%
}
\makeatother

\definecolor{Blue}{RGB}{86,180,233}
\definecolor{Orange}{RGB}{230,159,0}
\definecolor{Green}{RGB}{0,158,115}
\definecolor{GmailBlue}{RGB}{42, 93, 176} % for links
\usepackage[
	colorlinks=true,
    citecolor= GmailBlue,
	linkcolor=GmailBlue,
	urlcolor = GmailBlue,
]{hyperref}

\usepackage{pgfplots}
\usepgfplotslibrary{groupplots,colorbrewer}
\pgfplotsset{compat=newest}
\pgfplotsset{cycle list/Set1}
\usepackage{tikz}
\usetikzlibrary{matrix,calc,shapes,arrows.meta,positioning}
\tikzset{
    vertex/.style = {shape=circle,draw, minimum size = 1.8em, inner sep = 0pt},
    edge/.style = {->,> = latex}
}

\usepackage[capitalize,noabbrev,nameinlink]{cleveref}

\theoremstyle{plain}
  \newtheorem{theorem}{Theorem}
  \newtheorem{lemma}{Lemma}
  \newtheorem{proposition}{Proposition}
  
  \newtheorem{step}{Step}

\theoremstyle{definition}
  \newtheorem{assumption}{Assumption}

  \newtheorem{example}{Example}
  \newtheorem{remark}{Remark}

\crefname{equation}{equation}{equations}
\Crefname{equation}{Equation}{Equations}
\AtBeginEnvironment{appendices}{\crefalias{section}{appendix}}
\AtBeginEnvironment{appendices}{\crefalias{subsection}{appendix}}
\AtBeginEnvironment{appendices}{\crefalias{subsubsection}{appendix}}
\crefname{appsec}{appendix}{appendices}
\Crefname{appsec}{Appendix}{Appendices}
\Crefname{appendices}{Appendix}{Appendices}
\crefname{appendices}{appendix}{appendices}
\crefname{assumption}{assumption}{assumptions}
\Crefname{assumption}{Assumption}{Assumptions}
\crefname{condition}{condition}{conditions}
\Crefname{condition}{Condition}{Conditions}
\Crefname{lemma}{Lemma}{Lemmata}
\Crefname{footnote}{Footnote}{Footnotes}
\crefname{footnote}{footnote}{footnotes}

\renewcommand{\phi}{\varphi}

\def\R{\mathbb{R}}

\DeclareMathOperator{\E}{\mathbb{E}}
\DeclareMathOperator{\supp}{supp} % support

\DeclareMathOperator{\sgn}{sgn} % sign
\DeclareMathOperator{\Leb}{Leb}

\newcommand{\de}{\mathop{}\!\mathrm{d}}

\newcommand{\headercref}[2]{\texorpdfstring{\Cref{#2}}{#1 \ref{#2}}}

\hypersetup{
	pdftitle={Waiting for Trade in Markets with Aggregate Uncertainty},
	pdfauthor={Justus Preusser}
}

\begin{document}

\title{Waiting for Trade in Markets with Aggregate Uncertainty}%

\author{
Justus Preusser\thanks{\protect {\href{mailto:justus.preusser@unibocconi.it}{justus.preusser@unibocconi.it}}, Department of Economics and IGIER, Bocconi University..}\thanks{\protect I thank the editor, an associate editor, two referees,
Martino Banchio,
Christoph Carnehl,
Gregorio Curello,
Deniz Kattwinkel,
Teddy Kim,
Nenad Kos,
Stephan Lauermann,
Marco Ottaviani,
Elia Sartori,
and audiences at Bocconi, Naples, and SAET 2026
for their comments.
I gratefully acknowledge financial support from the European Research Council (HEUROPE 2022 ADG, GA No. 101055295 – InfoEcoScience).}
}

\date{
  This version: \today%
}

\maketitle

\begin{abstract}%
This paper studies learning in markets with aggregate uncertainty about whether trade is efficient. 
A long-lived seller offers prices to buyers, who are short-lived and arrive according to a Poisson process. 
A hidden state determines whether the buyers' common value exceeds the seller's reservation value. All parties observe noisy, private signals about the state. 
With small intertemporal frictions and when the seller has commitment power, the seller waits for a buyer with the most favorable signal to arrive up to an exit time that depends on the seller's private information.
This strategy profile maximizes both the seller's profit and the expected surplus.
Without commitment, the commitment profit is unattainable. Instead, there is an equilibrium in which the seller also waits for a buyer with the most favorable signal, but, relative to the commitment case, the seller exits inefficiently late, and the trade probability is inefficiently high. 


\end{abstract}

\vfill
\pagenumbering{gobble}
\pagebreak
\pagenumbering{arabic}

\section{Introduction}

This paper studies learning in markets with aggregate uncertainty about whether trade is efficient.
Consider a seller with a single good who encounters many potential buyers sequentially.
The buyers share a common value for the good, but it is uncertain whether this common value exceeds the seller's reservation value.
Further, the seller and the buyers each have noisy, private information about the state.
For example, the seller could be a homeowner who is uncertain whether their valuation for staying in their house exceeds the valuation of potential buyers.
The state could capture uncertain future infrastructure developments that differentially affect the seller's and the buyers' long-run valuation for the house.
For another example, the seller could hold a financial asset whose payoff is realized only in the future.
Depending on the state, the seller may or may not be the efficient holder of the asset, say due to liquidity constraints or heterogeneous risk attitudes.

In such markets, the seller gradually learns about the gains from trade while waiting for a buyer with a favorable signal. If the seller becomes sufficiently pessimistic, the seller may exit the market.\footnote{In housing markets, a significant portion of units remain unsold or are eventually withdrawn. \citet{anenberg2016information} reports that around 50\% of all listings are withdrawn or do not sell in a two-year sample from the San Francisco Bay Area and Los Angeles. \citet[footnote 12]{merlo2004bargaining} indicate that up to 25\% of all listings are withdrawn in their data set from England.} This paper studies a tractable model for understanding such exits and the extent to which trade reflects the information of market participants.

The paper shows that, in the setting described below, the seller may exit inefficiently late, resulting in an inefficiently high equilibrium trade probability.
The direction of this distortion is not obvious a priori. If the gains from trade can be negative, it may be efficient for the seller to exit, but exiting may also be in the seller's private interest.
The finding is thus that the seller's private incentive to exit is too weak. In contrast to this distortion, earlier work explains delay and impasses in trade via informational asymmetries in settings in which it would be efficient to trade in all states and, hence, the equilibrium trade probability can only be inefficiently low.

In the model, the seller has a single good, is long-lived, and posts prices. 
Buyers are short-lived, and their arrivals are unobservable and follow a Poisson process. 
A hidden state determines the buyers' common value and the seller's reservation value.
The buyers' common value increases in the state, but exceeds the seller's reservation value if and only if the state is sufficiently high; i.e., the ex post surplus from trading is single-crossing from below. 
While no one observes the state, the seller and the buyers have noisy, private, conditionally independent signals.
The seller's signal is called their type.
Importantly, there is a most-favorable buyer-signal about the state, and higher seller-types indicate higher states.
Each buyer observes only their private signal; they do not observe calendar time, the seller's type, or past actions. 
The seller can reveal calendar time and their type via messages.
Finally, there is an intertemporal friction: the market may break down at an exogenous Poisson rate; there is no further discounting or cost for holding the good on the market.

The analysis focuses on situations with a small breakdown rate. 
Thus, the model fits settings in which buyers arrive so quickly as to render breakdown risks negligible, and in which the stakes are so high as to render holding costs negligible.\footnote{In the example of selling a house, empirical estimates for holding costs seem to be small. \citet{anenberg2016information} estimates the weekly holding cost at around 0.2\% of the value of the house. See also \citet{merlo2015home}.}
The focus on small intertemporal frictions is in line with earlier work on information aggregation in decentralized markets. 
Conceptually, assuming small frictions focuses the analysis on the problem of learning about the state and paves the way for techniques for monotone decision problems (\citet{karlin1956theory}; \citet{lehmann1988comparing}).

To understand the direction of equilibrium distortions, the first step is to characterize efficient behavior.
To do so, the first part of the paper supposes that the seller has commitment power. The seller's problem then has a tractable solution that also maximizes the expected surplus.
This solution is the benchmark for the second part of the paper, which studies equilibria.

With commitment, the seller faces two problems. First, the seller seeks to minimize buyers' information rents stemming from their private signals.
Second---and this is the focus---the seller seeks to learn about the state.
Indeed, even if the seller could extract the expected surplus from trade, the seller must learn if trade is efficient.
The learning problem is complicated by the fact that the seller can only learn about the buyers' signals when the seller offers the good and buyers reject the seller's price. In particular, to learn the seller must risk inefficient trades before having fully resolved the uncertainty about the state. 

\Cref{thm:commitment_solution} shows that, with commitment power and a sufficiently small breakdown rate, both problems of the seller are optimally solved by trading with the first buyer who has the most favorable signal and who arrives before a certain date that depends on the seller's type.
At this date, the seller permanently exits the market.
To implement these trades, the seller can post a type- and time-invariant price that renders a buyer with the most favorable signal indifferent. 
The key step in this result is that waiting for a buyer with the most favorable signal maximizes the expected surplus from trading across all strategy profiles (\Cref{thm:maximizing_surplus}).
Note that waiting for the most favorable signal need not maximize the expected surplus conditional on trading since the ex post surplus is not necessarily increasing in the state (but only single-crossing).
Instead, this waiting strategy is optimal for learning whether the surplus is positive.
For the case without breakdowns, the proof is based on the information order of \citet{lehmann1988comparing}.
Each strategy profile induces a statistical experiment about the state, where the realization of the experiment is the random date at which trade would occur if the seller always offered the good.
Waiting for a buyer with the most favorable signal induces the Lehmann-most informative experiment within this class of experiments.
With breakdowns, the proof compares the information loss from trading with other signals versus the gain from avoiding a breakdown by also trading with other signals.

To illustrate the comparison of experiments in terms of Lehmann's order, suppose there are only two possible buyer-signals. Consider an alternative strategy profile under which trade occurs with the first buyer who has either signal and arrives before some cutoff date. This alternative strategy profile simply leads to trade with the first buyer who arrives to the market. Since buyers' arrivals are themselves uninformative about the state, the trading date is then also uninformative about the state. In contrast, the arrival rate of a buyer with the most favorable signal is higher if the state is high and, therefore, if the strategy profile entails waiting for such a buyer, the trading date is informative. Beyond this example, the model allows richer signals than binary ones. Moreover, in many strategy profiles the set of buyer-signals that would trade upon arrival varies from instant to instant. Consequently, not all strategy profiles can be compared in terms of Lehmann-informativeness of the induced trading date distribution. Nevertheless, waiting for the most favorable signal induces the most informative one, which suffices to characterize the surplus maximization problem.

\Cref{thm:commitment_solution} provides a tractable benchmark for how this market would operate efficiently or when the seller has commitment power. The remainder of the paper analyzes equilibrium distortions when the seller lacks commitment power.

The seller not only has initial private information (their type), but also acquires endogenous private information over time since buyers do not observe past prices or calendar time.
Emphasizing this endogenous informational advantage, \Cref{prop:inefficient_delay} shows that, with small breakdown rates, there exists no equilibrium in which the seller obtains the commitment profit in expectation.
To gain intuition, consider a candidate equilibrium that attains the commitment profit.
This equilibrium must maximize the expected surplus.
Waiting for trade is bad news about the state since buyers with the most favorable signal arrive more rapidly in higher states.
In particular, to maximize the expected surplus, the seller must eventually exit the market.
But buyers do not observe when the seller has reached the efficient exit date and, intuitively speaking, are too optimistic about their value for the good at this date.
In particular, some seller-types can obtain a price at which it is profitable to remain in the market.
Hence, such an equilibrium cannot exist.

Given that the commitment solution cannot be supported in equilibrium, many distortions are possible a priori since the strategy profile determines for each instant which seller-types and buyer-signals trade.
Under additional assumptions on the environment and for small breakdown rates, \Cref{prop:opaque_market:eq_example} shows that there is an equilibrium in which only the seller's exit time is distorted.
In this equilibrium, each seller-type posts a constant price and reveals nothing about their type or calendar time.
Moreover, each type waits for a buyer with the most favorable signal.
This waiting strategy is optimal for learning whether the constant price exceeds the seller's reservation value. The proof is again based on Lehmann-informativeness.
With small breakdown rates and since buyers do not observe calendar time or past prices, the learning motive dominates other potential incentives of different seller-types to separate.

In this equilibrium, the exit times differ from the efficient exit times.
Assuming that the seller's value is affine in the buyers' common value, \Cref{prop:comparison_eq_vs_commitment} shows that in equilibrium each seller-type exits inefficiently late. Thus, the equilibrium trade probability is higher than the one induced by surplus-maximizing strategy profiles. To gain intuition, recall that the seller cannot obtain the commitment profit. Hence, the constant price that the seller is posting in equilibrium cannot be too high. This price is derived from the buyers' beliefs about how long the seller has been waiting for trade. Since waiting depresses the buyers' beliefs, the fact that the price cannot be too high requires at least some seller-types to wait for a long time, i.e., exit inefficiently late. The stronger conclusion that all seller-types exit inefficiently late relies on the assumption that the seller's value is affine in the buyers' common value. 

The distortion from \Cref{prop:comparison_eq_vs_commitment} contrasts with settings in which trade is efficient in every state and in which, therefore, the trade probability can only be inefficiently low.
This distortion also suggests that the surplus generated in the market may improve through policies that discourage trade.
\Cref{prop:tax_equilibrium} confirms this intuition for a small tax on successful trades.

\paragraph*{Related literature.}
This paper is related to work on dynamic adverse selection, information aggregation, and pricing with interdependent values.
The main contribution is to take a step towards incorporating uncertainty about whether trade is efficient.

The literature on dynamic adverse selection typically considers lemons markets in which a long-lived informed seller receives offers from short-lived uninformed buyers, and it is efficient to trade in all states.\footnote{See 
\citet{horner2009public}, \citet{zhu2012finding}, \citet{kaya2015transparency}, \citet{fuchs2016transparency}, \citet{asriyan2017information}, \citet{kim2017information}, \citet{kim2017costly}, \citet{kaya2018trading}, \citet{fuchs2022time}, \citet{kaya2022market,kaya2020price,kaya2024repeated}, \citet{aghamolla2023endogenous}, and \citet{aghamolla2023mandatory}. For related models in which prices are set by the seller(s), see, e.g., \citet{kakhbod2020dynamic,kakhbod2024public}, \citet{barsanetti2022signaling}, and \citet{vairo2023transparency}. \citet{lauermann2016search} consider a model of price formation with random proposals. \citet{daley2012waiting,daley2020bargaining}, \citet{hwang2018dynamic}, and \citet{martel2022learning} consider related settings in which an informational asymmetry shrinks or develops over time via exogenous signals.}
With discounting, higher types of the seller have stronger incentives to wait for good offers and, thus, delay helps overcome the static lemons problem.
Instead, in the present setting, the seller delays trade (by waiting for buyers with favorable signals) to learn whether it is efficient to trade.
In the constructed equilibrium, this learning motive dominates other motives that some seller-types might have to separate.
Moreover, whereas in lemons markets with commonly known gains from trade the trade probability can only be inefficiently low, in the present setting the equilibrium trade probability is inefficiently high.

\citet{taylor1999time} considers a two-period setting in which trade may be inefficient and the seller has commitment power. Unlike the present paper, the seller knows the quality of the good, and each buyer's valuation is determined by quality and idiosyncratic taste. Thus, from the seller's perspective, the problem is not one of learning but one of searching for a buyer with high idiosyncratic taste. From the buyers' perspectives, the settings are similar in that buyers grow pessimistic if the good is not sold early.\footnote{See also \citet{martel2018quality} and \citet{preusser2022transparency} for settings with such declining beliefs.}

The literature on information aggregation in decentralized markets typically focuses on settings in which trade is commonly known to be efficient, and information aggregation means whether prices reveal the state (e.g., \citet{wolinsky1990information}; \citet{blouin2001decentralized}; \citet{lauermann2016search}; \citet{asriyan2021aggregation}).
The present paper instead considers settings with uncertainty, and information aggregation is in the sense of allocative efficiency. The state is not fully revealed, even for a patient seller with commitment power.
Most closely related within this literature, \citet{akkar2024mis} shows that, in an environment with an uncertain surplus and in which short-lived buyers offer prices, the expected surplus may converge to the no-information benchmark when there are many buyers.
By contrast, here, the expected surplus does not converge to the no-information benchmark since the long-lived seller offers prices and accounts for the long-run informativeness of the strategy profile.

The literature on dynamic pricing and mechanisms has studied related but distinct problems with learning, such as learning about the distribution of the buyers' valuations (\citet{gershkov2009learning}) or the arrival rate of buyers (\citet{mason2011learning}). Importantly, valuations are private and independent, setting the learning apart from the interdependent value setup studied here. See \citet{pavan2017dynamic} and \citet{bergemann2019dynamic} for surveys of the dynamic mechanism design literature.

In the social learning literature, \citet{bose2006dynamic,bose2008monopoly} consider a dynamic pricing problem with a hidden state and intertemporal frictions.
In this problem, it is challenging to find a notion of informativeness for understanding the seller's trade-off between learning about the state, minimizing buyers' information rents, and minimizing intertemporal frictions. 
The present paper obtains tractable characterizations by focusing on settings with small intertemporal frictions.
Bose et al. also investigate herding dynamics when the seller has multiple units. Such dynamics do not arise here since there is a single unit.

\citet{trefler1993ignorant} analyzes the value of information to a seller, focusing on repeated problems in which one can compare the Blackwell-informativeness of different actions within each period. Here, the problem is not repeated (there is a single unit) and, therefore, what matters is the overall inference across time.

\section{Model}\label{sec:model}

There are a seller with a single unit of a good and countably infinitely many buyers with a common value for the good.

\paragraph*{States and payoffs.}
A hidden state $\omega \in \mathbb{R}$ determines the buyers' common value $v_{B}(\omega)$ and the seller's reservation value $v_{S}(\omega)$.
The (ex post) surplus is $v = v_{B} - v_{S}$.
If the good is traded at price $p$ in state $\omega$ (according to the trade protocol described later), the trading buyer's payoff is $v_{B}(\omega) - p$, and all other buyers' payoffs are $0$. The seller's payoff from trading at price $p$ is $p$, and it is $v_{S}(\omega)$ if the seller does not trade with any buyer. 
Importantly, the state is hidden, and neither buyers nor the seller observe $v_{B}(\omega)$ or $v_{S}(\omega)$ prior to trading.
After normalizing, the seller's payoff from trading may be taken to be $p - v_{S}(\omega)$, while the seller's payoff from not trading is $0$.

The state is distributed according to a Borel probability measure $\mu$ with compact support.
The minimum and maximum, respectively, of the support are denoted $\ubar{\omega}$ and $\bar{\omega}$, respectively, and we denote $\Omega = [\ubar{\omega}, \bar{\omega}]$.

States are labeled such that the buyers' common value $v_{B}$ is strictly increasing.
As a substantive assumption, trade is efficient only in sufficiently high states: 
\begin{assumption}\label{assumption:SC}
The surplus $v$ is strictly single-crossing from below, and there exists $\omega_{0} \in (\ubar{\omega}, \bar{\omega})$ such that $v(\omega_{0}) = 0$; i.e., $v$ is strictly negative on $[\ubar{\omega}, \omega_{0})$ and strictly positive on $(\omega_{0}, \bar{\omega}]$.   
\end{assumption}
The seller's reservation value $v_{S}$ and the surplus $v$ may be non-monotonic.
Let both $v_{B}$ and $v_{S}$ be continuous.

\paragraph*{Information.}
The seller and the buyers observe private signals about the state.

The seller's signal, also called the seller's type, lies in a non-empty, compact interval $Y = [\ubar{y}, \bar{y}] \subseteq \R$.
Conditional on state $\omega$, the type has a density $g(\cdot\vert\omega)$ that is strictly positive on $Y$ and continuous in both arguments. Let $G(\cdot\vert\omega)$ be the associated CDF.

Each buyer's signal lies in a finite set $X$ with $\vert X\vert \geq 2$.
Conditional on state $\omega$, the buyers' signals are independently distributed according to a PMF $f(\cdot\vert \omega)$ and are independent of the seller's type. The PMF $f(\cdot\vert\omega)$ has support $X$ and is continuous in $\omega$.
The assumption that $X$ is finite is discussed in more detail in \Cref{sec:commitment:maximizing_surplus:breakdowns} in the context of a proof sketch for a key lemma.

As a second key assumption, there is a form of positive correlation between the seller's type, the buyers' signals, and the state:
\begin{assumption}
    \label{assumption:correlated}
    The type density $g(y\vert\omega)$ is strictly log-supermodular in $(y, \omega)$.\footnote{That is, if $y > y^{\prime}$ and $\omega>\omega^{\prime}$, then $g(y\vert \omega) g(y^{\prime}\vert \omega^{\prime}) > g(y\vert  \omega^{\prime}) g(y^{\prime}\vert \omega)$.}
    Further, there exists a \emph{most favorable} signal $\bar{x}\in X$ such that
    for all $x\in X\setminus\lbrace\bar{x}\rbrace$ 
    \begin{equation}\label{assumption:highest_signal}
        \frac{f(\bar{x}\vert \omega)}{f(x\vert \omega)} \quad\mbox{is strictly increasing in $\omega$.}
    \end{equation}
\end{assumption}
Strict log-supermodularity of $g$ means that higher types shift beliefs towards higher states in the sense of likelihood-ratio dominance. For intuition, it will sometimes be useful to consider a benchmark in which the seller has no private type $y$ (and then we ignore this part of \Cref{assumption:correlated}).

The second part of \Cref{assumption:correlated} says that there is a buyer-signal $\bar{x}$ that induces the most favorable belief about the state in the sense of likelihood-ratio dominance.
To give an example, suppose $X\subset \R$ and that $f$ is strictly log-supermodular. Then, \Cref{assumption:correlated} holds with $\bar{x} = \max X$. \Cref{assumption:correlated} is more permissive since the beliefs induced by signals in $X\setminus\lbrace\bar{x}\rbrace$ need not be ordered among each other.

For later reference, an important implication of \eqref{assumption:highest_signal} is that the probability $f(\bar{x}\vert \omega)$ is strictly increasing in $\omega$.\footnote{To see this, write $f(\bar{x}\vert \omega) = 1 / (\sum_{x\in X} f(x\vert \omega) / f(\bar{x}\vert \omega))$ and invoke \eqref{assumption:highest_signal}.}

Since trade can be inefficient, not all environments can lead to trade.
For there to be a strategy profile with a strictly positive expected surplus, one can show that it is necessary and sufficient that the posterior surplus be strictly positive conditional on the highest type $\bar{y}$ and the most favorable signal $\bar{x}$. Formally:
\begin{assumption}\label{assumption:trade_is_possible}
    It holds
    $\int_{\Omega} v(\omega) g(\bar{y}\vert\omega) f(\bar{x}\vert \omega)\de\mu(\omega)
    > 0    
    $.
\end{assumption}

\Cref{assumption:SC,assumption:correlated,assumption:trade_is_possible} are maintained throughout the entire paper.

\paragraph*{Arrivals.}
Time is continuous and there is a large finite time horizon $\bar{H}\in\R_{++}$.
The analysis will concern the limit $\bar{H}\to\infty$.
Nothing changes if $\bar{H}$ is finite and sufficiently large since all types of the seller will exit the market by a finite date.

Buyers arrive according to a Poisson process. The arrival rate is normalized to $1$, without loss.
Conditional on arriving, an arbitrary buyer's posterior belief about the state equals the prior $\mu$, and the buyer believes their arrival time to be uniform over $[0, \bar{H}]$.\footnote{To elaborate, consider a discrete time environment with an integer number $n$ of buyers and $n$ time intervals of length $\bar{H} / n$. Nature permutes the $n$ buyers uniformly at random. For all $i$, in the $i$'th time interval the $i$'th buyer of the permutation arrives with probability $\bar{H}/n$, independently across $i$. As $n\to\infty$, buyers' beliefs are as described.}
Each buyer's arrival is unobserved by the seller and all other buyers.

\paragraph*{Trade and breakdowns.}
The market may break down at Poisson rate $\lambda \in \R_{+}$. The arrival of a breakdown is independent of all other random variables in the environment. When a breakdown arrives, the game ends.
As indicated in the introduction, the results apply to small breakdown rates $\lambda$.

At each date before a breakdown, the seller sets a take-it-or-leave-it price and a costless message.
The space of messages is a measurable space $M$ that contains $\R_{+} \times Y $, which lets the seller potentially reveal calendar time and their type.

An arriving buyer observes their private signal, the current price, and the current message (and nothing else).
Then, the buyer decides whether to purchase the good at the current price.
Purchasing ends the game.
A buyer who declines to purchase leaves the market and does not return.
\paragraph*{Strategies.}
To describe the seller's strategy, recall that the seller does not observe buyers' arrivals. Moreover, once the seller's single unit is sold, the game ends. Thus, it suffices to describe the seller's action for each type and date before trading.
Formally, a strategy of the seller is a measurable function $(\beta, p, m)\colon \R_{+} \times Y\to \lbrace 0, 1\rbrace\times [\min v_{B}, \max v_{B}]\times M$.
Here, at time $t$, when the seller's type is $y$, the seller offers the good with probability $\beta(t, y)$ at price $p(t, y)$ (and otherwise does not offer the good) and sends message $m(t, y)$.
Nothing in the results would change if the seller could condition on a privately observed randomization device.

A strategy for the buyers is a measurable function $\alpha\colon X \times \R\times M\to [0, 1]$.
Here, $\alpha(x, p^{\prime}, m^{\prime})$ is the probability with which a buyer with private signal $x$ accepts price $p^{\prime}$ upon seeing message $m^{\prime}$.

Let $\Sigma$ be the set of strategy profiles.
\paragraph*{Expected utilities.}
Fix a strategy profile $\sigma = (\alpha, \beta, p, m)$.
Let $q_{\sigma}(t\vert y, \omega)$ be the Poisson rate at which trade happens in state $\omega$ at time $t$ when the seller's type is $y$:
\begin{equation}\label{eq:model:trade_arrival_rate}
    q_{\sigma}(t\vert y, \omega)
    =  \beta(t, y) \sum_{x\in X} f(x\vert \omega) \alpha(x, p(t, y), m(t, y)).
\end{equation}
In words: a buyer arrives (Poisson rate $1$), the seller offers the good ($\beta(t, y)$), and the buyer draws a signal and accepts (the sum).
Conditional on no breakdown, the probability of trading before a time $t$, denoted $Q_{\sigma}(t\vert y, \omega)$, is given by
\begin{equation*}
    Q_{\sigma}(t\vert y, \omega) = 1 - e^{- \int_{[0, t]} q_{\sigma}(s\vert y, \omega) \de s}
    .
\end{equation*}

Now consider the perspective of a buyer who arrived to the market, but before seeing their private signal and the seller's price and message.
The buyer initially believes calendar time to be uniform over $[0, \bar{H}]$.
Given $(y, \omega)$, the probability that the market has not broken down and the good is still available at a time $t$ equals $e^{-\lambda t - \int_{[0, t]} q_{\sigma}(s\vert y, \omega) \de s}$.
Hence, when accepting according to a strategy $\alpha^{\prime}$, this buyer's expected payoff equals
\begin{align*}
    \int_{\Omega}\int_{Y}\int_{[0, \bar{H}]}
    &(v_{B}(\omega) - p(t, y)) 
    \\
    &\times \beta(t, y) \sum_{x\in X}  f(x\vert \omega)  \alpha^{\prime}(x, p(t, y), m(t, y))
    \\ &\times  
    e^{-\lambda t - \int_{[0, t]} q_{\sigma}(s\vert y, \omega) \de s}
    \de t\de G(y\vert\omega)\de\mu(\omega)
    ,
\end{align*}
where the normalizing constant $1 / \bar{H}$ is dropped.
Let us say $\alpha$ is a \emph{best reply to $\sigma$} if $\alpha$ maximizes the above payoff across all buyer strategies.

When all follow $\sigma$, the buyers' expected utility is denoted $V_{B}(\sigma; \bar{H})$ and given by
\begin{equation*}
    V_{B}(\sigma; \bar{H}) = \int_{\Omega} \int_{Y}  \int_{[0, \bar{H}]} (v_{B}(\omega) - p(t, y))  e^{-\lambda t} \de Q_{\sigma}(t\vert y, \omega)  \de G(y\vert\omega) \de\mu(\omega)
    .
\end{equation*}
The seller's expected utility is denoted $V_{S}(\sigma; \bar{H})$ and given by
\begin{equation*}
    V_{S}(\sigma; \bar{H}) = \int_{\Omega} \int_{Y}  \int_{[0, \bar{H}]} (p(t, y) - v_{S}(\omega))  e^{-\lambda t} \de Q_{\sigma}(t\vert y, \omega)  \de G(y\vert\omega) \de\mu(\omega)
    .
\end{equation*}

As alluded to earlier, the analysis applies to the limit with an unbounded time horizon. Let $V_{B}(\sigma)$ and $V_{S}(\sigma)$, respectively, denote the limits of $V_{B}(\sigma;\bar{H})$ and $V_{S}(\sigma;\bar{H})$, respectively, as $\bar{H}\to\infty$. 
The buyers' best reply notion is also extended in the natural way.
Let $V(\sigma) = V_{B}(\sigma) + V_{S}(\sigma)$ be the \emph{expected surplus}. 

\paragraph*{Solution concept.}
The first part of the paper (\Cref{sec:commitment}) considers the case in which the seller has commitment power and the buyers' best reply is chosen favorably.
In this case, the seller maximizes $V_{S}$ across strategy profiles $\sigma \in\Sigma$ in which the buyer strategy is a best reply to $\sigma$. The \emph{commitment profit} is the value of this problem.
As mentioned, this part will also reveal how to maximize the expected surplus $V$.

The second part of the paper (\Cref{sec:equilibria}) relaxes commitment power.
Given a strategy profile $\sigma$, the strategy $(\beta, p, m)$ of the seller is a \emph{best reply to $\sigma$} if $(\beta, p, m)$ maximizes $V_{S}(\alpha, \cdot)$ across all seller strategies.
A strategy profile $\sigma$ is an \emph{equilibrium} if the seller's strategy and the buyers' strategy are both best replies to $\sigma$.
The results in \Cref{sec:equilibria} shall be more explicit about the supporting beliefs.
The seller's best reply notion already captures the seller's sequential rationality since, given buyers' stochastic unobserved arrivals, the seller expects to reach every date.

\paragraph*{Discussion.} 
The market is opaque: buyers do not observe calendar time or the seller's past actions. This feature has two advantages.
First, the seller with commitment power can reveal calendar time and past actions via messages, and we will see that the seller finds it (weakly) optimal to reveal this information; see \Cref{remark:other_commitment_imlementations}.
Thus, the solution for the opaque market is also a solution for more transparent markets.
Second, the opaque market also admits a tractable analysis when the seller does not have commitment power.
Similar notions of opaqueness are studied by, e.g., \citet{kim2017information,zhu2012finding,lauermann2016search}, though in somewhat different environments.

Individual arrivals are unobservable. In practice, this could mean that one cannot monitor who among those who see an offer are genuinely interested. Further, for a seller with commitment power who announces prices and messages in advance, it may be difficult to contract upon stochastic arrivals.
As an important implication, both with and without commitment, the seller cannot condition prices or messages on the signals of buyers who arrived to the market but chose not to trade. In particular, the seller is constrained to choose price-message pairs as a function of time and the type only. A similar constraint appears in the nearby setting of \citet{bose2008monopoly,bose2006dynamic}, as well as in more distant settings (e.g., \citet{garrett2016intertemporal,dilme2025dynamic,libgober2021informational}).

Intertemporal frictions are modeled as breakdowns. For the seller's utility, Poisson breakdowns have the same effect as exponential discounting. However, breakdowns affect the buyers' inferences about calendar time and, hence, indirectly about the state. In contrast, the buyers would not per se care about the seller's discounting future payoffs. In this way, modeling frictions as breakdowns enables a tractable way to relate the seller's and the buyers' utility in this interdependent value setting.

The seller does not observe the reservation value $v_{S}(\omega)$ prior to trading and, hence, does not learn the state from enjoying flow payoffs while holding the good. For example, the owner of a European-style option can sell the option before maturity but does not receive a payoff while waiting to trade. Similarly, a zero-coupon bond with an uncertain maturity payment generates no interim cash flows. In the example where the good is a house, the state may capture uncertain future developments, such as the construction of a nearby railway line, that affect the future value of the house without affecting or being revealed by the current flow payoff from living in the house.

\paragraph*{Benchmark: Trade is always efficient.}
The key feature of this model is that trade is not efficient in every state.
To have a contrasting benchmark, suppose for a moment that trade is always efficient and the seller's valuation is known to be zero, $v_{B}(\omega) \geq v_{S}(\omega)=0$ for all $\omega$. For simplicity, suppose also that there are no breakdowns, $\lambda = 0$. In this case, one can show that the following behavior is supported in an equilibrium.\footnote{The argument for this claim is a simplification of the argument for the upcoming \Cref{prop:opaque_market:eq_example}.} At all dates, all seller-types post a constant price equal to the prior buyer value $\int_{\Omega} v_{B}(\omega)\de\mu(\omega)$ and send the same (uninformative) message. This price is accepted by a buyer if and only if their private signal is the most favorable signal $\bar{x}$. Conjecturing this behavior, an arriving buyer only infers that they arrived before any buyer with signal $\bar{x}$. If the buyer's own signal is $\bar{x}$, they are indifferent to accepting the constant price and, otherwise, they prefer rejecting. This strategy profile leads to trade with probability one in all states---which is efficient---and the seller extracts the full surplus since buyers are indifferent whenever trading. Thus, there is no gap between the commitment solution and seller-preferred equilibria. Moreover, although the seller learns about the state while waiting for a buyer with signal $\bar{x}$ to arrive, this learning is inconsequential in that the seller never exits and the good is always traded.

When trade is inefficient in some states, we will see that the commitment profit is not attainable in equilibrium. Moreover, the seller's learning about the state matters in that the seller eventually exits the market.

\section{The seller's problem with commitment power}\label{sec:commitment}

This section characterizes the commitment problem in two steps.
In the first and main step, we consider the auxiliary problem of maximizing the expected surplus if the strategy profile could be chosen freely, ignoring incentives.
The expected surplus can be maximized type-by-type, and the resulting maximum is an upper bound on the seller's commitment profit.
Second, we verify that the seller attains the upper bound.

\subsection{Maximizing the expected surplus}\label{sec:commitment:maximizing_surplus}

The following notation is useful.
Let bold letters denote random variables. 
Let $\bm{\tau}_{d}$ be the random arrival time of a breakdown; this has the exponential distribution with rate $\lambda$ and is independent of all other random variables.
Let $\bm{\tau}_{\bar{x}}$ be the random first arrival time of a buyer with signal $\bar{x}$ (an ``$\bar{x}$-buyer''), where we recall that $\bar{x}$ is the particular buyer-signal from \Cref{assumption:correlated}.
Conditional on a type $y$ and a state $\omega$, the time $\bm{\tau}_{\bar{x}}$ has the exponential distribution with rate $ f(\bar{x}\vert \omega)$.
Let $\E$ be the expectation with respect to the joint distribution of $(\bm{\tau}_{d}, \bm{\tau}_{\bar{x}}, \bm{y}, \bm{\omega})$.

The next central lemma shows that, for all sufficiently small breakdown rates $\lambda$, the expected surplus is maximized when each type $y$ of the seller trades with the first $\bar{x}$-buyer who arrives before an exit time $\bar{s}(y)$ (provided no breakdown arrives first); trade is with no other buyer and never after $\bar{s}(y)$.
That is, trade is exactly in the event $\lbrace \bm{\tau}_{\bar{x}} \leq \bar{s}(\bm{y}) \wedge \bm{\tau}_{d}\rbrace$. 
The resulting expected surplus, denoted $\bar{V}$, is given by
\begin{equation}\label{eq:barV_def}
    \bar{V} = \E[v(\bm{\omega})\mathbbm{1}_{\bm{\tau}_{\bar{x}} \leq \bar{s}(\bm{y}) \wedge \bm{\tau}_{d}}].
\end{equation}
For all $y\in Y$, the candidate efficient exit time $\bar{s}(y)$ is defined by
\begin{equation}\label{eq:bars_def}
    \bar{s}(y) = \min\lbrace t\in\R_{+}\colon \E[v(\bm{\omega}) \vert \bm{\tau}_{\bar{x}} = t, \bm{y} = y] \leq 0\rbrace,
\end{equation}
i.e., the first time $t$ at which the posterior surplus is weakly negative conditional on $y$ and on the first $\bar{x}$-buyer arriving at $t$.
\Cref{appendix:posterior_surplus_from_trade} verifies that $\bar{s}$ is well-defined, continuous, and not constantly equal to zero. Moreover, $\bar{s}$ is strictly increasing whenever non-zero.
Thus, there is a cutoff $y_{0} \in [\ubar{y}, \bar{y})$ such that all types $y\in (y_{0}, \bar{y}]$ enter the market---$\bar{s}(y) > 0$---while all types $y\in [\ubar{y}, y_{0})$ abstain---$\bar{s}(y) = 0$.

\begin{lemma}\label{thm:maximizing_surplus}
    There exists $\bar{\lambda} > 0$ such that for all breakdown rates $\lambda\in [0, \bar{\lambda})$ it holds $\bar{V} = \sup_{\sigma\in\Sigma} V(\sigma)$.
\end{lemma}
In this setting, the strategy profile determines which seller-types and buyer-signals trade at each instant, which, in turn, determines the inference when the good does not trade by some date. The strategy profile also determines how long each seller-type waits for such buyer-signals to arrive.\footnote{In contrast, in a static environment with a single buyer, the only sources of information are the seller's type and the buyer's signal. The expected surplus is maximized by trading whenever the posterior surplus is positive conditional on this type and signal. With sequential arrivals, the fact that a buyer with a certain signal did not arrive by some date is an additional source of information.} 
The key assertion of \Cref{thm:maximizing_surplus} is that, to maximize the expected surplus, trade is optimally only with $\bar{x}$-buyers.
To clarify the economic interpretation, recall that the surplus $v$ is single-crossing but not necessarily increasing in the state. Thus, the expected surplus conditional on trade is not necessarily maximized by trading only with $\bar{x}$-buyers since observing signal $\bar{x}$ shifts beliefs towards the highest states.
Instead, trading only with $\bar{x}$-buyers is optimal for learning whether the surplus is positive, as explained next.

\begin{remark}
\Cref{appendix:comparative_statics} provides comparative statics of the maximized surplus $\bar{V}$ with respect to the buyers' and the seller's information structure. The appendix also shows that when the buyers' signals and seller's types are not necessarily conditionally independent, then the exit time $\bar{s}$ may be non-monotonic.
\end{remark}

\subsubsection{The proof idea without breakdowns.}
Let $\lambda = 0$. We shall use the information order of \citet{lehmann1988comparing} to analyze how the strategy profile shapes the inference about the state when the good does not trade.

To maximize the expected surplus, it suffices to consider strategy profiles $\sigma$ such that (i) each type $y$ of the seller offers the good exactly up to a time $T(y) \in\R_{+}\cup\lbrace\infty\rbrace$, and (ii) if the seller would always offer the good, then trade would happen with probability one in all states.
Intuitively, these strategy profiles suffice since the belief about the state is static whenever the seller does not offer the good.

Fix such a strategy profile $\sigma$.
The expected surplus can be maximized separately for each seller-type $y$, so let us also fix $y$.
Let $H_{\sigma}(t\vert \omega)$ be the probability of trading before time $t\in\mathbb{R}_{+}$ in state $\omega\in\Omega$ \emph{if} type $y$ always offered the good:
\begin{equation*}
    H_{\sigma}(t\vert \omega) = 1 - e^{-\int_{[0, t]}  \sum_{x\in X} f(x\vert \omega) \alpha(x, p(s, y), m(s, y)) \de s},
\end{equation*}
where the dependence on the fixed $y$ is suppressed in the notation.
Note that $H_{\sigma}(\cdot\vert \omega)$ is a CDF since, by the choice of $\sigma$, trade is certain if the good is always offered.
Since type $y$ of the seller offers the good up to time $T(y)$ and since there are no breakdowns, the expected surplus conditional on type $y$ is given by\footnote{Here, $\mu(\cdot\vert y)$ means the posterior belief of type $y$, i.e., $\mu(E\vert y) = \int_{E} g(y\vert\omega) \de\mu(\omega) / \int_{\Omega} g(y\vert\omega)\de\mu(\omega)$ for all measurable $E \subseteq \Omega$.\label{footnote:types_posterior_belief}}
\begin{equation*}
    V(\sigma\vert y) = 
    \int_{\Omega}\int_{\R_{+}}
    v(\omega) 
    \mathbbm{1}_{t \leq T(y)}
    \de H_{\sigma}(t\vert \omega) \de\mu(\omega\vert y)
    .
\end{equation*}
Let us re-interpret this integral.
A decision-maker (DM) draws a time $t$ according to the state-dependent CDF $H_{\sigma}(\cdot\vert \omega)$.
The DM then decides whether to ``approve'' (payoff $v$) or ``disapprove'' (payoff $0$), and here approves if and only if $t\leq T(y)$.

This re-interpretation suggests regarding strategy profiles as statistical experiments about the state and asking which strategy profiles induce more informative experiments.
Formally, an \emph{experiment} $\tilde H$ is an ordered collection 
$\langle \tilde H(\cdot\vert\omega) \rangle_{\omega\in\Omega}$ of CDFs.
The given type $y$ and strategy profile induce the experiment $\langle H_{\sigma}(\cdot\vert \omega)\rangle_{\omega\in\Omega}$.
Of particular interest is the experiment $\langle \bar{H}(\cdot\vert \omega)\rangle_{\omega\in\Omega}$ that is induced by a strategy profile under which trade is with the first arriving $\bar{x}$-buyer and no other buyer. Formally, let
\begin{equation*}
    \forall\omega\in\Omega, t\in\R_{+},\quad
    \bar{H}(t\vert \omega) = 1 - e^{-  f(\bar{x}\vert \omega) t}.
\end{equation*}
The following notion of informativeness shall be useful.
An experiment $\hat H$ is \emph{more informative in the sense of \citet{lehmann1988comparing}} than an experiment $\tilde{H}$ if\footnote{This definition departs from \citet{lehmann1988comparing}, who would require that the supremum increases in $\omega$.
Lehmann's definition is tailored towards problems in which the DM takes high actions after high experimental outcomes.
However, here the DM takes the high action of approving trade after low experimental outcomes---$t$ below $T(y)$---suggesting the modified definition.
}
\begin{equation*}
    \forall t\in\R_{+},\quad
    \hat{H}^{-1}(\tilde{H}(t\vert\omega)\vert\omega)
    \quad\mbox{is weakly decreasing in $\omega$,}
\end{equation*}
where $\hat{H}^{-1}(q\vert\omega) = \sup\lbrace s\in\R_{+}\colon \hat{H}(s\vert\omega)\leq q\rbrace$ for all $q \in [0, 1]$.

Not all experiments that are induced by some strategy profile can be ordered by Lehmann-informativeness.
Nevertheless---and this is the key observation---the experiment $\langle \bar{H}(\cdot\vert \omega)\rangle_{\omega\in\Omega}$ is Lehmann-more informative than the experiment $\langle H_{\sigma}(\cdot\vert \omega)\rangle_{\omega\in\Omega}$.
This observation is immediate from \Cref{assumption:correlated}.
Indeed, fix $t\in\R_{+}$.
It holds 
\begin{equation*}
    \bar{H}^{-1}(H_{\sigma}(t\vert \omega)\vert \omega)
     = \int_{[0, t]} \sum_{x\in X} \frac{f(x\vert \omega)}{f(\bar{x}\vert \omega)} \alpha(x, p(s, y), m(s, y)) \de s.
\end{equation*}
By \Cref{assumption:correlated}, for all $x\in X$ the ratio $f(x\vert \omega)/f(\bar{x}\vert \omega)$ decreases in $\omega$.
Thus, $\bar{H}^{-1}(H_{\sigma}(t\vert \omega)\vert \omega)$ decreases in $\omega$, as desired.

A Lehmann-more informative experiment raises utility in certain monotone problems (see e.g., \citet{lehmann1988comparing}; \citet{quah2009comparative}; \citet{kim2023comparing}), suggesting that the DM prefers the experiment $\langle \bar{H}(\cdot\vert \omega)\rangle_{\omega\in\Omega}$ to the experiment induced by any other strategy profile. Equivalently, the expected surplus is maximized by waiting for the first $\bar{x}$-buyer until some exit time, and not trading with any other buyer.
The formal argument, which is directly adapted from the literature on Lehmann's order, proceeds as follows.\footnote{One cannot directly apply existing results since these concern experiments that induce monotone decision rules, i.e., the DM is more eager to approve following smaller draws of $t$. For general strategy profiles $\sigma$, this need not be the case for the experiment $\langle H_{\sigma}(\cdot\vert \omega)\rangle_{\omega\in\Omega}$. This issue turns out to be superficial here. By property (i) above, we may assume the DM follows the monotone approval rule $t\mapsto \mathbbm{1}_{t\leq T(y)}$, which suffices for the argument shown here to go through.}
For all $\omega$ and $t$, let
$\xi(\omega, t) = \bar{H}^{-1}(H_{\sigma}(t\vert \omega)\vert \omega)$.
By a change of variables, we have
\begin{align*}
    V(\sigma\vert y)=
    &\int_{\Omega}\int_{\R_{+}} v(\omega)\mathbbm{1}_{t \leq T(y)} \de H_{\sigma}(t\vert \omega) \de \mu(\omega\vert y)
    \\
    =
    &\int_{\Omega}\int_{\R_{+}} v(\omega)\mathbbm{1}_{t \leq \xi(\omega, T(y))} \de \bar{H}(t\vert \omega) \de \mu(\omega\vert y)
    .
\end{align*}
Fix $t$. Recall that $\omega_{0}$ is the unique state at which $v$ single-crosses zero from below.
Since the experiments are ordered by Lehmann-informativeness, $\xi(\omega, t)$ is weakly decreasing in $\omega$.
Therefore, the difference $\mathbbm{1}_{t \leq \xi(\omega_{0}, T(y))} - \mathbbm{1}_{t \leq \xi(\omega, T(y))}$ is weakly positive if $\omega \geq \omega_{0}$, which is also exactly when $v(\omega)$ is weakly positive. Similarly, the difference is weakly negative whenever $v(\omega)$ is weakly negative. Therefore, all $\omega$ satisfy
    \begin{equation*}
    v(\omega) (\mathbbm{1}_{t \leq \xi(\omega_{0}, T(y))} - \mathbbm{1}_{t \leq \xi(\omega, T(y))})
    \geq
    0
    .
    \end{equation*}
Integrating across $t$ and $\omega$, it follows
\begin{equation*}
    V(\sigma\vert y) \leq 
    \int_{\Omega}\int_{\R_{+}}
    v(\omega)\mathbbm{1}_{t \leq \xi(\omega_{0}, T(y))} \de \bar{H}(t\vert \omega) \de \mu(\omega\vert y)
    .
\end{equation*}
This upper bound is precisely the DM's utility under the experiment $\langle \bar{H}(\cdot\vert \omega)\rangle_{\omega\in\Omega}$ when approving whenever the drawn $t$ lies below the threshold $\xi(\omega_{0}, T(y))$.
Equivalently, the upper bound is the expected surplus when type $y$ of the seller trades with the first $\bar{x}$-buyer who arrives before the exit time $\xi(\omega_{0}, T(y))$ and does not trade with any other buyer or after this time.
It only remains to optimize the exit time, which can be done straightforwardly by differentiating the above upper bound with respect to the exit time.
Different types $y$ differ in the optimal exit time $\bar{s}(y)$ as they hold different beliefs $\mu(\cdot\vert y)$ about the state, but all types optimally trade only with $\bar{x}$-buyers.

\subsubsection{Breakdowns.}\label{sec:commitment:maximizing_surplus:breakdowns}
The proof with breakdowns, $\lambda > 0$, is more difficult.
To trade before a breakdown, there is a benefit from also trading with non-$\bar{x}$-buyers if doing so realizes a positive expected surplus. However, trading with non-$\bar{x}$-buyers hampers efficiency, as sketched above.
Lehmann's order is inapplicable since the experimental draws of $t$ that the DM observes are now directly payoff-relevant.

The proof shows that for small $\lambda$ the gain from avoiding a breakdown is an order of magnitude smaller than the efficiency loss.
Specifically, the gain of avoiding a breakdown admits an upper bound that is proportional to $\lambda$ and the mass of times at which non-$\bar{x}$-buyers trade.
The efficiency loss admits a lower bound that is proportional to the same mass of times and the following rate $L$:
\begin{equation}\label{eq:efficiency_loss_bound_text}
    L =
    \min_{x\in X\setminus\lbrace \bar{x}\rbrace, y\in Y}
    \int_{\Omega} v(\omega) \left(\frac{f(x\vert \omega_{0})}{f(\bar{x}\vert \omega_{0})} - \frac{f(x\vert \omega)}{f(\bar{x}\vert \omega)}\right) f(\bar{x}\vert \omega) \de\mu(\omega\vert y)
    ,
\end{equation}
where $\omega_{0}$ is the state at which $v$ strictly single-crosses $0$ from below, and where we recall that the set $X$ is finite while $Y$ is compact.
The rate $L$ is strictly positive since the difference of the likelihood ratios
\begin{equation*}
    \frac{f(x\vert \omega_{0})}{f(\bar{x}\vert \omega_{0})} - \frac{f(x\vert \omega)}{f(\bar{x}\vert \omega)}
\end{equation*}
is positive if and only if $\omega$ is above $\omega_{0}$, which is also exactly when $v$ is positive.
Intuitively, the difference of the likelihood ratios reflects how much more favorable $\bar{x}$ is about the state than all other signals. The fact that $\bar{x}$ is more favorable than all other signals is what implies that waiting for an $\bar{x}$-signal is optimal for efficiency, as sketched above for the case without breakdowns. 

The key comparison in the proof is then between the relative sizes of $\lambda$ and $L$. 
The conclusion of \Cref{thm:maximizing_surplus} obtains if $\lambda$ is below a critical level $\bar\lambda$ that depends on $L$.
All else equal, the rate $L$ is high if $\bar{x}$-signals arrive quickly---$f(\bar{x}\vert\cdot)$ is high---and are much more favorable than other signals---the difference in likelihood ratios is high.

The rate $L$ tends to vanish if the discrete set of signals $X$ is fine.
Suppose that the buyers' signal PMF is obtained by discretizing a continuous, strictly positive, strictly log-supermodular density on $[0, 1]$ to an even grid of $n\in\mathbb{N}$ points.\footnote{That is, denoting the density by $\phi(\cdot\vert \omega)$, let $X_{n} = \lbrace 1/n, \ldots, (n-1) / n, 1\rbrace$ and $f_{n}(x\vert \omega) = \int_{x-n^{-1}}^{x} \phi(z\vert \omega) \de z$ for all $x\in X_{n}$.} Let $L_{n}$ be the resulting rate \eqref{eq:efficiency_loss_bound_text}.
Then, $L_{n}$ vanishes as $n\to\infty$.
Now let $(\lambda_{n})_{n\in\mathbb{N}}$ be a vanishing sequence of breakdown rates.
Along this sequence of environments, one can show that the conclusion of \Cref{thm:maximizing_surplus} holds for all sufficiently large $n$, provided that
\begin{equation*}
    \lim_{n\to\infty }
    \frac{\lambda_{n}}{L_{n}} = 0
\end{equation*}
holds. That is, if the breakdown rate $\lambda_{n}$ is eventually of smaller order than the efficiency loss rate $L_{n}$, then trading only with the most favorable buyer-signal maximizes the expected surplus.
In the limit with no breakdowns and an absolutely continuous signal distribution, there exist approximate but not exact solutions to the problem of maximizing the expected surplus.

\subsection{The commitment profit}\label{sec:commitment:profit}

This section leverages the characterization of the optimal expected surplus to characterize the seller's commitment profit.
When buyers are best replying, the seller cannot do better than maximizing and extracting the expected surplus.
In many environments, there is a tension between maximizing the surplus and extracting it.
In this environment and for small breakdown rates, there is no tension since the maximized surplus $\bar{V}$ obtains by only trading with $\bar{x}$-buyers (\Cref{thm:maximizing_surplus}).
Specifically, the next theorem shows that the seller extracts $\bar{V}$ via a price that renders $\bar{x}$-buyers indifferent and that is rejected by all other buyers.
\begin{theorem}\label{thm:commitment_solution}
    For all $\lambda\in [0, \bar{\lambda})$, the seller's commitment profit equals $\bar{V}$ as defined in \eqref{eq:barV_def}.
    There is a strategy profile such that the seller's utility equals $\bar{V}$, buyers are best replying, and such that the following hold.
    \begin{itemize}
        \item For all $y\in Y$ and at all times $t\in\R_{+}$, type $y$ of the seller offers the good if and only if $t \leq \bar{s}(y)$, and then at a price of 
        \begin{equation*}
            \bar{p} = \E[v_{B}(\bm{\omega})\vert \bm{\tau}_{\bar{x}} \leq \bar{s}(\bm{y})\wedge \bm{\tau}_{d}].
        \end{equation*}  
        At all times, all types of the seller send the same (uninformative) message.
        \item The price $\bar{p}$ is accepted following a private signal $x$ if and only if $x = \bar{x}$.
    \end{itemize}
\end{theorem}

In light of \Cref{thm:maximizing_surplus}, the proof reduces to verifying that only $\bar{x}$-buyers find it optimal to trade.
Consider a buyer who is offered $\bar{p}$ and sees the seller's uninformative message.
Conjecturing the described behavior of the other buyers, this buyer infers that they arrived at a time before any $\bar{x}$-buyer and breakdown.
The price $\bar{p}$ equals the buyers' posterior value if their own private signal is $\bar{x}$ since, in that case, this buyer represents the first $\bar{x}$-buyer.
Since $\bar{x}$ induces the most favorable belief, the buyer has a best reply of accepting if and only if their private signal is $\bar{x}$.

\begin{remark}\label{remark:other_commitment_imlementations}
    There are other strategy profiles that implement the commitment profit, including ones where the seller reveals calendar time and their type.
    For example, for each $t$ and $y$ define $\bar{p}(t, y) = \E[v_{B}(\bm{\omega})\vert \bm{\tau}_{\bar{x}} = t, \bm{y} = y]$ as the posterior buyer value conditional on type $y$ and the first $\bar{x}$-buyer's arriving at time $t$. Suppose at each instant $t$ each type $y$ posts price $\bar{p}(t, y)$ (unless $t > \bar{s}(y)$) and also reveals the initial price $\bar{p}(0, y)$ via a message. The initial price $\bar{p}(0, y)$ can be shown to be strictly increasing in $y$, while for fixed $y$ the price $\bar{p}(t, y)$ is strictly decreasing in $t$. Thus, buyers can infer calendar time and the seller's type. In particular, a buyer who infers $(t, y)$ finds it optimal to accept $\bar{p}(t, y)$ if and only if their signal is $\bar{x}$. 
\end{remark}

\section{Equilibria}\label{sec:equilibria}
This section investigates the consequences of relaxing the seller's commitment power.
We will see that the seller cannot obtain the commitment profit in equilibrium.
There is an equilibrium which looks qualitatively similar to the commitment solution in that only $\bar{x}$-buyers are induced to trade and the seller eventually exits the market, but the exit times differ from the efficient exit times $\bar{s}$.

\subsection{The commitment profit is unattainable in equilibrium}
\begin{theorem}\label{prop:inefficient_delay}
    If $\lambda < \bar{\lambda}$, then there does not exist an equilibrium in which the seller's utility equals the commitment profit $\bar{V}$.
\end{theorem}
To explain this result, first consider the benchmark where the seller has no private type $y$ and there are no breakdowns.
Let us sketch why no equilibrium supports the behavior described in \Cref{thm:commitment_solution}, adapted to the case without a private seller-type. That is, suppose the seller stays on the market up to an exit time $\bar{s}\in\R_{++}$ while offering a constant price of $\bar{p} = \E[v_{B}(\bm{\omega})\vert \bm{\tau}_{\bar{x}} \leq \bar{s}]$, and buyers accept $\bar{p}$ if and only if their private signal is $\bar{x}$.
Since $\bar{s}$ maximizes the expected surplus, the posterior surplus conditional on trading at $\bar{s}$ must equal zero, $\E[v_{B}(\bm{\omega})\vert \bm{\tau}_{\bar{x}} = \bar{s}] = \E[v_{S}(\bm{\omega})\vert \bm{\tau}_{\bar{x}} = \bar{s}]$, since else it would be efficient to exit slightly earlier or later.
However, the price $\bar{p}$ equals the posterior buyer value conditional on trading \emph{before} $\bar{s}$.
Since $\bar{x}$-buyers arrive more rapidly in high states, the event $\lbrace\bm{\tau}_{\bar{x}} = \bar{s}\rbrace$ implies a strictly lower posterior buyer value than the event $\lbrace\bm{\tau}_{\bar{x}} \leq \bar{s}\rbrace$, i.e., $\bar{p} = \E[v_{B}(\bm{\omega})\vert \bm{\tau}_{\bar{x}} \leq \bar{s}] > \E[v_{B}(\bm{\omega})\vert \bm{\tau}_{\bar{x}} = \bar{s}]$.
Thus, also $\bar{p} > \E[v_{S}(\bm{\omega})\vert \bm{\tau}_{\bar{x}} = \bar{s}]$.
Consequently, the seller cannot resist the temptation to keep the good on the market at price $\bar{p}$ at time $\bar{s}$.

When the seller has a private type, the argument for \Cref{prop:inefficient_delay} is more nuanced. While \Cref{prop:inefficient_delay} asserts that the seller cannot obtain $\bar{V}$ in expectation across all types, some types may benefit from a lack of commitment, as we will see in \Cref{sec:opaque_markets:comparison}.
Moreover, equilibria may conceivably take more complex forms than the one assumed in the above sketch.
Consider a candidate equilibrium that attains the commitment profit.
Since the commitment profit entails extracting the maximized expected surplus, one can show that it is necessary that only $\bar{x}$-buyers are induced to trade, and that each seller-type $y$ exits at time $\bar{s}(y)$. Thus, there must be a price-message pair $(\tilde p, \tilde m)$ that is only accepted by $\bar{x}$-buyers. The proof then shows that some types of the seller have a profitable deviation of the form sketched above, i.e., these types $y$ constantly offer $(\tilde p , \tilde m)$ but stay on the market beyond the efficient exit time $\bar{s}(y)$.

\Cref{prop:inefficient_delay} uses that in this opaque market buyers do not observe calendar time or past prices.
Reconsider the case without a private seller-type.
At time $\bar{s}$ there is a gap between the price $\bar{p}$ that the seller can charge to $\bar{x}$-buyers and the posterior buyer value conditional on trading at time $\bar{s}$. This gap arises since the inference from waiting for an $\bar{x}$-buyer depends on the waiting duration, but buyers do not observe calendar time. To gain a contrasting perspective, let us informally consider a market without breakdowns in which calendar time and all past prices are public (but the seller cannot commit to prices). Since the seller has no private type and the market is transparent, the seller has no informational advantage over buyers. In particular, the buyers' expected utility must be weakly positive even if the seller deviates from some conjectured strategy and, therefore, any deviation yields the seller at most the commitment profit. One can show that this market has an equilibrium in which the seller obtains the commitment profit. 

\subsection{An equilibrium with passive beliefs}\label{sec:no_commitment:opaque_markets:equilibria}
Given that the commitment profit is unattainable, what do equilibria look like?
Many distortions are possible a priori since different seller-types could offer different prices and messages, thereby trading with different sets of buyer-signals at each instant.
The next result characterizes a tractable equilibrium that is qualitatively similar to the commitment solution: each seller-type trades with the first $\bar{x}$-buyer who arrives before an exit time $s^{\ast}(y)$, supported by posting a price $p^{\ast}$ that does not depend on the type or calendar time. This equilibrium highlights the distortion on a single margin---the exit times.
We will compare the exit times in \Cref{sec:opaque_markets:comparison}.

The existence proof uses the following assumption.
\begin{assumption}\label{assumption:regularity_for_equilibrium}
    The seller's reservation value $v_{S}(\omega)$ is strictly decreasing in $\omega$. It holds $\E[v_{B}(\bm{\omega})] < \max_{\omega\in \Omega} v_{S}(\omega)$.
\end{assumption}
As explained below, strict monotonicity of the seller's reservation value $v_{S}$ implies that in the constructed equilibrium the seller faces a non-trivial learning problem, while the inequality implies that the seller eventually exits the market.

\begin{theorem}\label{prop:opaque_market:eq_example}
    Let \Cref{assumption:regularity_for_equilibrium} hold.
    There exists $\lambda^{\ast} > 0$ such that for all breakdown rates $\lambda\in [0, \lambda^{\ast})$
    there exists a price $p^{\ast}\in\R$, exit times $s^{\ast}\colon Y\to \R_{+}$, and an equilibrium such that:
    \begin{enumerate}
            \item For all $y\in Y$ and at all times $t\in\R_{+}$, type $y$ of the seller offers the good if and only if $t \leq s^{\ast}(y)$, and then at a price of $p^{\ast}$. At all times, all types of the seller send the same (uninformative) message.
            The function $s^{\ast}$ is continuous, not constantly zero, and strictly increasing whenever non-zero.
            \item 
            The price $p^{\ast}$ is accepted following a private signal $x$ if and only if $x = \bar{x}$.
            Prices strictly above $p^{\ast}$ are never accepted.
            A general price is accepted following a private signal $x$ if and only if the price is weakly less than
            \begin{equation}
            \label{eq:prop_opaque_market:eq_example:accept_deviation}
                \frac{\int_{\Omega}\int_{Y}\int_{[0, s^{\ast}(y)]} v_{B}(\omega) f(x\vert \omega) e^{-(\lambda +  f(\bar{x}\vert \omega)) t} \de t\de G(y\vert\omega)\de\mu(\omega)}
                {\int_{\Omega}\int_{Y}\int_{[0, s^{\ast}(y)]} f(x\vert \omega) e^{-(\lambda +  f(\bar{x}\vert \omega)) t} \de t\de G(y\vert\omega)\de\mu(\omega)}
                .
            \end{equation}
        \end{enumerate}
        Moreover, the price $p^{\ast}$ and the exit times $s^{\ast}$ satisfy
        \begin{subequations}
        \begin{align}
            \label{eq:opaque_equilibrium_price}
            p^{\ast} &= \E[v_{B}(\bm{\omega})\vert \bm{\tau}_{\bar{x}} \leq s^{\ast}(\bm{y})\wedge \bm{\tau}_{d}];
            \\
            \label{eq:opaque_equilibrium_stopping}
            \forall y\in Y, \quad
            s^{\ast}(y) &= \min\lbrace t\in\R_{+}\colon \E[(p^{\ast} - v_{S}(\bm{\omega}))\vert \bm{\tau}_{\bar{x}} = t, \bm{y} = y] \leq 0\rbrace.
        \end{align}
        \end{subequations}
\end{theorem}
\paragraph*{The buyers' incentives.}
The strategy of the buyers is supported by passive beliefs: the belief when offered $p^{\ast}$ equals the belief when offered every other price.
Precisely, the value in \eqref{eq:prop_opaque_market:eq_example:accept_deviation} equals the posterior value of a buyer with private signal $x$ who infers that they arrived before another $\bar{x}$-buyer and a breakdown.
The posterior value is maximized at $x = \bar{x}$, and it equals $p^{\ast}$ if $x = \bar{x}$.
Hence, buyers are best replying.

\paragraph*{The seller's incentives.}
The seller faces the problems of trading at high prices and of learning about the state.
Deviating from the candidate strategy makes the seller worse off in both problems.
Buyers never accept a price strictly above $p^{\ast}$.
Therefore, a deviation of the seller yields a state-dependent payoff of at most $p^{\ast} - v_{S}$.
Now think of $p^{\ast} - v_{S}$ as the payoff of a fictitious planner who freely chooses for which seller-types and buyer-signals the good is traded, as in the earlier analysis of surplus-maximizing strategy profiles (\Cref{thm:maximizing_surplus}).
Since $v_{S}$ is strictly decreasing in the state (\Cref{assumption:regularity_for_equilibrium}), the payoff $p^{\ast} - v_{S}$ single-crosses zero from below.
Thus, paralleling \Cref{thm:maximizing_surplus}, the fictitious planner optimally trades with the first $\bar{x}$-buyer who arrives before a type-dependent exit time $s^{\ast}$, provided $\lambda$ is small. 

Market opaqueness is important for this argument.
If buyers observed calendar time, prices would presumably vary over time as buyers update about the state.
By waiting for an $\bar{x}$-buyer, the seller risks trading at a late date where prices may be low.

The inequality $\E[v_{B}(\bm{\omega})] < \max_{\omega\in \Omega} v_{S}(\omega)$ is used to prove existence of a price $p^{\ast}$ that is consistent with buyers' beliefs given $s^{\ast}$ (i.e., \eqref{eq:opaque_equilibrium_price}) and such that $s^{\ast}$ is optimal for the seller given $p^{\ast}$ (i.e., \eqref{eq:opaque_equilibrium_stopping}).
This inequality requires the prior $\mu$ not to be too favorable. 
If instead $\E[v_{B}(\bm{\omega})] \geq \max_{\omega\in \Omega} v_{S}(\omega)$ holds, similar arguments show that for small breakdown rates $\lambda$ there is an equilibrium in which all types of the seller stay in the market forever, offering a price equal to $\E[v_{B}(\bm{\omega})\vert \bm{\tau}_{\bar{x}} \leq \bm{\tau}_{d}]$.

\paragraph*{Equilibrium uniqueness.}
Equilibrium need not be unique as unexpected price offers could be punished via extreme beliefs.
However, when $v_{S}$ is strictly decreasing and $\lambda=0$, in all equilibria in which belief punishments are ``not too harsh,'' almost surely the seller trades with an $\bar{x}$-buyer or does not trade at all; for a precise statement, see \Cref{appendix:opaque_eq:uniqueness_within_class}.
The equilibria described by \Cref{prop:opaque_market:eq_example} are then natural focal equilibria. (\Cref{prop:opaque_market:eq_example} does not assert there is a unique such equilibrium.)

\subsection{Inefficiently late exits}\label{sec:opaque_markets:comparison}
In the equilibrium described by \Cref{prop:opaque_market:eq_example}, the seller's utility must be strictly less than under commitment (by \Cref{prop:inefficient_delay}).
The buyers' utility is zero since the only buyers who trade are indifferent.
Thus, the expected surplus must also be strictly less than the maximum $\bar{V}$.
This section seeks to identify the source of this inefficiency.
We will obtain the strongest result when both $v_{B}$ and $v_{S}$ are affine functions of the state, an assumption discussed further ahead. 
In this case, we will see that all seller-types exit the market inefficiently late in equilibrium. Moreover, if some types efficiently abstain from entering, some of these types inefficiently enter the market.
Recall that efficiently all types above a threshold $y_{0}$ enter the market (possibly, all types enter, $y_{0} = \ubar{y}$).
\begin{proposition}\label{prop:comparison_eq_vs_commitment}
    Let \Cref{assumption:regularity_for_equilibrium} hold.
    Let $\lambda < \min\lbrace\bar{\lambda}, \lambda^{\ast}\rbrace$.
    Let $s^{\ast}$ be a function that meets the conclusion of \Cref{prop:opaque_market:eq_example} for some price and equilibrium.
    If both $v_{B}$ and $v_{S}$ are affine functions of the state on $\supp \mu$, then $s^{\ast}(y) > \bar{s}(y)$ for all $y\in [y_{0}, \bar{y}]$ (and $s^{\ast}(y) \geq \bar{s}(y) = 0$ for all $y\in [\ubar{y}, y_{0})$).
    For possibly non-affine $v_{B}$ and $v_{S}$, 
    it holds $s^{\ast}(y) > \bar{s}(y)$ for all $y$ in a non-empty open set.
\end{proposition}
When all types exit inefficiently late, the trade probability is strictly higher in equilibrium than when maximizing the expected surplus. This result contrasts with settings in which trade is efficient in every state and, therefore, the equilibrium trade probability can only be distorted downwards.

To explain \Cref{prop:comparison_eq_vs_commitment}, consider the benchmark in which the seller has no private type.
In equilibrium and under the commitment solution, respectively, the seller exits at times $s^{\ast}\in\R_{+}$ and $\bar{s}\in\R_{+}$, respectively.
Let us argue that the equilibrium exit time is inefficiently high, $s^{\ast} > \bar{s}$.
Towards a contradiction, suppose $s^{\ast}\leq \bar{s}$.
Since trading late depresses beliefs, the equilibrium price $p^{\ast} = \E[v_{B}(\bm{\omega})\vert \bm{\tau}_{\bar{x}} \leq s^{\ast}\wedge \bm{\tau}_{d}]$ is then weakly greater than $\bar{p} = \E[v_{B}(\bm{\omega})\vert \bm{\tau}_{\bar{x}} \leq \bar{s}\wedge\bm{\tau}_{d}]$, the price from \Cref{thm:commitment_solution} that can be used to obtain the commitment profit.
Intuitively, the equilibrium price $p^{\ast}$ is too high. Specifically, since $p^{\ast}\geq\bar{p}$, if the seller deviates from the equilibrium strategy by exiting only at time $\bar{s}$ while offering $p^{\ast}$, the seller obtains a utility of at least
\begin{equation*}
    \E[(\bar{p} - v_{S}(\bm{\omega}))\mathbbm{1}_{\bm{\tau}_{\bar{x}} \leq \bar{s} \wedge \bm{\tau}_{d}}]
    = 
    \E[v_{B}(\bm{\omega})\mathbbm{1}_{\bm{\tau}_{\bar{x}} \leq \bar{s} \wedge \bm{\tau}_{d}}] - \E[v_{S}(\bm{\omega})\mathbbm{1}_{\bm{\tau}_{\bar{x}} \leq \bar{s} \wedge \bm{\tau}_{d}}]
    = \bar{V}.
\end{equation*}
The equilibrium utility is at least the utility from this deviation and, hence, at least the commitment profit $\bar{V}$.
This contradicts \Cref{prop:inefficient_delay}.
Without a private seller-type, the argument does not require $v_{B}$ or $v_{S}$ to be affine.

When the seller has a private type, the argument is more nuanced since the posterior buyer values conditional on trading in equilibrium and under commitment, respectively, depend on the entire profile of exit times, $s^{\ast}$ and $\bar{s}$, respectively.
Therefore, the prices $p^{\ast}$ and $\bar{p}$ cannot be compared as easily.
As a second difficulty, \Cref{prop:inefficient_delay} does not rule out that some types get strictly more than their commitment profit. Indeed, if some types efficiently abstain---$y_{0} > \ubar{y}$---then the lowest types $[\ubar{y}, y_{0}]$ all have a commitment profit of zero. However, \Cref{prop:comparison_eq_vs_commitment} implies that types near $y_{0}$ enter the market in equilibrium and make a profit.

The general argument proceeds as follows.
When both $v_{B}$ and $v_{S}$ are affine, one can show that $s^{\ast}$ is distorted away from $\bar{s}$ in the same direction for all types.
Towards a contradiction, suppose $s^{\ast}(y) \leq \bar{s}(y)$ for all $y$.
Let $\bar{\phi}(y) = \E [v_{B}(\bm{\omega})\vert \bm{\tau}_{\bar{x}} \leq \bar{s}(y) \wedge \bm{\tau}_{d}\vert\bm{y}=y]$ be the posterior buyer value conditional on type $y$'s trading under commitment. Intuitively, $\bar{\phi}(y)$ is the price that would yield type $y$ their commitment profit if type $y$ were the only type in the market.
On the one hand, the equilibrium price $p^{\ast}$ cannot be too high: since the seller does not obtain their commitment profit in expectation, one can show that $p^{\ast} < \bar{\phi}(y)$ must hold for some types $y$.
On the other hand, since early exits ($s^{\ast} \leq \bar{s}$) raise the posterior buyer value, it cannot be that $p^{\ast} < \bar{\phi}(y)$ holds for all $y$.
Using that $p^{\ast} < \bar{\phi}(y)$ holds strictly for some but not all types, one can show that also $p^{\ast} > \bar{\phi}(y)$ holds strictly for at least one type $y$.
For this type $y$, the equilibrium exit time $s^{\ast}(y)$ must be strictly above the efficient exit time $\bar{s}(y)$, contradicting the assumption $s^{\ast} \leq \bar{s}$.
When $v_{B}$ and $v_{S}$ are non-affine, this argument delivers that at least some types must exit inefficiently late in equilibrium.

Both $v_{B}$ and $v_{S}$ are trivially affine on $\supp\mu$ if the state is binary. Generally, it is without loss to redefine the state as the buyers' common value. Hence, the content of assuming affinity is that the seller's value is affine in the buyers' common value.

\subsection{Encouraging early exits}
\Cref{prop:comparison_eq_vs_commitment} suggests that the equilibrium surplus can be raised via policies that encourage the seller to exit earlier. Let us briefly discuss a tax as one such policy.

For fixed tax $\psi \in \R_{++}$, when trade occurs at price $p$ in state $\omega$, the seller's payoff is $p - \psi - v_{S}(\omega)$ and the trading buyer's payoff is $v_{B}(\omega) - p$. When not trading, all payoffs are $0$, as before.
An equilibrium with tax $\psi$ is defined analogously to an equilibrium without tax.
For the sake of simplicity, suppose there are no breakdowns---$\lambda = 0$---and under commitment all types enter the market---$\bar{s}(\ubar{y}) > 0$.

If $\psi$ is sufficiently close to $0$ and \Cref{assumption:regularity_for_equilibrium} holds, one can show there is an equilibrium in which, again, each type $y$ of the seller posts a constant price $p^{\ast}_{\psi}$ and trades with the first $\bar{x}$-buyer who arrives before some exit time $s^{\ast}_{\psi}(y)$.
As before, the price $p^{\ast}_{\psi}$ renders $\bar{x}$-buyers indifferent.
The resulting expected surplus from trading is $\E[v(\bm{\omega}) \mathbbm{1}_{\bm{\tau}_{\bar{x}} \leq s^{\ast}_{\psi}(\bm{y})}]$ (since there are no breakdowns).
The next result shows that this surplus is raised by a small tax. If one regards taxes as welfare-neutral transfers, it follows that also the overall welfare is raised by such a tax.
\begin{proposition}\label{prop:tax_equilibrium}
    Let \Cref{assumption:regularity_for_equilibrium} hold.
    Let both $v_{B}$ and $v_{S}$ be affine functions of the state on $\supp\mu$.
    Let $\lambda = 0$ and $\bar{s}(\ubar{y}) > 0$.
    There exists $\psi \in \R_{++}$ and an equilibrium with tax $\psi$ in which the expected surplus $\E[v(\bm{\omega}) \mathbbm{1}_{\bm{\tau}_{\bar{x}} \leq s^{\ast}_{\psi}(\bm{y})}]$ is strictly higher than in every equilibrium without a tax that is of the form described by \Cref{prop:opaque_market:eq_example}.
\end{proposition}
The intuition is straightforward. Assuming for a moment that the seller constantly posts a fixed price before exiting, imposing a tax induces the seller to exit earlier since the tax shrinks the set of states in which trade is profitable. Since the exit times without tax are inefficiently high (\Cref{prop:comparison_eq_vs_commitment}), slightly reducing exit times raises the expected surplus.
The full argument is more cumbersome since the seller's exits affect each buyer's inference from being offered the good (see \eqref{eq:opaque_equilibrium_price}) and, hence, which fixed price the seller can constantly post to render $\bar{x}$-buyers indifferent.

\appendix

\numberwithin{equation}{section}
\numberwithin{figure}{section}

\section{Proofs: commitment}

\subsection{A fundamental single-crossing lemma}
The following lemma is known at least since \citet{karlin1956theory}.
The arguments that establish the lemma will be used again and again.
\begin{lemma}\label{lemma:basic_SCP}
    Let $h\colon\Omega\to\R$ be continuous and single-crossing from below.
    Let $\psi\colon\Omega\times \R \to\R_{++}$ be continuous and log-supermodular.
    Then $r\mapsto \int_{\Omega} h(\omega)\psi(\omega, r) \de\mu(\omega)$ is single-crossing from below.
\end{lemma}
\begin{proof}[Proof of \Cref{lemma:basic_SCP}]
    Assume $h$ changes sign, as otherwise there is nothing to prove.
    Let $\omega_{0}^{\prime} = \sup\lbrace\omega\in\Omega\colon h(\omega)\leq 0\rbrace$.
    Take $r, r^{\prime}\in \R$ such that $r > r^{\prime}$.
    Since $h$ is single-crossing from below and $\psi$ is log-supermodular, it holds 
    \begin{equation*}
        h(\omega)\left(\frac{\psi(\omega, r)}{\psi(\omega, r^{\prime})} - \frac{\psi(\omega_{0}^{\prime}, r)}{\psi(\omega_{0}^{\prime}, r^{\prime})}\right) \geq 0
    \end{equation*}
    for all $\omega$.
    Hence,
    \begin{equation*}
        \int_{\Omega} h(\omega)\psi(\omega, r) \de\mu(\omega)
    \geq
    \frac{\psi(\omega_{0}^{\prime}, r)}{\psi(\omega_{0}^{\prime}, r^{\prime})}  \int_{\Omega} h(\omega)\psi(\omega, r^{\prime}) \de\mu(\omega),
    \end{equation*}
    which implies the single-crossing property.
\end{proof}

\subsection{The posterior surplus and optimal exit times}\label{appendix:posterior_surplus_from_trade}

This appendix derives basic properties of the exit time $\bar{s}$. We recall the definition $\bar{s}(y) = \min\lbrace t\in\R_{+}\colon \E [v(\bm{\omega})\vert \bm{\tau}_{\bar{x}} = t, \bm{y} = y] \leq 0\rbrace$ for all $y\in Y$.
For all $(t, y)$, the posterior surplus $\E [v(\bm{\omega})\vert \bm{\tau}_{\bar{x}} = t, \bm{y} = y]$ has the same sign as $\hat{v}(t, y)$, where
\begin{equation}\label{eq:posterior_gains_definition}
    \hat{v}(t, y)
    =
    \int_{\Omega} v(\omega) g(y\vert\omega) f(\bar{x}\vert \omega) e^{- f(\bar{x}\vert \omega) t} \de \mu(\omega)
    .
\end{equation}
\begin{lemma}\label{lemma:hatv_properties}
    The exit time $\bar{s}(y)$ is well-defined, continuous, weakly increasing, and there exists $y_{0}\in[\ubar{y}, \bar{y})$ such that $\bar{s}(y) = 0$ for all $y \in [\ubar{y}, y_{0})$ and $\bar{s}(y) > 0$ for all $y\in (y_{0}, \bar{y}]$.
\end{lemma}

\begin{proof}[Proof of \Cref{lemma:hatv_properties}]
    We make two preliminary observations.
    First, since $\omega\mapsto f(\bar{x}\vert \omega)$ is strictly increasing (as noted in \Cref{sec:model}), $e^{- f(\bar{x}\vert \omega) t}$ is strictly log-submodular in $(\omega, t)$.
    Combined with the fact that $v$ strictly single-crosses $0$ from below, it follows that $t\mapsto\hat{v}(t, y)$ strictly single-crosses $0$ from above for any fixed $y$.
    Second, since $g$ is strictly log-supermodular while $v$ strictly single-crosses $0$ from below, the function $y\mapsto \hat{v}(t, y)$ strictly single-crosses $0$ from below for any fixed $t$.

    The sign of $\hat{v}(t, y)$ equals the sign of 
    \begin{equation*}
        \int_{\Omega} v(\omega) g(y\vert\omega) f(\bar{x}\vert \omega) e^{ (f(\bar{x}\vert \omega_{0}) - f(\bar{x}\vert \omega)) t} \de \mu(\omega)
        ,
    \end{equation*}
    where we recall that $v$ strictly single-crosses $0$ from below at state $\omega_{0}$.
    Since $\omega\mapsto f(\bar{x}\vert \omega)$ is strictly increasing, as $t\to\infty$ the term $e^{ (f(\bar{x}\vert \omega_{0}) - f(\bar{x}\vert \omega)) t}$ vanishes if $\omega > \omega_{0}$, and diverges to $\infty$ if $\omega < \omega_{0}$.
    Using also that $v$ is continuous and strictly single-crosses $0$ from below at $\omega_{0}$, it follows $\hat{v}(t, y) < 0$ for all sufficiently large $t$.

    For all $y$, since $\hat{v}(t, y) < 0$ for sufficiently large $t$, the set in the definition of $\bar{s}(y)$ is non-empty. The minimum over this set is attained by continuity of $g$ and $f$, and so $\bar{s}$ is well-defined.
    Continuity of $\bar{s}$ is now inherited from continuity of $\hat{v}$, and the fact that $t\mapsto\hat{v}(t, y)$ is strictly single-crossing for fixed $y$.
    
    Finally, since $y\mapsto \hat{v}(t, y)$ is strictly single-crossing from below for any fixed $t$, it follows that $\bar{s}$ is increasing, strictly so if non-zero.
    \Cref{assumption:trade_is_possible} asserts $\hat{v}(0, \bar{y}) > 0$, implying $\bar{s}(\bar{y}) > 0$.
    Since $\bar{s}$ is increasing, there exists $y_{0} \in [\ubar{y}, \bar{y})$ as desired.
\end{proof}

\subsection{Maximizing the expected surplus: proof of \headercref{Lemma}{{thm:maximizing_surplus}}}\label{appendix:maximizing_surplus}
\Cref{thm:maximizing_surplus} follows from the slightly more general \Cref{lemma:delta_maximization} below (apply that lemma with $u_{n} = v$ for all $n$).
The added generality is useful for the equilibrium analysis.

We consider the following problem. 
Fix a sequence $(\lambda_{n}, u_{n})_{n\in\mathbb{N}}$, where $\lambda_{n}$ represents a breakdown rate and $u_{n}$ a surplus from trading. 
Given a strategy profile $\sigma\in\Sigma$ and type $y\in Y$, define the expected surplus
\begin{equation*}
    U_{n}(\sigma, y) = \int_{\Omega}\int_{\R_{+}} u_{n}(\omega) e^{-\lambda_{n} t} \de Q_{\sigma}(t\vert y, \omega) \de\mu(\omega\vert y),
\end{equation*}
where $Q_{\sigma}(t\vert y, \omega) = 1 - e^{-\int_{[0, t]} q_{\sigma}(s\vert y, \omega)\de s}$ is the probability of trading before time $t$ given $\omega$, and where $\mu(\cdot\vert y)$ is type $y$'s posterior (see \Cref{footnote:types_posterior_belief}).
Let $Q_{\sigma}(y, \omega) = \lim_{t\to\infty} Q_{\sigma}(t\vert y, \omega)$ be the ex ante trade probability in state $\omega$.

The next lemma provides conditions under which each $U_{n}(\cdot, y)$ is maximized by waiting for an $\bar{x}$-buyer up to a specific exit date.
To that end, let $\bar{Q}(t\vert \omega) = 1 - e^{- f(\bar{x}\vert \omega) t}$ be the probability that an $\bar{x}$-buyer arrives before $t$ given $\omega$.
Given $s\in\R_{+}\cup\lbrace\infty\rbrace$, let
\begin{equation*}
    \bar{U}_{n}(s, y) = \int_{\Omega}\int_{\R_{+}} u_{n}(\omega) e^{-\lambda_{n} t} \mathbbm{1}_{t \leq s} \de \bar{Q}(t\vert \omega) \de\mu(\omega\vert y)
\end{equation*}
be the expected surplus when type $y$ waits for the first $\bar{x}$-buyer until time $s$.

\begin{lemma}\label{lemma:delta_maximization}
    Let $(\lambda_{n})_{n\in\mathbb{N}}$ be a sequence in $\R_{+}$ converging to $0$.
    For all $n$, let $u_{n}\colon\Omega\to\R$ be a continuous function that is strictly single-crossing from below at a state $\omega_{0, n}^{\dagger} \in (\ubar{\omega}, \bar{\omega})$.
    Suppose there is $M\in\R$ such that $\vert u_{n}(\omega)\vert \leq M$ for all $n$ and $\omega$.
    Suppose $(u_{n})_{n\in\mathbb{N}}$ converges pointwise to a continuous function $u\colon\Omega\to\R$, and $(\omega_{0, n}^{\dagger})_{n\in\mathbb{N}}$ converges to $\omega_{0}^{\dagger} \in (\ubar{\omega}, \bar{\omega})$ such that $u$ is strictly single-crossing from below at $\omega_{0}^{\dagger}$.
    For all $y\in Y$ and $n\in\mathbb{N}$, let\footnote{For all $n$, both $s^{\dagger}_{n}$ and $s^{\dagger}$ are well-defined and continuous, by the same arguments as in \Cref{lemma:hatv_properties}.}
    \begin{align*}
        s^{\dagger}_{n}(y) =& \min\lbrace t\in\R_{+}\colon \E[u_{n}(\bm{\omega})\vert \bm{\tau}_{\bar{x}} = t, \bm{y} = y] \leq 0\rbrace,
        \\
        s^{\dagger}(y) =& \min\lbrace t\in\R_{+}\colon \E[u(\bm{\omega})\vert \bm{\tau}_{\bar{x}} = t, \bm{y} = y] \leq 0\rbrace
        .
    \end{align*}
    Suppose $(s^{\dagger}_{n})_{n\in\mathbb{N}}$ converges uniformly to $s^{\dagger}$.

    Then, there exists $n^{\dagger}$ such that for all $n\geq n^{\dagger}$ and $y\in Y$, it holds 
    \begin{equation}\label{eq:eq_delta_best_reply:objective}
         \sup_{\sigma\in\Sigma} U_{n}(\sigma, y) 
         =
         \bar{U}_{n}(s^{\dagger}_{n}(y), y)
        .
    \end{equation}
\end{lemma}

\begin{proof}[Proof of \Cref{lemma:delta_maximization}]
\setcounter{step}{0}

To derive several bounds, it helps to focus on strategy profiles that do not induce unnecessary delay.\footnote{With an abuse of terminology, in this proof let us allow for strategy profiles in which the seller's strategy is only required to be measurable with respect to time for each fixed seller-type. All bounds derived in the argument, as well as the claimed \eqref{eq:eq_delta_best_reply:objective}, are type-wise. Finally, the claimed bound on the right side of \eqref{eq:eq_delta_best_reply:objective} is evidently attainable by a measurable strategy profile.} We next formalize and justify this focus.

For all $\sigma\in\Sigma$, $t\in\R_{+}$, $x\in X$, and $y\in Y$, let $\gamma_{\sigma}(t, x, y) = \beta(t, y) \alpha(x, p(t, y), m(t, y))$.
Next, let
$\Gamma_{\sigma}(t, x, y) = \int_{[0, t]} \gamma_{\sigma}(s, x, y) \de s$, and let $\Gamma_{\sigma}(x, y) = \lim_{t\to\infty}\Gamma_{\sigma}(t, x, y)$ (possibly $\Gamma_{\sigma}(x, y) = \infty$).
In words, $\Gamma_{\sigma}(t, x, y)$ is the mass of times before $t$ at which a buyer with signal $x$ trades with type $y$.
These masses describe the probability of trade without breakdowns, once weighted with the signal PMF: $Q_{\sigma}(t\vert y, \omega) = 1 - e^{-\sum_{x\in X} \Gamma_{\sigma}(t, x, y) f(x\vert \omega)}$.

A strategy profile is \emph{front-loaded} if for all $y\in Y$ there exists a cutoff $r(y) \in \mathbb{R}_{+}\cup\lbrace\infty\rbrace$ such that, for almost all $t\in \R_{+}$, if $t < r(y)$, then $\sum_{x\in X} \gamma_{\sigma}(t, x, y) \geq 1$, and if $t>r(y)$, then $\sum_{x\in X} \gamma_{\sigma}(t, x, y) = 0$.
Let $\Sigma_{FL}$ be the set of front-loaded strategy profiles.
In words, at each instant before $r(y)$, at least one buyer-signal leads to trade; after $r(y)$, there is no trade.

\begin{step}
    For all $n\in\mathbb{N}$ and all $\sigma\in\Sigma$, there exists a front-loaded $\tilde\sigma\in\Sigma_{FL}$ such that all $y\in Y$ satisfy $U_{n}(\tilde\sigma, y) \geq U_{n}(\sigma, y)$. 
\end{step}
\begin{proof}
    We construct $\tilde\sigma$ separately for each type $y$. So, fix $y$.
    The idea is as follows.
    For each signal $x$, we shift the mass $\Gamma_{\sigma}(t, x, y)$ of times before $t$ at which buyer-signal $x$ trades forward to a time $\zeta(t)$, with $\zeta(t) \leq t$.
    Since in any strategy profile these masses must always grow with rate at most $1$, we cannot shift the masses forward too much. 
    The specific choice of $\zeta$ ensures this constraint.
    The shifted mass is denoted $\tilde{\Gamma}(\cdot, x)$ below.
    
    For all $t\in\R_{+}\cup\lbrace\infty\rbrace$, let $\zeta(t) = \int_{[0, t]} \max_{x\in X} \gamma_{\sigma}(s, x, y) \de s$ (possibly, $\zeta(\infty) = \infty$).
    For later reference, since $\gamma_{\sigma}\in [0, 1]$, we have $\zeta(t) \leq t$ for all $t$, and $\zeta(t)$ and $t - \zeta(t)$ are weakly increasing in $t$.
    Since $\zeta$ is continuous, the image $\zeta(\R_{+})$ is a (possibly unbounded) interval $Z$ with $\min Z = 0$.
    For all $z\in Z$, define the generalized inverse $\zeta^{-1}(z) = \sup\lbrace t\in\R_{+}\colon \zeta(t) \leq z\rbrace$.
    Finally, for all $x\in X$ and $z\in Z$, let $\tilde{\Gamma}(z, x) = \int_{[0, \zeta^{-1}(z)]} \gamma_{\sigma}(s, x, y) \de s$.

  Then, for each $x$, the function $\tilde{\Gamma}(\cdot, x)$ is $1$-Lipschitz. Indeed, since $\zeta$ is continuous and weakly increasing, it holds $\zeta(\zeta^{-1}(z))=z$ for all $z\in Z$. Hence, $\tilde{\Gamma}(z, x) - \tilde{\Gamma}(z', x) = \int_{\zeta^{-1}(z')}^{\zeta^{-1}(z)} \gamma_{\sigma}(s, x, y) \de s \leq \int_{\zeta^{-1}(z')}^{\zeta^{-1}(z)} \max_{x'\in X} \gamma_{\sigma}(s, x', y) \de s = \zeta(\zeta^{-1}(z)) - \zeta(\zeta^{-1}(z')) = z - z'$. Thus, $\tilde{\Gamma}(\cdot, x)$ admits a derivative $\tilde{\gamma}(\cdot, x)$ (with respect to Lebesgue measure) such that $\tilde{\gamma}(\cdot, x) \leq 1$. Of course, $\tilde{\gamma}(\cdot, x) \geq 0$ since $\tilde{\Gamma}(\cdot, x)$ is increasing.

    For later reference, for all $z, z' \in Z$ such that $z' \leq z$ it holds 
    \begin{equation*}
        \sum_{x\in X} \tilde{\Gamma}(z, x) - \tilde{\Gamma}(z', x) \geq \int_{\zeta^{-1}(z')}^{\zeta^{-1}(z)} \max_{x\in X} \gamma_{\sigma}(s, x, y) \de s = z - z'.
    \end{equation*}
    Thus, for almost all $z$ it holds $\sum_{x\in X} \tilde{\gamma}(z, x) \geq 1$.

    For all $z\in Z$ and $\omega\in\Omega$, let $\tilde{Q}(z\vert\omega) = 1 - e^{-\sum_{x\in X} f(x\vert\omega) \tilde{\Gamma}(z, x)}$.
    By the definition of $\tilde{\Gamma}$, it holds $\tilde{Q}(z\vert\omega) = 1 - e^{-\int_{[0, \zeta^{-1}(z)]}\sum_{x\in X} f(x\vert\omega) \gamma_{\sigma}(s, x, y) \de s} = Q_{\sigma}(\zeta^{-1}(z)\vert y, \omega)$.
    Moreover, if $\zeta$ is constant on an interval $[t, t']$, then $Q_{\sigma}(\cdot\vert y, \omega)$ assigns no mass to $[t, t']$. Therefore, $t = \zeta^{-1}(\zeta(t))$ for $Q_{\sigma}(\cdot\vert y, \omega)$-almost all $t$. 
    Thus, by a change of variables,
    \begin{align*}
        U_{n}(\sigma, y) 
        = &\int_{\Omega}\int_{\R_{+}} u_{n}(\omega) e^{-\lambda_{n} t} \de Q_{\sigma}(t\vert y, \omega) \de \mu(\omega\vert y)\\
        = &\int_{\Omega}\int_{\R_{+}} u_{n}(\omega) e^{-\lambda_{n} \zeta^{-1}(\zeta(t))} \de Q_{\sigma}(t\vert y, \omega) \de \mu(\omega\vert y) \\
        = &\int_{\Omega}\int_{Z} u_{n}(\omega) e^{-\lambda_{n} \zeta^{-1}(z)} \de \tilde{Q}(z\vert \omega) \de \mu(\omega\vert y) \\
        = &\int_{\Omega}\int_{Z} u_{n}(\omega) e^{-\lambda_{n} z} e^{-\lambda_{n}(\zeta^{-1}(z) - z)} \de \tilde{Q}(z\vert \omega) \de \mu(\omega\vert y).
    \end{align*}
    As observed earlier, $t\mapsto \zeta(t) - t$ is weakly decreasing. Thus, $z\mapsto \zeta^{-1}(z) - z$ is weakly increasing. In particular, $z\mapsto e^{-\lambda_{n}(\zeta^{-1}(z) - z)}$ is weakly decreasing and its image is contained in $[0, 1]$. Such a weakly decreasing function can be represented as a mixture of step functions, i.e., there is a Borel probability measure $\nu$ on the extended positive real line $[0, \infty]$ such that $e^{-\lambda_{n}(\zeta^{-1}(z) - z)} = \int_{[0, \infty]} \mathbbm{1}_{z \leq r} \de \nu(r)$ for Lebesgue-almost all $z\in Z$.
    Since $\tilde{Q}(\cdot\vert\omega)$ is absolutely continuous with respect to Lebesgue measure, Fubini's theorem shows
    \begin{equation*}
        U_{n}(\sigma, y) = \int_{[0, \infty]} \int_{\Omega}\int_{Z} u_{n}(\omega) e^{-\lambda_{n} z} \mathbbm{1}_{z \leq r} \de \tilde{Q}(z\vert \omega) \de \mu(\omega\vert y)\de\nu(r).
    \end{equation*}
    By linearity, there exists $r\in [0, \infty]$ such that
    \begin{equation*}
        U_{n}(\sigma, y) \leq \int_{\Omega}\int_{Z} u_{n}(\omega) e^{-\lambda_{n} z} \mathbbm{1}_{z \leq r} \de \tilde{Q}(z\vert \omega) \de \mu(\omega\vert y).
    \end{equation*}
    The right side equals $U_{n}(\tilde{\sigma}, y)$ for a strategy profile $\tilde\sigma$ in which, at each time $z$ before $r$, the fixed type $y$ and any signal $x$ trade at rate $\tilde{\gamma}(z, x)$, and in which no trade occurs after $r$. 
    This rate is feasible since $\tilde{\gamma}(z, x) \in [0, 1]$ holds, as shown above.
    Since $\sum_{x\in X} \tilde{\gamma}(z, x) \geq 1$ for almost all $z$ (as shown above), this strategy profile is front-loaded following type $y$.
    Since $y$ was arbitrary, we obtain $\tilde{\sigma}$ with the desired properties.
\end{proof}

For all $s\in\R_{+}\cup\lbrace\infty\rbrace$ and $y\in Y$, let $\bar{U}(s, y) = \int_{\Omega} u(\omega)\bar{Q}(s\vert \omega) \de\mu(\omega\vert y)$ be the expected surplus from waiting for an $\bar{x}$-buyer until $s$ when there are no breakdowns.
The next step asserts that among strategies that wait for $\bar{x}$-buyers the exit times $s^{\dagger}_{n}(y)$ and $s^{\dagger}(y)$ are optimal in their respective objectives.
\begin{step}
For all $n\in\mathbb{N}$ and $y\in Y$, the function $\bar{U}_{n}(\cdot, y)$ is uniquely maximized at $s_{n}^{\dagger}(y)$, and $\bar{U}(\cdot, y)$ is uniquely maximized at $s^{\dagger}(y)$.
\end{step}
\begin{proof}
    The derivative of $\bar{U}_{n}(s, y)$ with respect to $s$ equals 
    \begin{equation}\label{eq:eq_delta_best_reply:optimal_exit_objective_derivative}
    e^{-\lambda_{n} s} \int_{\Omega} u_{n}(\omega)  f(\bar{x}\vert \omega) e^{- f(\bar{x}\vert \omega) s}\de\mu(\omega\vert y)
    .
    \end{equation}
    Note $e^{- f(\bar{x}\vert \omega) s}$ is strictly log-submodular in $(\omega, s)$ since $f(\bar{x}\vert \omega)$ is strictly increasing in $\omega$.
    Since $u_{n}$ is strictly single-crossing from below at an interior state $\omega_{0, n}^{\dagger}$, it follows that the derivative is strictly single-crossing from above.
    Thus, the function $\bar{U}_{n}(\cdot, y)$ is uniquely maximized at the smallest point $s$ for which its derivative is weakly negative.
    This point equals $s^{\dagger}_{n}(y)$ since the sign of the derivative equals the sign of the expectation in the definition of $s^{\dagger}_{n}(y)$.
    An analogous argument shows the claim for $\bar{U}(\cdot, y)$.
\end{proof}

The content of the next step is that it suffices to consider front-loaded strategy profiles $\sigma$ that trade with probability bounded away from one. This bound is expressed using the masses $\Gamma_{\sigma}$.

\begin{step}
    There exists $T > 0$ and $n^{\dagger}_{1} \in \mathbb{N}$ such that for all $y\in Y$, $\sigma\in\Sigma_{FL}$, and $n\in\mathbb{N}$, if $\sum_{x\in X} \Gamma_{\sigma}(x, y) \geq T$ and $n\geq n^{\dagger}_{1}$, then $\bar{U}_{n}(s^{\dagger}_{n}(y), y) > U_{n}(\sigma, y)$.
\end{step}

For the proof, given $\sigma\in\Sigma$ and $y\in Y$, it is useful to rewrite $U_{n}(\sigma, y)$ as:
\begin{subequations}\label{eq:delta_maximization_objective}
\begin{align}
U_{n}(\sigma, y) =& 
\int_{\Omega} u_{n}(\omega)Q_{\sigma}(y, \omega) \de\mu(\omega\vert y) 
    \label{eq:delta_maximization_objective:1}
    \\
    &-
    \int_{\Omega}\int_{\R_{+}} u_{n}(\omega) (1 - e^{-\lambda_{n} t}) \de Q_{\sigma}(t\vert y, \omega) \de\mu(\omega\vert y) 
    .
    \label{eq:delta_maximization_objective:2}
\end{align}
\end{subequations}

\begin{proof}
    As an intermediate claim, we argue that as $n\to\infty$ the integral \eqref{eq:delta_maximization_objective:2} vanishes uniformly across all front-loaded $\sigma\in\Sigma_{FL}$ and $y\in Y$.
    Let $\sigma\in\Sigma_{FL}$ and $y\in Y$.
    Put $\ubar{f} = \min_{x\in X, \omega\in\Omega} f(x\vert\omega)$ (recalling that $\Omega$ is compact and $f$ continuous).
    By assumption, the signal PMF has full support in every state, i.e., $\ubar{f} > 0$.
    Let $r(y)\in \R_{+}\cup\lbrace\infty\rbrace$ be the cutoff of the front-loaded $\sigma$ at type $y$.
    For all $t\leq r(y)$, the Poisson rate of trade satisfies $q_{\sigma}(t\vert y, \omega) = \sum_{x\in X} f(x\vert\omega) \gamma_{\sigma}(t, x, y) \geq \ubar{f} \sum_{x\in X} \gamma_{\sigma}(t, x, y) \geq \ubar{f}$.
    For $t > r(y)$, the rate of trade is zero.
    Therefore, for all $\omega$,
    \begin{align*}
        \int_{\R_{+}} (1 - e^{-\lambda_{n} t}) \de Q_{\sigma}(t\vert y, \omega) 
        = &\int_{\R_{+}} (1 - e^{-\lambda_{n} t}) q_{\sigma}(t\vert y, \omega) e^{-\int_{[0, t]} q_{\sigma}(s\vert y, \omega) \de s} \de t
        \\
        \leq &\int_{\R_{+}} (1 - e^{-\lambda_{n} t}) e^{-t \ubar{f}} \de t.
    \end{align*}
    Since $t\mapsto e^{-t\ubar{f}}$ is integrable, Dominated Convergence implies that this upper bound vanishes as $n\to\infty$.
    Since also $u_{n}(\omega)$ is uniformly bounded across $n$ and $\omega$, it follows that \eqref{eq:delta_maximization_objective:2} vanishes uniformly across $\sigma\in\Sigma_{FL}$ and $y\in Y$.

    We now prove the claim.
    Towards a contradiction, suppose the claim is false. That is, there are a subsequence $(n_{k})_{k\in\mathbb{N}}$, a sequence $(y_{k})_{k\in\mathbb{N}}$ in $Y$, and a sequence $(\sigma_{k})_{k\in\mathbb{N}}$ in $\Sigma_{FL}$ such that $\bar{U}_{n_{k}}(s^{\dagger}_{n_{k}}(y_{k}), y_{k}) \leq U_{n_{k}}(\sigma_{k}, y_{k})$ for all $k$ and such that $\sum_{x} \Gamma_{\sigma_{k}}(x, y_{k})$ diverges as $k\to\infty$. 
    By possibly passing to a subsequence, let $(y_{k})_{k\in\mathbb{N}}$ converge to $y\in Y$ (recall $Y$ is compact).

    Since $\sum_{x} \Gamma_{\sigma_{k}}(x, y_{k})$ diverges, for all $\omega$ we get $\lim_{k\to\infty} Q_{\sigma_{k}}(y_{k}, \omega) = 1$.
    By Dominated Convergence, the integral \eqref{eq:delta_maximization_objective:1} converges to $\int_{\Omega} u(\omega) \de\mu(\omega\vert y)$.
    For front-loaded strategy profiles, the intermediate claim implies that the integral in \eqref{eq:delta_maximization_objective:2} vanishes.
    Thus, $U_{n_{k}}(\sigma_{k}, y_{k})$ converges to $\int_{\Omega} u(\omega) \de\mu(\omega\vert y)$ as $k\to\infty$.
    Meanwhile, since $(s^{\dagger}_{n})_{n\in\mathbb{N}}$ converges uniformly to $s^{\dagger}$, a similar argument shows $\lim_{k\to\infty} \bar{U}_{n_{k}}(s^{\dagger}_{n_{k}}(y_{k}), y_{k}) = \bar{U}(s^{\dagger}(y), y)=\int_{\Omega} u(\omega) (1 -e^{- f(\bar{x}\vert \omega) s^{\dagger}(y)})\de\mu(\omega\vert y)$.
    From Step 2, we know 
    \begin{equation*}
    \int_{\Omega} u(\omega) \de\mu(\omega\vert y) < \int_{\Omega} u(\omega) \bar{Q}(s^{\dagger}(y)\vert \omega) \de \mu(\omega\vert y) = \bar{U}(s^{\dagger}(y), y).
    \end{equation*}
    Since $\lim_{k\to\infty} \bar{U}_{n_{k}}(s^{\dagger}_{n_{k}}(y_{k}), y_{k}) = \bar{U}(s^{\dagger}(y), y)$, for $k$ sufficiently large we have a contradiction to $\bar{U}_{n_{k}}(s^{\dagger}_{n_{k}}(y_{k}), y_{k}) \leq U_{n_{k}}(\sigma_{k}, y_{k})$. 
\end{proof}

Let $n^{\dagger}_{1}$ and $T$ be as in the previous step.
Let $\Sigma_{FL}^{\dagger}$ be the set of front-loaded $\sigma$ such that $\sum_{x\in X} \Gamma_{\sigma}(x, y) \leq T$ for all $y$.
For $n \geq n^{\dagger}_{1}$, Steps 1 and 3 imply that it suffices to consider strategy profiles in $\Sigma_{FL}^{\dagger}$.
Let us henceforth take the entire sequence to mean the sequence above $n^{\dagger}_{1}$.

To state the remaining proof strategy, define the following. For all $\sigma\in\Sigma$, $t\in\R_{+}$, $y\in Y$, $\omega\in\Omega$, let
\begin{equation*}
    \xi_{\sigma}(t, y, \omega) = \sum_{x\in X}\frac{f(x\vert\omega)}{f(\bar{x}\vert\omega)} \Gamma_{\sigma}(t, x, y).
\end{equation*}
and $\xi_{\sigma}(y, \omega) = \lim_{t\to\infty} \xi_{\sigma}(t, y, \omega)$ (possibly $\xi_{\sigma}(y, \omega) = \infty$).
By construction, $Q_{\sigma}(t\vert y, \omega) = \bar{Q}(\xi_{\sigma}(t, y, \omega)\vert \omega)$.

Step 2 shows that $\bar{U}_{n}(\cdot, y)$ is maximized at $s_{n}^{\dagger}(y)$, for all $n$ and $y$.
Hence, for given $n, y, \sigma$, we can show $\bar{U}_{n}(s_{n}^{\dagger}(y), y) \geq U_{n}(\sigma, y)$ by showing $\bar{U}_{n}(\xi_{\sigma}(y, \omega_{0, n}^{\dagger}), y) \geq U_{n}(\sigma, y)$.
The left side corresponds to the surplus from a strategy that waits for an $\bar{x}$-buyer until the specific time $\xi_{\sigma}(y, \omega_{0, n}^{\dagger})$.

By the construction of $\xi_{\sigma}$, a change of variables yields
\begin{equation*}
    \int_{\Omega}
    u_{n}(\omega)\left(
    \int_{\R_{+}} e^{-\lambda_{n} t} \mathbbm{1}_{t\leq \xi_{\sigma}(y, \omega)} \de \bar{Q}(t\vert \omega) - \int_{\R_{+}}  e^{-\lambda_{n} \xi_{\sigma}(t, y, \omega)} \de Q_{\sigma}(t\vert y, \omega) \right)\de\mu(\omega\vert y) = 0.
\end{equation*}
By adding this zero, we can write the difference $\bar{U}_{n}(\xi_{\sigma}(y, \omega_{0, n}^{\dagger}), y) - U_{n}(\sigma, y)$ as
\begin{align}
    \nonumber
    & \bar{U}_{n}(\xi_{\sigma}(y, \omega_{0, n}^{\dagger}), y) - U_{n}(\sigma, y)
    \\
    = & \int_{\Omega} u_{n}(\omega) \left(\int_{\R_{+}} e^{-\lambda_{n} t} \mathbbm{1}_{t\leq \xi_{\sigma}(y, \omega_{0, n}^{\dagger})} \de \bar{Q}(t\vert \omega) - \int_{\R_{+}} e^{-\lambda_{n} t} \de Q_{\sigma}(t\vert y, \omega) \right) \de\mu(\omega \vert y)
    \nonumber
    \\
    = & \int_{\Omega} \int_{\R_{+}} u_{n}(\omega) e^{-\lambda_{n} t} \left(\mathbbm{1}_{t\leq \xi_{\sigma}(y, \omega_{0, n}^{\dagger})} - \mathbbm{1}_{t\leq \xi_{\sigma}(y, \omega)}\right) \de \bar{Q}(t\vert \omega)\de\mu(\omega\vert y) 
    \label{eq:delta_maximization:decomp1}
    \\
    & + \int_{\Omega} \int_{\R_{+}} u_{n}(\omega) \left(e^{-\lambda_{n} \xi_{\sigma}(t, y, \omega)} - e^{-\lambda_{n} t}\right) \de Q_{\sigma}(t\vert y, \omega) \de\mu(\omega\vert y)
    \label{eq:delta_maximization:decomp2}
    .
\end{align}
Let $\Delta_{n, y, \sigma}^{1}$ be the integral in \eqref{eq:delta_maximization:decomp1}, and let $\Delta_{n, y, \sigma}^{2}$ be the integral in \eqref{eq:delta_maximization:decomp2}.
For each $n$, we provide bounds on $\Delta_{n, y, \sigma}^{1}$ and $\Delta_{n, y, \sigma}^{2}$ that are uniform across all $y\in Y$ and $\sigma\in\Sigma_{FL}^{\dagger}$.
These bounds will show that if $n$ is sufficiently large, then $\bar{U}_{n}(\xi_{\sigma}(y, \omega_{0, n}^{\dagger}), y) \geq U_{n}(\sigma, y)$ for all $\sigma\in\Sigma_{FL}^{\dagger}$ and $y\in Y$, completing the proof.\footnote{For later reference, this argument will in fact show $\bar{U}_{n}(s^{\dagger}_{n}(y), y) > U_{n}(\sigma, y)$ if $\sum_{x\neq \bar{x}} \Gamma_{\sigma}(x, y) > 0$.\label{footnote:delta_maximization_gamma_uniqueness}}

We next describe the bounds. 
Find $K \geq 1$ such that $f(x\vert \omega) / f(\bar{x}\vert \omega) \leq K$ for all $x$ and $\omega$; this is possible since $f$ is continuous in the state and has full support in each state.
Next, for all $n$, define
    \begin{equation*}
        L_{n} =
        \min_{x\in X\setminus\lbrace \bar{x}\rbrace, y\in Y}
        \int_{\Omega} u_{n}(\omega) \left(\frac{f(x\vert \omega_{0, n}^{\dagger})}{f(\bar{x}\vert \omega_{0, n}^{\dagger})} f(\bar{x}\vert \omega) - f(x\vert \omega)\right) \de\mu(\omega\vert y)
        .
    \end{equation*}
    The minimum is attained since $X$ is finite, $Y$ is compact, and $f$ and $g$ are continuous ($g$ enters $\mu(\cdot\vert\cdot)$).
    Berge's Maximum Theorem implies that $(L_{n})_{n\in\mathbb{N}}$ converges to
    \begin{equation*}
        L =
        \min_{x\in X\setminus\lbrace \bar{x}\rbrace, y\in Y}
        \int_{\Omega} u(\omega) \left(\frac{f(x\vert \omega_{0}^{\dagger})}{f(\bar{x}\vert \omega_{0}^{\dagger})} f(\bar{x}\vert \omega) - f(x\vert \omega)\right) \de\mu(\omega\vert y)
        .
    \end{equation*}
    For all fixed $(x, y, \omega)$ such that $\omega\neq\omega_{0}^{\dagger}$, the integrand is strictly positive since $u$ is strictly single-crossing from below at $\omega_{0}^{\dagger}\in (\ubar{\omega}, \bar{\omega})$, and in view of \eqref{assumption:highest_signal} of \Cref{assumption:correlated}; for $\omega = \omega_{0}^{\dagger}$, the integrand equals zero.
    Since $X$ is finite, $Y$ is compact, and $f$ and $g$ are continuous, it holds $L > 0$.
    In particular, $\lim_{n\to\infty} \lambda_{n} / L_{n} = 0$, which we record for future reference.
    Next, find $L^{\prime} \geq  1$ such that
    \begin{equation*}
        L^{\prime} \geq 
        f(\bar {x}\vert \omega) \frac{f(x\vert \omega^{\prime})}{f(\bar{x}\vert \omega^{\prime})}
    \end{equation*}
    for all $x\in X$ and $\omega, \omega^{\prime}\in\Omega$; this is possible since $f$ always has full support, is continuous in all arguments, and since $X$ is finite while $\Omega$ is compact.
    Finally, recall $M$ is such that $\vert u_{n}(\omega)\vert \leq M$ for all $n$ and $\omega$.

The next steps will show that for all $\sigma\in\Sigma_{FL}^{\dagger}$ and $y\in Y$,
\begin{equation*}
    \Delta_{n, y, \sigma}^{1} + \Delta_{n, y, \sigma}^{2} \geq \left(e^{-(\lambda_{n} K + L') T} L_{n} - K M \lambda_{n}\right) \sum_{x\neq \bar{x}} \Gamma_{\sigma}(x, y).
\end{equation*}
Since $\lim_{n\to\infty}\lambda_{n} / L_{n} = 0$, the difference in parentheses is eventually strictly positive.
Since this difference does not depend on $\sigma$ or $y$, it follows that if $n$ is sufficiently large, then $\bar{U}_{n}(\xi_{\sigma}(y, \omega_{0, n}^{\dagger}), y) \geq U_{n}(\sigma, y)$ for all $\sigma\in\Sigma_{FL}^{\dagger}$ and $y\in Y$, completing the proof.

It remains to prove the bounds.
\begin{step}
    For all $\sigma\in\Sigma_{FL}^{\dagger}$, $y\in Y$ and $n\in\mathbb{N}$, it holds $\Delta_{n, y, \sigma}^{1} \geq  e^{-(\lambda_{n} K + L') T} L_{n} \sum_{x\neq\bar{x}}\Gamma_{\sigma}(x, y)$.
\end{step}
\begin{proof}
    Fix such $\sigma$, $y$, and $n$.
    Notice that $\xi_{\sigma}(y, \omega)$ is weakly decreasing in $\omega$, and recall that $u_{n}$ single-crosses $0$ from below at $\omega_{0, n}^{\dagger}$.
    Therefore, for all $t$ and $\omega$, it holds 
    \begin{equation*}
    u_{n}(\omega)\left(\mathbbm{1}_{t \leq \xi_{\sigma}(y, \omega_{0, n}^{\dagger})} - \mathbbm{1}_{t \leq \xi_{\sigma}(y, \omega)}\right) \geq 0.
    \end{equation*}
    By definition of $K$, we also have $t \leq K \sum_{x\in X} \Gamma_{\sigma}(x, y) \leq K T$ whenever $t \leq \xi_{\sigma}(y, \omega)$ for at least one $\omega$.
    Therefore,
    \begin{align*}
        \Delta_{n, y, \sigma}^{1} \geq &\int_{\Omega} \int_{\R_{+}} u_{n}(\omega) e^{-\lambda_{n} K T} \left(\mathbbm{1}_{t \leq \xi_{\sigma}(y, \omega_{0, n}^{\dagger})} - \mathbbm{1}_{t \leq \xi_{\sigma}(y, \omega)}\right) \de \bar{Q}(t\vert\omega) \de \mu(\omega\vert y)
        \nonumber
        \\
        =&
        e^{-\lambda_{n} K T}
        \int_{\Omega} u_{n}(\omega) \left(e^{- \sum_{x\in X} \Gamma_{\sigma}(x, y) f(x\vert\omega)} - e^{- \xi_{\sigma}(y, \omega_{0, n}^{\dagger}) f(\bar{x}\vert \omega)}\right) \de\mu(\omega\vert y).
    \end{align*}
    We shall further bound the integral with respect to $\omega$.
    For the remainder of the argument, since $n$, $\sigma$, and $y$ are fixed, let us abbreviate $\xi^{\dagger} = \xi_{\sigma}(y, \omega_{0, n}^{\dagger})$ and $\Gamma \cdot f(\omega) = \sum_{x\in X} \Gamma_{\sigma}(x, y) f(x\vert\omega)$ for any $\omega$.
    The integral of interest is therefore given by
    \begin{equation}\label{eq:discount_commitment:lower_bound:1:spelled_out}
       \int_{\Omega} u_{n}(\omega) \left(e^{- \Gamma \cdot f(\omega)} - e^{- \xi^{\dagger} f(\bar{x}\vert \omega)}\right) \de\mu(\omega\vert y).
    \end{equation}
    By convexity of the exponential function, it holds
    \begin{align*}
        &
        e^{- \xi^{\dagger} f(\bar{x}\vert \omega)}\left(\xi^{\dagger} f(\bar{x}\vert \omega) - \Gamma \cdot f(\omega)\right)
        \\
        \leq
        & 
        e^{- \Gamma \cdot f(\omega)} - e^{- \xi^{\dagger} f(\bar{x}\vert \omega)}
        \\
        \leq
        & 
        e^{- \Gamma \cdot f(\omega)}\left(\xi^{\dagger} f(\bar{x}\vert \omega) - \Gamma \cdot f(\omega)\right)
        .
    \end{align*}
    With these inequalities and recalling that $u_{n}$ single-crosses $0$ from below at $\omega_{0, n}^{\dagger}$, we obtain the following lower bound on \eqref{eq:discount_commitment:lower_bound:1:spelled_out}:
    \begin{align}\label{eq:discount_commitment:lower_bound:1:further_bound}
        \eqref{eq:discount_commitment:lower_bound:1:spelled_out} \geq
        &
        \int_{\omega\geq\omega_{0, n}^{\dagger}} u_{n}(\omega) e^{- \xi^{\dagger} f(\bar{x}\vert \omega)}\left(\xi^{\dagger} f(\bar{x}\vert \omega) - \Gamma \cdot f(\omega)\right)
        \de\mu(\omega\vert y)
        \\
        \label{eq:discount_commitment:lower_bound:1:further_bound2}
        &
        +
        \int_{\omega<\omega_{0, n}^{\dagger}} u_{n}(\omega) e^{- \Gamma \cdot f(\omega)}\left(\xi^{\dagger} f(\bar{x}\vert \omega) - \Gamma \cdot f(\omega)\right)
        \de\mu(\omega\vert y)
        .
    \end{align}
    Notice that, for all $\omega$, the definition of $\xi_{\sigma}$ yields
    \begin{equation}
        u_{n}(\omega)\left(\xi^{\dagger} f(\bar{x}\vert \omega) - \Gamma \cdot f(\omega)\right)
        =
        \sum_{x\neq\bar{x}} \Gamma_{\sigma}(x, y) u_{n}(\omega) \left(\frac{f(x\vert \omega_{0, n}^{\dagger})}{f(\bar{x}\vert \omega_{0, n}^{\dagger})} f(\bar{x}\vert \omega) - f(x\vert \omega)\right).
        \label{eq:delta_maximization:lower_bound:1:summands}
    \end{equation}
    Since $u_{n}$ single-crosses from below at $\omega_{0, n}^{\dagger}$, \Cref{assumption:correlated} implies that each summand in \eqref{eq:delta_maximization:lower_bound:1:summands} is weakly positive.
    Further, the choice of $L^{\prime}$ implies
    \begin{equation*}
        e^{-\xi^{\dagger} f(\bar{x}\vert \omega)} \geq e^{-\sum_{x\in X}\Gamma_{\sigma}(x, y) L^{\prime}}
        \quad\mbox{and}\quad
        e^{-\Gamma \cdot f(\omega)}
        \geq e^{-\sum_{x\in X}\Gamma_{\sigma}(x, y) L^{\prime}}
        .
    \end{equation*}
    Thus, since each summand in \eqref{eq:delta_maximization:lower_bound:1:summands} is weakly positive, we obtain:
    \begin{align*}
        \eqref{eq:discount_commitment:lower_bound:1:further_bound} + \eqref{eq:discount_commitment:lower_bound:1:further_bound2}
        \geq
        &
        e^{-\sum_{x\in X}\Gamma_{\sigma}(x, y)L^{\prime}}  \sum_{x\neq\bar{x}}\Gamma_{\sigma}(x, y) \int_{\Omega} u_{n}(\omega) \left(\frac{f(x\vert \omega_{0, n}^{\dagger})}{f(\bar{x}\vert \omega_{0, n}^{\dagger})} f(\bar{x}\vert \omega) - f(x\vert \omega)\right) \de\mu(\omega\vert y)
        \\
        \geq
        &
        L_{n} e^{- T L^{\prime}}  \sum_{x\neq\bar{x}}\Gamma_{\sigma}(x, y)
        ,
    \end{align*}
    where the final inequality holds by definition of $L_{n}$.
    Collecting inequalities, we conclude
    $\Delta_{n, y, \sigma}^{1} \geq e^{-(\lambda_{n} K + L') T} L_{n} \sum_{x\neq\bar{x}}\Gamma_{\sigma}(x, y)$.
\end{proof}

\begin{step}
    For all $n\in\mathbb{N}$, $\sigma\in\Sigma_{FL}^{\dagger}$, and $y\in Y$ it holds $\Delta_{n, y, \sigma}^{2} \geq - K M \lambda_{n} \sum_{x\neq \bar{x}} \Gamma_{\sigma}(x, y)$.
\end{step}

\begin{proof}
Fix $\sigma\in\Sigma_{FL}^{\dagger}$, $y\in Y$, and $n\in\mathbb{N}$.
Recall the definition:
\begin{equation*}
    \Delta_{n, y, \sigma}^{2} = 
    \int_{\Omega} \int_{\R_{+}} u_{n}(\omega) \left(e^{-\lambda_{n} \xi_{\sigma}(t, y, \omega)} - e^{-\lambda_{n} t}\right) \de Q_{\sigma}(t\vert y, \omega) \de\mu(\omega\vert y).
\end{equation*}
Fix $\omega$. We shall bound the inner integral with respect to $t$.
Distinguish two cases for $\omega$.
First, let $u_{n}(\omega) \geq 0$. We shall derive a lower bound on the integral with respect to $t$.
Indeed, by convexity of the exponential function, 
\begin{equation*}
    \int_{\R_{+}} \left(e^{- \lambda_{n} \xi_{\sigma}(t, y, \omega)} - e^{-\lambda_{n} t}\right) \de Q_{\sigma}(t\vert y, \omega) 
    \geq 
    \int_{\R_{+}} e^{-\lambda_{n} t} \lambda_{n}(t - \xi_{\sigma}(t, y, \omega))  \de Q_{\sigma}(t\vert y, \omega)
    .
\end{equation*}
Further, for all $t$, it holds
\begin{equation*}
t - \xi_{\sigma}(t, y, \omega) 
= t - \Gamma_{\sigma}(t, \bar{x}, y) - \sum_{x\neq \bar{x}} \Gamma_{\sigma}(t, x, y) \frac{f(x\vert \omega)}{f(\bar{x}\vert \omega)}
\geq 
- K \sum_{x\neq \bar{x}} \Gamma_{\sigma}(x, y)
\end{equation*}
since $\Gamma_{\sigma}(t, x, y)$ is weakly increasing in $t$ and at most equal to $t$ for all fixed $(x, y)$.
Thus, we obtain the lower bound
\begin{align*}
        \int_{\R_{+}} e^{-\lambda_{n} t} \lambda_{n}(t - \xi_{\sigma}(t, y, \omega))  \de Q_{\sigma}(t\vert y, \omega)
        \geq
        &
        - K \lambda_{n} \sum_{x\neq \bar{x}} \Gamma_{\sigma}(x, y)
        \int_{\R_{+}} e^{-\lambda_{n} t}\de Q_{\sigma}(t\vert y, \omega)
        \\
        \geq
        &
        - K \lambda_{n} \sum_{x\neq \bar{x}} \Gamma_{\sigma}(x, y).
\end{align*}
Second, let $u_{n}(\omega) < 0$. We shall derive an upper bound on the integral with respect to $t$.
Again by convexity of the exponential function,
\begin{equation*}
    \int_{\R_{+}} \left(e^{- \lambda_{n} \xi_{\sigma}(t, y, \omega)} - e^{-\lambda_{n} t}\right) \de Q_{\sigma}(t\vert y, \omega) 
    \leq 
    \int_{\R_{+}} e^{-\lambda_{n} \xi_{\sigma}(t, y, \omega)} \lambda_{n}(t - \xi_{\sigma}(t, y, \omega))  \de Q_{\sigma}(t\vert y, \omega)
    .
\end{equation*}
Since $\sigma$ is front-loaded, it holds $t \leq \sum_{x\in X} \Gamma_{\sigma}(t, x, y)$ for $Q_{\sigma}(\cdot\vert y, \omega)$-almost all $t$.
Hence, 
\begin{align*}
t - \xi_{\sigma}(t, y, \omega) \leq \sum_{x\in X} \Gamma_{\sigma}(t, x, y)\left(1 - \frac{f(x\vert \omega)}{f(\bar{x}\vert \omega)}\right) &= \sum_{x\neq \bar{x}} \Gamma_{\sigma}(t, x, y)\left(1 - \frac{f(x\vert \omega)}{f(\bar{x}\vert \omega)}\right)\\ &\leq \sum_{x\neq\bar{x}} \Gamma_{\sigma}(t, x, y).
\end{align*}
Therefore,
\begin{align*}
    & \int_{\R_{+}} e^{-\lambda_{n} \xi_{\sigma}(t, y, \omega)} \lambda_{n}(t - \xi_{\sigma}(t, y, \omega))  \de Q_{\sigma}(t\vert y, \omega)
    \\
    &
    \leq 
    \lambda_{n} \sum_{x\neq\bar{x}} \int_{\R_{+}} \Gamma_{\sigma}(t, x, y) e^{-\lambda_{n} \xi_{\sigma}(t, y, \omega)} \de Q_{\sigma}(t\vert y, \omega)
    \leq 
    \lambda_{n} K \sum_{x\neq\bar{x}} \Gamma_{\sigma}(x, y),
\end{align*}
where the final inequality uses $K \geq 1$ and that $\Gamma_{\sigma}(t, x, y)$ weakly increases in $t$.
Combining the bounds for the two cases and integrating across $\omega$, the claim follows.
\end{proof}
Steps 4 and 5 provide the bounds that were needed to complete the proof.
\end{proof}

\subsection{Proof of \headercref{Theorem}{{thm:commitment_solution}}}
In every strategy profile $\sigma$ in which buyers are best replying, the buyer utility $V_{B}(\sigma)$ is weakly positive since buyers could decline all prices.
Thus, $V_{S}(\sigma) = V(\sigma) - V_{B}(\sigma) \leq V(\sigma)$.
In particular, \Cref{thm:maximizing_surplus} implies $V_{S}(\sigma) \leq \E\left[v(\bm{\omega}) \mathbbm{1}_{\bm{\tau}_{\bar{x}} \leq \bar{s}(\bm{y}) \wedge \bm{\tau}_{d}}\right]$.
A moment's thought implies that the seller's profit equals $\E\left[v(\bm{\omega}) \mathbbm{1}_{\bm{\tau}_{\bar{x}} \leq \bar{s}(\bm{y}) \wedge \bm{\tau}_{d}}\right]$ under the strategy profile described in \Cref{thm:commitment_solution}.
It remains to verify that buyers are best replying in the strategy profile described in \Cref{thm:commitment_solution}.
In this strategy profile and for an arbitrary private signal $x$, an arriving buyer infers only that they arrived at a time before another buyer with signal $\bar{x}$, before a breakdown, and before the seller has left the market. 
This buyer's posterior value equals
\begin{equation*}
    \frac{\int_{\Omega}\int_{Y}\int_{0}^{\bar{s}(y)} v_{B}(\omega) f(x\vert \omega) e^{-(\lambda +  f(\bar{x}\vert \omega)) t} \de t\de G(y\vert\omega)\de\mu(\omega)}
    {\int_{\Omega}\int_{Y}\int_{0}^{\bar{s}(y)} f(x\vert \omega) e^{-(\lambda +  f(\bar{x}\vert \omega)) t} \de t \de G(y\vert\omega)\de\mu(\omega)}
    \\
    = \frac{\int_{\Omega} v_{B}(\omega) f(x\vert\omega)  \nu(\omega) \de\mu(\omega)}
    {\int_{\Omega}f(x\vert\omega) \nu(\omega) \de\mu(\omega)}
    ,
\end{equation*}
where $\nu(\omega) = \int_{Y}\int_{[0, \bar{s}(y)]} e^{-(\lambda +  f(\bar{x}\vert \omega)) t} \de t\de G(y\vert\omega)$ for all $\omega$.
Importantly, $\nu$ does not depend on the buyer's signal $x$.
Since $v_{B}$ is strictly increasing, and in view of \eqref{assumption:highest_signal}, Theorem 2.2 of \citet{karlin1980mtp} implies that the posterior value is maximized at $x = \bar{x}$. For $x = \bar{x}$, the value equals $\bar{p}$. Thus, buyers are best replying.
\qed

\subsection{The posterior buyer value}\label{appendix:buyer_value_decreases_in_date}
This part of the appendix records an auxiliary result on the posterior buyer value.

\begin{lemma}\label{lemma:strict_monotone_buyer_value_cond_trade_interval}
    For all $y\in Y$ and $t, T\in\R_{+}$ such that $t < T$, the expectation
    \begin{equation*}
        \E [v_{B}(\bm{\omega})\vert \bm{\tau}_{\bar{x}}\in [t, T], \bm{\tau}_{\bar{x}} \leq \bm{\tau}_{d}, \bm{y} = y]
    \end{equation*}
    is strictly decreasing in both $t$ and $T$.  
\end{lemma}

\begin{proof}[Proof of \Cref{lemma:strict_monotone_buyer_value_cond_trade_interval}]
    Fixing $y$, abbreviate $\phi(\omega) =  f(\bar{x}\vert \omega)$ for all $\omega$, and let $\mu(\cdot\vert y)$ be type $y$'s posterior belief (see \cref{footnote:types_posterior_belief}).
    We prove only that the ratio is strictly decreasing in $t$ on a neighborhood where $t < T$; the other claim is proven similarly.
    The argument is adapted from \citet[Theorem 2.2]{karlin1980mtp}.
    We have
    \begin{equation*}
        \E [v_{B}(\bm{\omega})\vert \bm{\tau}_{\bar{x}}\in [t, T], \bm{\tau}_{\bar{x}} \leq \bm{\tau}_{d}, \bm{y} = y]
        =
        \frac{\int_{\Omega}\int_{[t, T]} v_{B}(\omega) \phi(\omega) e^{-(\lambda + \phi(\omega))s}  \de s\de\mu(\omega \vert y)}{\int_{\Omega}\int_{[t, T]} \phi(\omega) e^{-(\lambda + \phi(\omega))s} \de s\de\mu(\omega \vert y)}.
    \end{equation*}
    The derivative with respect to $t$ has the same sign as (here, from the first equation onwards, the integrals are taken with respect to the measure $\Leb\times\mu(\cdot\vert y)\times\mu(\cdot\vert y)$)
    \begin{align*}
        &\int_{\Omega} \int_{t}^{T}v_{B}(\omega) \phi(\omega)e^{- (\lambda + \phi(\omega))s} \de s \de\mu(\omega\vert y)\int_{\Omega} \phi(\omega)e^{-(\lambda + \phi(\omega))t} \de\mu(\omega\vert y)
        \\
        &
        -
        \int_{\Omega} \int_{t}^{T}\phi(\omega) e^{- (\lambda + \phi(\omega))s} \de s \de\mu(\omega\vert y)\int_{\Omega}  v_{B}(\omega)\phi(\omega)e^{-(\lambda + \phi(\omega))t} \de\mu(\omega\vert y)      
        \\
        =
        &
        \int\limits_{\substack{s\in [t, T],\\ \omega > \omega^{\prime}}}
        \phi(\omega)\phi(\omega^{\prime}) e^{-\lambda(s + t)}\left(v_{B}(\omega) e^{-\phi(\omega)s - \phi(\omega^{\prime})t} + v_{B}(\omega^{\prime}) e^{-\phi(\omega^{\prime})s - \phi(\omega)t}\right) 
        \\
        &
        -
        \int\limits_{\substack{s\in [t, T],\\ \omega > \omega^{\prime}}}
        \phi(\omega)\phi(\omega^{\prime}) e^{-\lambda(s + t)} \left(v_{B}(\omega) e^{-\phi(\omega^{\prime})s - \phi(\omega)t} + v_{B}(\omega^{\prime}) e^{-\phi(\omega)s - \phi(\omega^{\prime})t}\right)
        \\
        =
        &
        \int\limits_{\substack{s\in [t, T],\\ \omega > \omega^{\prime}}}
        \phi(\omega)\phi(\omega^{\prime}) e^{-\lambda(s + t)} \left(v_{B}(\omega) - v_{B}(\omega^{\prime})\right) 
        \left(e^{-\phi(\omega)s - \phi(\omega^{\prime})t} - e^{-\phi(\omega^{\prime})s - \phi(\omega)t}\right)
        .
    \end{align*}
    The sign of $e^{-\phi(\omega)s - \phi(\omega^{\prime})t} - e^{-\phi(\omega^{\prime})s - \phi(\omega)t}$ equals the sign of $(\phi(\omega^{\prime}) - \phi(\omega))(s - t)$.
    Both $v_{B}(\omega)$ and $\phi(\omega) =  f(\bar{x}\vert \omega)$ are strictly increasing in $\omega$.
    Hence, for all $\omega >\omega^{\prime}$ and $s\in (t, T)$, the integrand in the final line is strictly negative.
\end{proof}

\section{Proofs: equilibria}\label{appendix:opaque_market}

\subsection{Necessary conditions for surplus maximization}\label{appendix:necessary_conditions_for_surplus_maximization}

The next lemma, which will be used to prove \Cref{prop:inefficient_delay}, gives necessary conditions for a strategy profile to maximize the expected surplus: trade is never with non-$\bar{x}$-buyers, and every type $y$ of the seller is on the market for a total time of $\bar{s}(y)$.

\begin{lemma}\label{lemma:quasi_uniqueness}
    Let $\lambda < \bar{\lambda}$.
    Let $\sigma\in\Sigma$ satisfy $V(\sigma) = \bar{V}$.
    Then $\Leb$-almost all $(t, y)$ satisfy
    \begin{equation}\label{eq:lemma:quasi_uniqueness:1}
        \beta(t, y) \sum_{x\in X\colon x\neq \bar{x}} \alpha(x, p(t, y), m(t, y)) = 0,
    \end{equation}
    and $\Leb$-almost all $y$ satisfy
    \begin{equation}\label{eq:lemma:quasi_uniqueness:2}
        \int_{\R_{+}} \beta(r, y) \alpha(\bar{x}, p(r, y), m(r, y)) \de r
        = \bar{s}(y)
        .
    \end{equation}
    If, additionally, $\lambda > 0$, then $\Leb$-almost all $(t, y)$ such that $t < \bar{s}(y)$ satisfy
    \begin{equation}\label{eq:lemma:quasi_uniqueness:3}
        \int_{[0, t]} \beta(r, y) \alpha(\bar{x}, p(r, y), m(r, y)) \de r = t
        .
    \end{equation}

\end{lemma}
\begin{proof}[Proof of \Cref{lemma:quasi_uniqueness}]
    The proof of \Cref{lemma:delta_maximization} implies \Cref{eq:lemma:quasi_uniqueness:1}.\footnote{
    Specifically, $V(\sigma) = \bar{V}$ implies $V(\sigma\vert y) = \bar{V}(y)$ for almost all $y$, where $V(\sigma\vert y)$ is the surplus generated by $\sigma$ conditional on type $y$, and $\bar{V}(y)$ is the surplus generated by the surplus-maximizing strategy profile from \Cref{sec:commitment:maximizing_surplus} conditional on type $y$.
    \Cref{footnote:delta_maximization_gamma_uniqueness} observes that if $\sigma$ is front-loaded (as defined in the proof of \Cref{lemma:delta_maximization}), then \eqref{eq:lemma:quasi_uniqueness:1} holds. Now fix $y$ such that $V(\sigma\vert y) = \bar{V}(y)$. Step 1 of the proof of \Cref{lemma:delta_maximization} shows that $V(\sigma\vert y)$ can be represented as a mixture over $V(\tilde\sigma\vert y)$, where each $\tilde\sigma$ is front-loaded. Informally, each $\tilde\sigma$ is obtained by truncating $\sigma$ at some cutoff date and shifting all trading intensities to earlier dates; in particular, if $\sigma$ leads to trade between type $y$ and a buyer with some signal $x$ before some date $t$, then so does $\tilde\sigma$. Now, it holds $V(\tilde\sigma\vert y) \leq \bar{V}(y)$ for each $\tilde\sigma$ in the mixture. Thus, almost all $\tilde\sigma$ in the mixture satisfy \eqref{eq:lemma:quasi_uniqueness:1} for $y$, i.e., almost surely do not lead type $y$ to trade with non-$\bar{x}$-buyers. By the construction of the mixture, also $\sigma$ satisfies \eqref{eq:lemma:quasi_uniqueness:1}.}
    Step 2 of that proof implies \Cref{eq:lemma:quasi_uniqueness:2}.
    
    Now let $\lambda > 0$. Briefly, the idea is as follows. For all types $y$ and dates $t$ before the exit time $\bar{s}(y)$, the posterior surplus conditional on this type and the first $\bar{x}$-buyer arriving at $t$ is strictly positive (by choice of $\bar{s}(y)$). Hence, any breakdown at $t$ would diminish the surplus. To avoid breakdowns, the seller should offer the good with probability one and an arriving $\bar{x}$-buyer should accept with probability one, which yields \eqref{eq:lemma:quasi_uniqueness:3}.
    
    Formally, in view of \eqref{eq:lemma:quasi_uniqueness:1}, for all $\omega$ and almost all $(t, y)$, we have 
    \begin{equation*}
    Q_{\sigma}(t\vert y, \omega) = 1 - e^{- f(\bar{x}\vert \omega) \int_{[0, t]} \beta(r, y) \alpha(\bar{x}, p(r, y), m(r, y)) \de r}
    .
    \end{equation*}
    For all $(t, y)$ such that $t < \bar{s}(y)$, let
    \begin{equation*}
        \xi(t, y) = \min\left\lbrace s\in\R_{+}\colon \int_{[0, s]} \beta(r, y) \alpha(\bar{x}, p(r, y), m(r, y)) \de r = t\right\rbrace
        ,
    \end{equation*}
    where the minimum is well-defined in view of \eqref{eq:lemma:quasi_uniqueness:2}.
    Note $\xi(t, y) \geq t$.
    With this definition, $Q_{\sigma}(\xi(t, y)\vert y, \omega) = 1 - e^{- f(\bar{x}\vert \omega) t} = \bar{Q}(t\vert \omega)$ for almost all $(t, y)$ such that $t<\bar{s}(y)$.
    We now claim the following bound on $V(\sigma)$:
    \begin{align*}
        V(\sigma) &=
        \int_{\Omega}\int_{Y}\int_{\R_{+}} v(\omega) e^{-\lambda t} \de Q_{\sigma}(t\vert y, \omega)\de G(y\vert\omega)\de\mu(\omega)
        \\
        &=
        \int_{\Omega}\int_{Y}\int_{\R_{+}} v(\omega) e^{-\lambda \xi(t, y)} \mathbbm{1}_{t \leq \bar{s}(y)} \de \bar{Q}(t\vert \omega)\de G(y\vert\omega)\de\mu(\omega)
        \\
        &=
        \int_{\R_{+}}\int_{Y}\int_{\Omega} v(\omega) g(y\vert\omega)f(\bar{x}\vert \omega) e^{-f(\bar{x}\vert \omega) t} e^{-\lambda \xi(t, y)} \mathbbm{1}_{t \leq \bar{s}(y)} \de\mu(\omega)\de y\de t
        \\
        &\leq
        \int_{\R_{+}}\int_{Y}\int_{\Omega} v(\omega) g(y\vert\omega)f(\bar{x}\vert \omega) e^{-f(\bar{x}\vert \omega) t} e^{-\lambda t} \mathbbm{1}_{t \leq \bar{s}(y)} \de\mu(\omega)\de y\de t
        = \bar{V}
        .
    \end{align*}
    Indeed, the first equalities follow from a change of variables and changing the order of integration.
    The inequality holds since $\xi(t, y) \geq t$ holds almost everywhere and since, by choice of $\bar{s}(y)$, all $t \in [0, \bar{s}(y))$ satisfy 
    \begin{equation*}    
    \sgn \int_{\Omega} v(\omega) g(y\vert\omega)f(\bar{x}\vert \omega) e^{-f(\bar{x}\vert \omega) t} \de\mu(\omega) 
    =
    \sgn \E[v(\bm{\omega})\vert \bm{\tau}_{\bar{x}} = t, \bm{y} = y]
    > 0.
    \end{equation*}
    The inequality $V(\sigma) \leq \bar{V}$ is strict if there is a non-zero set of $(t, y)$ such that $t < \bar{s}(y)$ and $\xi(t, y) > t$.
    Thus, \eqref{eq:lemma:quasi_uniqueness:3} holds.
\end{proof}

\subsection{Proof of \headercref{Theorem}{{prop:inefficient_delay}}}

We first prove the claim assuming $\lambda > 0$. The case $\lambda = 0$ obtains via a small modification, discussed later.

Towards a contradiction, let $\sigma$ be an equilibrium such that $V_{S}(\sigma) = \bar{V}$.

For all $(y, \omega)$, let the probability of trading before a time $t\in\R_{+}$ be $Q_{\sigma}(t\vert y, \omega)$.
Let $Q_{\sigma}(y, \omega) = \lim_{t\to\infty} Q_{\sigma}(t\vert y, \omega)$ be the ex ante trade probability.
For all types $y$, let $V_{S}(y; \sigma)$ be type $y$'s equilibrium utility, i.e.,
\begin{equation*}
    V_{S}(y; \sigma) =
    \int_{\Omega} \int_{\R_{+}}
    (p(t, y) - v_{S}(\omega)) e^{-\lambda t} \de Q_{\sigma}(t\vert y, \omega)  \de\mu(\omega\vert y)
    ,
\end{equation*}
where $\mu(\cdot\vert y)$ is type $y$'s posterior belief (see \cref{footnote:types_posterior_belief}).

For $\lambda \neq 0$, \Cref{lemma:quasi_uniqueness} implies that for almost all $(t, y)$, it holds
\begin{equation}\label{eq:inefficient_delay:tradeprob_with_discounting}
    \forall x\in X,\quad
    \beta(t, y) \alpha(x, p(t, y), m(t, y)) = \mathbbm{1}_{t\leq\bar{s}(y)}\mathbbm{1}_{x = \bar{x}}.
\end{equation} 
That is, the good is offered exactly up to date $\bar{s}(y)$ and an arriving buyer accepts if and only if their signal is $\bar{x}$.
In particular, there exists a pair $(\tilde{p}, \tilde{m})$ such that $\alpha(x, \tilde{p}, \tilde{m}) = \mathbbm{1}_{x = \bar{x}}$.
Let $A$ denote the set of such pairs $(\tilde{p}, \tilde{m})$, and let $p^{\ast} = \sup\lbrace \tilde{p}\colon \exists \tilde{m}\in M\mbox{ s.t. } (\tilde{p}, \tilde{m}) \in A\rbrace$.

We will next establish an identity for the seller's equilibrium utility $V_{S}(y;\sigma)$, valid for almost all $y$.
We begin with an upper bound.
Let $Y'$ be the set of $y$ such that \eqref{eq:inefficient_delay:tradeprob_with_discounting} holds for almost all $t$, so that $Y\setminus Y'$ has measure zero.
For $y\in Y'$, conditional on type $y$'s trading, the price is at most $p^{\ast}$ almost surely.
For such a type $y$, the equilibrium utility is bounded as follows:
\begin{align}
    V_{S}(y; \sigma)
    \leq 
    &\int_{\Omega} \int_{\R_{+}}  (p^{\ast} - v_{S}(\omega)) e^{-\lambda t} \de Q_{\sigma}(t\vert y, \omega) 
    \de\mu(\omega\vert y)
    \nonumber
    \\
    =
    &\E\left[(p^{\ast} - v_{S}(\bm{\omega}))\mathbbm{1}_{\bm{\tau}_{\bar{x}} \leq \bar{s}(y) \wedge \bm{\tau}_{d}}\vert \bm{y} = y\right]
    \label{eq:inefficient_delay:vs_upper_bound}
    ,
\end{align}
where the equality uses \eqref{eq:inefficient_delay:tradeprob_with_discounting}.

We will next derive a lower bound.
Let $\varepsilon > 0$ and find $(\tilde{p}, \tilde{m}) \in A$ such that $\tilde{p} \geq p^{\ast} - \varepsilon$ (which is possible by definition of $p^{\ast}$).
Suppose a type $y\in Y^{\prime}$ of the seller offers $(\tilde{p}, \tilde{m})$ up to time $\bar{s}(y)$.
Since $(\tilde{p}, \tilde{m}) \in A$, this offer is accepted by the first arriving $\bar{x}$-buyer.
Since $\sigma$ is an equilibrium, for almost all $y\in Y^{\prime}$ and all $\varepsilon > 0$,
\begin{equation}\label{eq:inefficient_delay:vs_lower_bound}
    V_{S}(y; \sigma)\geq
    \E\left[(p^{\ast} - v_{S}(\bm{\omega}))\mathbbm{1}_{\bm{\tau}_{\bar{x}} \leq \bar{s}(y) \wedge \bm{\tau}_{d}}\vert \bm{y} = y\right] - \varepsilon
    .
\end{equation}
Let $Y^{\prime\prime}$ be the set of $y\in Y^{\prime}$ satisfying \eqref{eq:inefficient_delay:vs_lower_bound} for all $\varepsilon$. Hence, $Y \setminus Y^{\prime\prime}$ has measure zero.

Combining the bounds, for all types $y\in Y^{\prime\prime}$ the equilibrium utility is given by $\E\left[(p^{\ast} - v_{S}(\bm{\omega}))\mathbbm{1}_{\bm{\tau}_{\bar{x}} \leq \bar{s}(y) \wedge \bm{\tau}_{d}}\vert \bm{y} = y\right]$.
By assumption, the seller's ex ante expected utility also equals $\bar{V}$, where $\bar{V} =\E [(v_{B}(\bm{\omega}) - v_{S}(\bm{\omega})) \mathbbm{1}_{\bm{\tau}_{\bar{x}} \leq \bar{s}(\bm{y}) \wedge \bm{\tau}_{d}}] $.
Since $Y \setminus Y^{\prime\prime}$ has measure zero, integrating over $Y^{\prime\prime}$ delivers 
\begin{equation*}
    \E[(p^{\ast} - v_{S}(\bm{\omega}))\mathbbm{1}_{\bm{\tau}_{\bar{x}} \leq \bar{s}(\bm{y}) \wedge \bm{\tau}_{d}}]
    =
    \E [(v_{B}(\bm{\omega}) - v_{S}(\bm{\omega})) \mathbbm{1}_{\bm{\tau}_{\bar{x}} \leq \bar{s}(\bm{y}) \wedge \bm{\tau}_{d}}].
\end{equation*}
Thus, $0 = \E [(p^{\ast} - v_{B}(\bm{\omega}))  \mathbbm{1}_{\bm{\tau}_{\bar{x}} \leq \bar{s}(\bm{y}) \wedge \bm{\tau}_{d}}]$.
Since $\bm{\tau}_{\bar{x}}$ is strictly positive almost surely, and since $\bar{s}$ is not constantly $0$ (\Cref{lemma:hatv_properties}), also $0 = \E [(p^{\ast} - v_{B}(\bm{\omega}))  \mathbbm{1}_{\bm{\tau}_{\bar{x}} \leq \bar{s}(\bm{y}) \wedge \bm{\tau}_{d}}\vert 0 < \bar{s}(\bm{y})]$.

We next use this equation for $p^{\ast}$ to find a candidate set of types with a profitable deviation.
Let $Y^{\ast}$ be the set of $y$ such that $\bar{s}(y) > 0$ and $0 \leq \E [(p^{\ast} - v_{B}(\bm{\omega}))  \mathbbm{1}_{\bm{\tau}_{\bar{x}} \leq \bar{s}(y) \wedge \bm{\tau}_{d}}\vert \bm{y} = y]$.
Note that $Y^{\ast}$ has non-zero measure, since, otherwise, the equation $0 = \E [(p^{\ast} - v_{B}(\bm{\omega}))  \mathbbm{1}_{\bm{\tau}_{\bar{x}} \leq \bar{s}(\bm{y}) \wedge \bm{\tau}_{d}}\vert 0 < \bar{s}(\bm{y})]$ is contradicted.
Since also $Y\setminus Y^{\prime\prime}$ has measure zero, the intersection $Y^{\ast}\cap Y^{\prime\prime}$ has non-zero measure.

We shall show that all types $y\in Y^{\ast}\cap Y^{\prime\prime}$ have a profitable deviation, which yields the desired contradiction.
Fix $y\in Y^{\ast}\cap Y^{\prime\prime}$.
Let $\varepsilon > 0$ and $\eta > 0$.
Find $(\tilde{p}, \tilde{m}) \in A$ such that $\tilde{p} \geq p^{\ast} - \eta$. Thus, $\alpha(\bar{x}, \tilde{p}, \tilde{m}) = 1$, and $(\tilde{p}, \tilde{m})$ is not accepted by non-$\bar{x}$ buyers.
Suppose type $y$ offers $(\tilde{p}, \tilde{m})$ up to time $(\bar{s}(y) + \varepsilon)$, so that absent a breakdown the probability of trade in an arbitrary state $\omega$ equals $1 - \exp(- f(\bar{x}\vert \omega)(\bar{s}(y) + \varepsilon))$.
The profit from doing so is at least $\E[(p^{\ast} - v_{S}(\bm{\omega})) \mathbbm{1}_{\bm{\tau}_{\bar{x}} \leq (\bar{s}(y) + \varepsilon) \wedge \bm{\tau}_{d}} \vert \bm{y} = y] - \eta$.
We show that for $\varepsilon > 0$ and $\eta > 0$ sufficiently small, it holds
\begin{equation*}
\E[(p^{\ast} - v_{S}(\bm{\omega}))\mathbbm{1}_{\bm{\tau}_{\bar{x}} \leq (\bar{s}(y) + \varepsilon) \wedge \bm{\tau}_{d}}\vert \bm{y} = y] - \eta > \E[(p^{\ast} - v_{S}(\bm{\omega}))\mathbbm{1}_{\bm{\tau}_{\bar{x}} \leq \bar{s}(y) \wedge \bm{\tau}_{d}}\vert \bm{y} = y],
\end{equation*}
which gives the desired contradiction since the right side is type $y$'s utility in the candidate equilibrium (by choice of $Y^{\prime\prime}$).
Clearly, it suffices to show 
\begin{equation*}
\E[(p^{\ast} - v_{S}(\bm{\omega}))\mathbbm{1}_{\bm{\tau}_{\bar{x}} \leq (\bar{s}(y) + \varepsilon) \wedge \bm{\tau}_{d}}\vert \bm{y} = y] > \E[(p^{\ast} - v_{S}(\bm{\omega}))\mathbbm{1}_{\bm{\tau}_{\bar{x}} \leq \bar{s}(y) \wedge \bm{\tau}_{d}}\vert \bm{y} = y]
\end{equation*}
for sufficiently small $\varepsilon > 0$.
For $\varepsilon = 0$, the two expectations coincide.
Thus, it suffices to show that the derivative of $\E[(p^{\ast} - v_{S}(\bm{\omega}))\mathbbm{1}_{\bm{\tau}_{\bar{x}} \leq (\bar{s}(y) + \varepsilon) \wedge \bm{\tau}_{d}}\vert \bm{y} = y]$ with respect to $\varepsilon$ at $\varepsilon = 0$ is strictly positive.
The sign of the derivative equals the sign of 
\begin{equation}\label{eq:inefficient_delay:derivative_sign}
    \E[p^{\ast} - v_{S}(\bm{\omega})\vert \bm{\tau}_{\bar{x}} = \bar{s}(y) \leq \bm{\tau}_{d}, \bm{y} = y].
\end{equation}
Before bounding this term, we note the equation:
\begin{equation}\label{eq:inefficient_delay:bars_definition}
    \E[v_{B}(\bm{\omega}) - v_{S}(\bm{\omega})\vert \bm{\tau}_{\bar{x}} = \bar{s}(y) \leq \bm{\tau}_{d}, \bm{y} = y]
    =
    \E[v_{B}(\bm{\omega}) - v_{S}(\bm{\omega})\vert \bm{\tau}_{\bar{x}} = \bar{s}(y), \bm{y} = y] = 0
    ;
\end{equation}
here, the first equality holds since breakdowns are independent of all other random variables and since $\bm{\tau}_{\bar{x}}$ is already conditioned on a particular realization; the second equality follows from the definition of $\bar{s}(y)$ and since $\bar{s}(y) > 0$ holds.
Returning to \eqref{eq:inefficient_delay:derivative_sign}, we now claim,
\begin{align*}
    \eqref{eq:inefficient_delay:derivative_sign}
    =
    & \E[p^{\ast} - v_{B}(\bm{\omega})\vert \bm{\tau}_{\bar{x}} = \bar{s}(y) \leq \bm{\tau}_{d}, \bm{y} = y]
    \\
    >
    & \E[p^{\ast} - v_{B}(\bm{\omega})\vert \bm{\tau}_{\bar{x}} \leq \bar{s}(y) \wedge \bm{\tau}_{d}, \bm{y} = y]
    \geq
    0
    .
\end{align*}
Indeed, the equality follows from \eqref{eq:inefficient_delay:bars_definition}, the strict inequality follows from \Cref{lemma:strict_monotone_buyer_value_cond_trade_interval} and $\bar{s}(y) > 0$, and the weak inequality follows from the choice of $Y^{\ast}$.
This completes the proof for the case $\lambda > 0$.

Let us now discuss how to modify the proof for the case $\lambda = 0$.
\Cref{lemma:quasi_uniqueness} now delivers that for almost all $(t, y)$ and all $x\in X\setminus\lbrace \bar{x}\rbrace$ it holds $\beta(t, y) \alpha(x, p(t, y), m(t, y)) = 0$; i.e., whenever the seller offers the good, the offer is only possibly accepted by $\bar{x}$-buyers.
\Cref{lemma:quasi_uniqueness} also delivers that the ex ante trade probability equals $1 - \exp(- f(\bar{x}\vert \omega)\bar{s}(y))$ for almost all $(y, \omega)$.\footnote{However, \Cref{lemma:quasi_uniqueness} does not characterize the probability of trading before each given time $t$, as in \eqref{eq:inefficient_delay:tradeprob_with_discounting}; this characterization is not needed without breakdowns since, intuitively, the ex ante trade probability pins down the surplus.}
Thus, there must be pairs $(\tilde{p},\tilde{m})$ that are accepted with non-zero probability by $\bar{x}$-buyers, but surely rejected by all other buyers.
We now define $A$ as the set of such pairs, and define $p^{\ast}$ exactly as above.
With this, one can again show that for almost all $y$ the equilibrium utility of type $y$ equals $\E[(p^{\ast} - v_{S}(\bm{\omega}))\mathbbm{1}_{\bm{\tau}_{\bar{x}} \leq \bar{s}(y)}\vert \bm{y} = y]$.
(This is the same expression as before, now with $\lambda = 0$.)
The rest of the argument is very similar.
\qed

\subsection{Proof of \headercref{Theorem}{{prop:opaque_market:eq_example}}}\label{proof:opaque_market:eq_example}
\setcounter{step}{0}

It suffices to show that if $(\lambda_{n})_{n\in\mathbb{N}}$ is an arbitrary sequence in $\R_{+}$ converging to $0$, then there is a subsequence along which eventually the described strategy profile is an equilibrium.
Fix $(\lambda_{n})_{n\in\mathbb{N}}$.
Since some expectations involve the arrival of a breakdown, we now notate the expectation via $\E_{\lambda_{n}}$ when the breakdown rate is $\lambda_{n}$, and $\E_{0}$ is for the case without breakdowns.

By \Cref{assumption:regularity_for_equilibrium}, it holds $ \E_{0}[v_{B}(\bm{\omega}) ] < \max v_{S}$.
Therefore and since $(\lambda_{n})_{n\in\mathbb{N}}$ vanishes, for sufficiently large $n$ 
\begin{equation}\label{eq:equilibrium_construction:infinite_stopping_price}
    \E_{\lambda_{n}}[v_{B}(\bm{\omega})\vert \bm{\tau}_{\bar{x}} \leq \bm{\tau}_{d}]
    \quad\mbox{is less than and bounded away from}\quad \max v_{S}
    .
\end{equation}
In what follows, we consider such sufficiently large $n$.

The next lemma addresses the existence of candidate prices and stopping times. The proof, presented later, is a routine application of the Intermediate Value Theorem.
\begin{lemma}\label{lemma:candidate_equilibrium:existence}
There exists $\eta > 0$ such that for all $n$ there exists $p^{\ast}_{n}\in\R$ and $s^{\ast}_{n}\colon Y\to\R_{+}$ such that:
\begin{subequations}\label{eq:candidate_eq_construction}
\begin{align}
    \label{eq:candidate_eq_construction:1}
    \forall y\in Y, \quad &s^{\ast}_{n}(y) = \min\left\lbrace t\in\R_{+}\colon \E_{\lambda_{n}}[v_{S}(\bm{\omega}) \vert \bm{\tau}_{\bar{x}} = t, \bm{y} = y]\geq p^{\ast}_{n}\right\rbrace;
    \\
    \label{eq:candidate_eq_construction:2}
    &p^{\ast}_{n} = \E_{\lambda_{n}}[v_{B}(\bm{\omega}) \vert \bm{\tau}_{\bar{x}} \leq s^{\ast}_{n}(\bm{y}) \wedge \bm{\tau}_{d}];
    \\
    \label{eq:candidate_eq_construction:3}
    &\E_{\lambda_{n}}[v_{S}(\bm{\omega}) \vert \bm{\tau}_{\bar{x}} = 0, \bm{y} = \bar{y}] + \eta
    \leq p^{\ast}_{n} \leq \max v_{S} - \eta
    .
\end{align}
\end{subequations}
Moreover, $s^{\ast}_{n}$ is (well-defined and) not constantly zero, but is continuous and weakly increasing, and $s^{\ast}_{n}$ is strictly increasing when non-zero.
\end{lemma}

Using that $p^{\ast}_{n}$ is bounded across $n$ (it lies in the compact interval $[v_{B}(\ubar{\omega}), v_{B}(\bar{\omega})]$), also $s^{\ast}_{n}(y)$ is bounded across $y$ and $n$, as inspection of \eqref{eq:candidate_eq_construction:1} shows.
Since $s^{\ast}_{n}$ is weakly increasing, Helly's Selection Theorem yields a subsequence along which $(p^{\ast}_{n})_{n\in\mathbb{N}}$ converges to a number $p^{\ast}$ while $(s^{\ast}_{n})_{n\in\mathbb{N}}$ converges to a weakly increasing function $s^{\ast}$.
We show that the described strategy profile is an equilibrium far enough along this subsequence.
By possibly relabeling, let this subsequence be the entire sequence.

Routine arguments show that $(p^{\ast}, s^{\ast})$ satisfies the analogue of \eqref{eq:candidate_eq_construction}, i.e., 
\begin{subequations}\label{eq:candidate_eq_construction_limit}
\begin{align}
    \label{eq:candidate_eq_construction_limit:1}
    \forall y\in Y, \quad &s^{\ast}(y) = \min\left\lbrace t\in\R_{+}\colon \E_{0}[v_{S}(\bm{\omega}) \vert \bm{\tau}_{\bar{x}} = t, \bm{y} = y]\geq p^{\ast}\right\rbrace;
    \\
    \label{eq:candidate_eq_construction_limit:2}
    &p^{\ast} = \E_{0}[v_{B}(\bm{\omega}) \vert \bm{\tau}_{\bar{x}} \leq s^{\ast}(\bm{y})];
    \\
    \label{eq:candidate_eq_construction_limit:3}
    &\E_{0}[v_{S}(\bm{\omega}) \vert \bm{\tau}_{\bar{x}} = 0, \bm{y} = \bar{y}] + \eta
    \leq p^{\ast} \leq \max v_{S} - \eta
    .
\end{align}
\end{subequations}

As a technical fact, the convergence of $(s^{\ast}_{n})_{n\in\mathbb{N}}$ to $s^{\ast}$ is uniform.
This fact follows from, e.g., \citet[Chapter 0.1]{resnick2008extreme} by noting that $s^{\ast}$ is continuous (this can be shown from \eqref{eq:candidate_eq_construction_limit:1}), and since for all $n$ the function $s^{\ast}_{n}$ is weakly increasing.
We shall use uniform convergence when invoking \Cref{lemma:delta_maximization} further ahead.

For later reference, note that $\omega\mapsto p^{\ast} - v_{S}(\omega)$ is strictly single-crossing from below at an interior state $\omega_{0}^{\ast}\in (\ubar{\omega}, \bar{\omega})$. 
Indeed, $v_{S}$ is strictly decreasing and continuous, and \eqref{eq:candidate_eq_construction_limit:3} implies that $p^{\ast} - v_{S}$ changes sign.
Since $(p^{\ast}_{n})_{n}$ converges to $p^{\ast}$, for sufficiently large $n$ also $p^{\ast}_{n} - v_{S}$ is strictly single-crossing from below at an interior state $\omega_{0, n}^{\ast} \in (\ubar{\omega}, \bar{\omega})$, where $(\omega_{0, n}^{\ast})_{n\in\mathbb{N}}$ converges to $\omega_{0}^{\ast}$.
In what follows, we consider $n$ such that the single-crossing property holds.

Now let $u_{n} = p^{\ast}_{n} - v_{S}$ and $u = p^{\ast} - v_{S}$.
Applying \Cref{lemma:delta_maximization} to $(\lambda_{n}, u_{n}, u)$, we conclude there exists $n^{\ast}$ such that for all $n\geq n^{\ast}$ and $y\in Y$,
\begin{align}\label{eq:candidate_eq_construction:maximization}
    \nonumber
    & \sup_{\sigma\in\Sigma} 
    \int_{\Omega}\int_{\R_{+}} (p^{\ast}_{n} - v_{S}(\omega)) e^{-\lambda_{n} t} \de Q_{\sigma}(t\vert y, \omega) \de\mu(\omega\vert y)
    \\
    = &
    \int_{\Omega}\int_{\R_{+}} (p^{\ast}_{n} - v_{S}(\omega)) e^{-\lambda_{n} t} \mathbbm{1}_{t\leq s^{\ast}_{n}(y)} \de\bar{Q}(t\vert \omega) \de\mu(\omega\vert y)
    \nonumber
    \\
    = &
    \E_{\lambda_{n}}[(p^{\ast}_{n} - v_{S}(\bm{\omega}))\mathbbm{1}_{\bm{\tau}_{\bar{x}} \leq s^{\ast}_{n}(y) \wedge \bm{\tau}_{d}}\vert \bm{y} = y].
\end{align}
Here, we used that, in light of \eqref{eq:candidate_eq_construction:1}, the function $s^{\ast}_{n}$ equals $s^{\dagger}_{n}$ as defined in \Cref{lemma:delta_maximization}.

Now let $n\geq n^{\ast}$.
We claim for $n\geq n^{\ast}$, there is an equilibrium as described in \Cref{prop:opaque_market:eq_example} with $p^{\ast}_{n}$ and $s^{\ast}_{n}$, respectively, describing the price and exit times, respectively.
Note that the last expression in \eqref{eq:candidate_eq_construction:maximization} is precisely type $y$'s utility in the candidate equilibrium since the seller trades at price $p^{\ast}_{n}$ with the first $\bar{x}$-buyer arriving before the exit time $s^{\ast}_{n}(y)$, provided there is no breakdown, and otherwise the seller does not trade.

Consider the seller.
We have to show that each type $y$ of the seller finds it optimal to offer $p^{\ast}_{n}$ up to time $s^{\ast}_{n}(y)$, yielding the utility on the right side of \eqref{eq:candidate_eq_construction:maximization}.
Let $(\beta, p, m)$ be an arbitrary seller strategy.
Since buyers reject all prices strictly above $p^{\ast}_{n}$, assume that the strategy $p$ never sets a price strictly greater than $p^{\ast}_{n}$ (the seller may as well not offer the good).
Thus, type $y$'s utility from $(\beta, p, m)$ is at most
\begin{equation*}
    \int_{\Omega}\int_{\R_{+}} (p^{\ast}_{n} - v_{S}(\omega)) e^{-\lambda_{n} t} \de Q_{\sigma}(t\vert y, \omega) \de\mu(\omega\vert y)
    .
\end{equation*}
In view of \eqref{eq:candidate_eq_construction:maximization}, this upper bound is at most type $y$'s utility in the candidate equilibrium.

To verify the buyers' best replies, recall that, in the candidate equilibrium and for a given private signal $x$, an arriving buyer infers only that they arrived at a time before another buyer with signal $\bar{x}$, before a breakdown, and before the seller has left the market. 
This buyer's posterior value equals
\begin{equation*}
    \frac{\int_{\Omega}\int_{Y}\int_{0}^{s^{\ast}_{n}(y)} v_{B}(\omega) f(x\vert \omega) e^{-(\lambda_{n} +  f(\bar{x}\vert \omega)) t} \de t\de G(y\vert\omega)\de\mu(\omega)}
    {\int_{\Omega}\int_{Y}\int_{0}^{s^{\ast}_{n}(y)} f(x\vert \omega) e^{-(\lambda_{n} +  f(\bar{x}\vert \omega)) t} \de t \de G(y\vert\omega)\de\mu(\omega)}
    = \frac{\int_{\Omega} v_{B}(\omega) f(x\vert\omega)  \nu(\omega) \de\mu(\omega)}
    {\int_{\Omega}f(x\vert\omega) \nu(\omega) \de\mu(\omega)}
    ,
\end{equation*}
where $\nu(\omega) = \int_{Y}\int_{[0, s^{\ast}_{n}(y)]} e^{-(\lambda_{n} +  f(\bar{x}\vert \omega)) t} \de t\de G(y\vert\omega)$ for all $\omega$.
Importantly, $\nu$ does not depend on the buyer's signal $x$.
(This is where conditional independence is used.)
Since $v_{B}$ is strictly increasing, and in view of \Cref{assumption:correlated}, Theorem 2.2 of \citet{karlin1980mtp} implies that the posterior value is maximized at $x = \bar{x}$. For $x = \bar{x}$, the value equals $p_{n}^{\ast}$. Thus, the described buyer strategy is a best reply.
\qed

\begin{proof}[Proof of \Cref{lemma:candidate_equilibrium:existence}]
    For all $y\in Y$ and $t \in \R_{+}$, let $\hat{v}_{S}(t, y) = \E_{0}[v_{S}(\bm{\omega}) \vert \bm{\tau}_{\bar{x}} = t, \bm{y} = y]$.
    Notice that $\hat{v}_{S}$ does not depend on $n$, i.e., on the breakdown rate $\lambda_{n}$.
    The function $\hat{v}_{S}$ is continuous in both arguments, strictly increasing in $t$, and strictly decreasing in $y$.
    Moreover, for fixed $y$ the value $\hat{v}_{S}(t, y)$ converges to $v_{S}(\ubar{\omega}) = \max v_{S}$.
    These properties can be shown analogously to arguments given in other appendices using that $v_{S}$ is strictly decreasing in the state while $g$ is strictly log-supermodular.
    
    For all $p\in (\hat{v}_{S}(0, \bar{y}), \max v_{S})$ and $y\in Y$, let $\hat{s}(p, y)$ be the smallest $t\in \R_{+}$ such that $p \leq \hat{v}_{S}(t, y)$; such a time exists since $\hat{v}_{S}(t, y)\to \max v_{S}$ as $t\to\infty$.
    The value $\hat{s}(p, y)$ is continuous and weakly increasing in $p$ and $y$ since $\hat{v}_{S}(t, y)$ is strictly increasing and continuous in $t$, while $\hat{v}_{S}(t, y)$ is strictly decreasing and continuous in $y$. By the same argument, $\hat{s}(p, y)$ is strictly increasing in $y$ when non-zero.
    
    Finally, for all $n\in\mathbb{N}$ and $\tilde{s}\colon Y \to\R_{+}$ such that $\tilde{s}(y) > 0$ for a non-zero set of $y$, let $\hat{p}_{n}(\tilde{s}) = \E_{\lambda_{n}}[v_{B}(\bm{\omega}) \vert \bm{\tau}_{\bar{x}} \leq \tilde{s}(\bm{y})\wedge \bm{\tau}_{d}]$.

    Fixing $n$, we next show there exists $p^{\ast}_{n} \in (\hat{v}_{S}(0, \bar{y}), \max v_{S})$ such that $p^{\ast}_{n} = \hat{p}_{n}(\hat{s}(p^{\ast}_{n}, \cdot))$.
    The candidate function $s^{\ast}_{n}$ obtains by letting $s^{\ast}_{n} = \hat{s}(p^{\ast}_{n}, \cdot)$.
    Along the way, we argue that $s^{\ast}_{n}$ is non-zero for a non-zero set of types.
    In a final step, we show there is $\eta > 0$ such that \eqref{eq:candidate_eq_construction:3} holds for sufficiently large $n$.
    
    First, we argue $\hat{p}_{n}(\hat{s}(p, \cdot)) < p$ for $p$ sufficiently close to $\max v_{S}$.
    For $p \to \max v_{S}$, we get $\hat{s}(p, y) \to \infty$ for all fixed $y\in Y$ since $\hat{v}_{S}(t, y)$ strictly increases in $t$ and converges to $\max v_{S}$ as $t\to\infty$.
    By Dominated Convergence, $\lim_{p\to \max v_{S}}\hat{p}_{n}(\hat{s}(p, \cdot)) = \E_{\lambda_{n}}[v_{B}(\bm{\omega})\vert \bm{\tau}_{\bar{x}} \leq \bm{\tau}_{d}] < \max v_{S}$, where we recall \eqref{eq:equilibrium_construction:infinite_stopping_price}.
    
    Second, we argue $p < \hat{p}_{n}(\hat{s}(p, \cdot))$ for $p$ sufficiently close to $\hat{v}_{S}(0, \bar{y})$.
    For $p \in (\hat{v}_{S}(0, \bar{y}), \max v_{S})$ sufficiently close to $\hat{v}_{S}(0, \bar{y})$, there exists $y_{p} \in Y$ such that $\hat{v}_{S}(0, y_{p}) = p$; for such a price $p$, we have $\hat{s}(p, y) > 0$ if and only if $y > y_{p}$. 
    Thus, $\hat{p}_{n}(\hat{s}(p, \cdot)) = \E_{\lambda_{n}}[v_{B}(\bm{\omega})\vert \bm{\tau}_{\bar{x}} \leq \hat{s}(p, \bm{y}) \wedge \bm{\tau}_{d}, \bm{y} \geq y_{p}]$.
    Moreover, $0\leq \hat{s}(p, y) \leq \hat{s}(p, \bar{y})$ for all $y > y_{p}$.
    The expectation $\E_{\lambda_{n}}[v_{B}(\bm{\omega})\vert \bm{\tau}_{\bar{x}} = t, \bm{y}  = y]$ is continuous in $(t, y)$, and the price $\hat{p}_{n}(\hat{s}(p, \cdot))$ is an average over this function on a subset of $(t, y)\in [0, \hat{s}(p, \bar{y})]\times [y_{p}, \bar{y}]$.
    Since $y_{p}\to\bar{y}$ and $\hat{s}(p, \bar{y})\to 0$ as $p\to \hat{v}_{S}(0, \bar{y})$, we get $\hat{p}_{n}(\hat{s}(p, \cdot)) \to \E_{\lambda_{n}}[v_{B}(\bm{\omega})\vert \bm{\tau}_{\bar{x}} = 0, \bm{y}  = \bar{y}]$ as $p\to\hat{v}_{S}(0, \bar{y})$. 
    \Cref{assumption:trade_is_possible} asserts $\hat{v}_{S}(0, \bar{y}) = \E_{\lambda_{n}}[v_{S}(\bm{\omega})\vert \bm{\tau}_{\bar{x}} = 0, \bm{y} = \bar{y}] < \E_{\lambda_{n}}[v_{B}(\bm{\omega})\vert \bm{\tau}_{\bar{x}} = 0, \bm{y} = \bar{y}]$.
    Thus, $p < \hat{p}_{n}(\hat{s}(p, \cdot))$ for $p$ sufficiently close to $\hat{v}_{S}(0, \bar{y})$.

    The above argument also shows that for $p$ close to but greater than $\hat{v}_{S}(0, \bar{y})$ the function $\hat{s}(p, \cdot)$ is non-zero for a non-zero set of types.
    Since $\hat{s}(p, \cdot)$ is pointwise increasing in $p$, it follows that for all $p\in (\hat{v}_{S}(0, \bar{y}), \max v_{S})$ the function $\hat{s}(p, \cdot)$ is non-zero for a non-zero set of types.
    Equipped with this fact, Dominated Convergence verifies that $p\mapsto \hat{p}_{n}(\hat{s}(p, \cdot))$ is continuous on $(\hat{v}_{S}(0, \bar{y}), \max v_{S})$.
    Thus, there exists $p^{\ast}_{n}$ satisfying $p^{\ast}_{n} = \hat{p}_{n}(\hat{s}(p^{\ast}_{n}, \cdot))$, and $\hat{s}(p^{\ast}_{n}, \cdot)$ is non-zero for a non-zero set of types.
    
    Finally, we argue that there exists $\eta > 0$ such that \eqref{eq:candidate_eq_construction:3} holds for all but finitely many $n$, i.e., $\E_{0}[v_{S}(\bm{\omega}) \vert \bm{\tau}_{\bar{x}} = 0, \bm{y} = \bar{y}] + \eta
        \leq p^{\ast}_{n} \leq \max v_{S} - \eta$.
    If $p^{\ast}_{n} \to \max v_{S}$ along a subsequence, then along this subsequence $s^{\ast}_{n}$ diverges pointwise so that, similar to the above argument, one can show $\hat{p}_{n}(s^{\ast}_{n}) \to \E_{0}[v_{B}(\bm{\omega})]$ along the subsequence; but this limit is strictly less than $\max v_{S}$ (by \eqref{eq:equilibrium_construction:infinite_stopping_price}), contradicting $p^{\ast}_{n} = \hat{p}_{n}(s^{\ast}_{n})$ far enough along the subsequence.
    Likewise, if $p^{\ast}_{n}\to \E_{0}[v_{S}(\bm{\omega}) \vert \bm{\tau}_{\bar{x}} = 0, \bm{y} = \bar{y}]$ along a subsequence, then, similar to the above argument, $\hat{p}_{n}(s^{\ast}_{n}) \to \E_{0}[v_{B}(\bm{\omega})\vert \bm{\tau}_{\bar{x}} = 0, \bm{y} = \bar{y}]$; but this limit is strictly larger than $\E_{0}[v_{S}(\bm{\omega}) \vert \bm{\tau}_{\bar{x}} = 0, \bm{y} = \bar{y}]$, contradicting $p^{\ast}_{n} = \hat{p}_{n}(s^{\ast}_{n})$ far enough along the subsequence.
    \end{proof}

\subsection{Proof of \headercref{Proposition}{{prop:comparison_eq_vs_commitment}}}
    \setcounter{step}{0}
    
    Without using affinity, we first show:

    \begin{step}
        It holds $s^{\ast}(y) > \bar{s}(y)$ for all $y$ in a non-empty open set.
    \end{step}

    Since $s^{\ast}$ and $\bar{s}$ are continuous, it suffices to rule out that $s^{\ast}(y) \leq \bar{s}(y)$ holds for all $y$.
    Towards a contradiction, suppose $s^{\ast}(y) \leq \bar{s}(y)$ for all $y$.
    Recall that $\bar{s}$ and $s^{\ast}$ are both weakly increasing, and neither is constantly $0$. 
    Since $s^{\ast}\leq\bar{s}$, also $y_{0}\leq y^{\ast}_{0}$, where we define $y_{0} = \inf\lbrace y \in Y\colon \bar{s}(y) > 0\rbrace$ and $y_{0}^{\ast} = \inf\lbrace y \in Y\colon s^{\ast}(y) > 0\rbrace$.
    
    For a later step, define for all types $y\in Y$:
    \begin{align*}
        \phi^{\ast}(y) &= \E[v_{B}(\bm{\omega})\vert \bm{\tau}_{\bar{x}} \leq s^{\ast}(y) \wedge \bm{\tau}_{d}, \bm{y} = y]
        ,
        \\
        \bar{\phi}(y) &= \E[v_{B}(\bm{\omega})\vert \bm{\tau}_{\bar{x}} \leq \bar{s}(y)\wedge \bm{\tau}_{d}, \bm{y} = y]
    \end{align*}
    provided the respective exit times, $s^{\ast}(y)$ and $\bar{s}(y)$, are non-zero.
    Further, let $V^{\ast}(y) = \E[(p^{\ast} - v_{S}(\bm{\omega}))\mathbbm{1}_{\bm{\tau}_{\bar{x}} \leq s^{\ast}(y) \wedge \bm{\tau}_{d}}\vert \bm{y} = y]$ and $\bar{V}(y) = \E[(v_{B}(\bm{\omega}) - v_{S}(\bm{\omega}))\mathbbm{1}_{\bm{\tau}_{\bar{x}}\leq \bar{s}(y) \wedge \bm{\tau}_{d}}\vert \bm{y} = y]$, respectively, be type $y$'s equilibrium and commitment expected utility, respectively.

    Since $s^{\ast}(y) \leq \bar{s}(y)$ for all $y$, \Cref{lemma:strict_monotone_buyer_value_cond_trade_interval} implies $\phi^{\ast}(y) \geq \bar{\phi}(y)$ for all $y$ such that $s^{\ast}(y)>0$ (for these $y$, both $\phi^{\ast}(y)$ and $\bar{\phi}(y)$ are defined).

    We first claim that there is a set $Y'$ having non-zero measure such that $p^{\ast} < \bar{\phi}(y)$ and $s^{\ast}(y) > 0$ for all $y\in Y'$.
    From the construction in \Cref{prop:opaque_market:eq_example}, we have $s^{\ast}(y) > 0$ for an open set of $y$.
    Towards a contradiction, let $p^{\ast} \geq \bar{\phi}(y)$ for almost all $y$ such that $s^{\ast}(y) > 0$.
    Since $s^{\ast} \leq \bar{s}$, also $\bar{s}(y) > 0$.
    Type $y$'s equilibrium utility $V^{\ast}(y)$ is at least the utility from offering $p^{\ast}$ until time $\bar{s}(y)$.
    Thus:
    \begin{align*}
        V^{\ast}(y) \geq 
        & \E[(p^{\ast} - v_{S}(\bm{\omega}))\mathbbm{1}_{\bm{\tau}_{\bar{x}} \leq \bar{s}(y) \wedge \bm{\tau}_{d}}\vert \bm{y} = y] \\ \geq 
        & \E[(\bar{\phi}(y) - v_{S}(\bm{\omega}))\mathbbm{1}_{\bm{\tau}_{\bar{x}} \leq \bar{s}(y) \wedge \bm{\tau}_{d}}\vert \bm{y} = y] \\ 
        = & \E[(v_{B}(\bm{\omega}) - v_{S}(\bm{\omega}))\mathbbm{1}_{\bm{\tau}_{\bar{x}} \leq \bar{s}(y) \wedge \bm{\tau}_{d}}\vert \bm{y} = y] = \bar{V}(y),
    \end{align*}
    where the penultimate equality is by definition of $\bar{\phi}(y)$.
    In particular, almost all types $y$ such that $s^{\ast}(y) > 0$ obtain at least $\bar{V}(y)$ in equilibrium.
    By continuity, also $V^{\ast}(y_{0}^{\ast}) \geq \bar{V}(y_{0}^{\ast})$ for the cutoff type $y_{0}^{\ast}$ who enters in equilibrium.
    Consequently, the cutoff types for entry coincide, $y_{0} = y_{0}^{\ast}$.\footnote{Given the assumption $s^{\ast} \leq \bar{s}$, necessarily $y_{0} \leq y_{0}^{\ast}$. If $y_{0} < y_{0}^{\ast}$, then $\bar{V}(y_{0}^{\ast}) > 0$. But since $V^{\ast}(y_{0}^{\ast}) \geq \bar{V}(y_{0}^{\ast})$, it would follow that type $y_{0}^{\ast}$ finds it strictly optimal to enter in equilibrium.}
    In particular, all types $y$ who enter in equilibrium obtain at least $\bar{V}(y)$, and these are also exactly the types who enter under the commitment solution. Therefore, the seller's expected equilibrium profit across all types is at least $\bar{V}$, contradicting \Cref{prop:inefficient_delay}.

    Next, we claim that there exists $y\in Y$ such that $s^{\ast}(y) > 0$ and $p^{\ast} > \phi^{\ast}(y)$.
    Let $Y'$ be as above.
    Since $p^{\ast} < \bar{\phi}(y)$ for all $y\in Y'$, and since $\bar{\phi} \leq \phi^{\ast}$ (as observed earlier), also $p^{\ast} < \phi^{\ast}(y)$ and $s^{\ast}(y) > 0$ for all $y\in Y'$.
    Now recall $p^{\ast} = \E[v_{B}(\bm{\omega})\vert \bm{\tau}_{\bar{x}}\leq s^{\ast}(\bm{y})\wedge \bm{\tau}_{d}]$ from \Cref{prop:opaque_market:eq_example}.
    By iterated expectations and since arrival times are almost surely strictly positive, it follows $p^{\ast} = \E[\phi^{\ast}(\bm{y})\vert \bm{\tau}_{\bar{x}} \leq s^{\ast}(\bm{y})\wedge \bm{\tau}_{d}, s^{\ast}(\bm{y}) > 0]$.
    Since $Y'$ has non-zero measure and $p^{\ast} < \phi^{\ast}(y')$ for all $y'\in Y'$, there must exist $y$ such that $s^{\ast}(y) > 0$ and $p^{\ast} > \phi^{\ast}(y)$.

    Now, fix $y\in Y$ such that $s^{\ast}(y) > 0$ and $p^{\ast} > \phi^{\ast}(y)$, which was just proven to exist.
    We argue that $s^{\ast}(y) \leq \bar{s}(y)$ is contradicted for type $y$.
    Note $0 < s^{\ast}(y) \leq \bar{s}(y)$, where the first inequality holds by the choice of $y$, and the second is by assumption.
    Thus, $\phi^{\ast}(y)$ and $\bar{\phi}(y)$ are defined.
    We now infer $p^{\ast} > \phi^{\ast}(y) \geq \bar{\phi}(y)$, where the first inequality is by choice of $y$, and the second inequality was observed earlier to follow from $s^{\ast}(y) \leq \bar{s}(y)$.
    In particular, $p^{\ast} > \bar{\phi}(y)$.
    We further bound the right side of this inequality.
    \Cref{lemma:strict_monotone_buyer_value_cond_trade_interval} implies $\bar{\phi}(y) \geq \E[v_{B}(\bm{\omega}) \vert \bm{\tau}_{\bar{x}} = \bar{s}(y), \bm{y}=y]$.
    Further, since $\bar{s}(y) > 0$ it holds $\E[v_{B}(\bm{\omega}) \vert \bm{\tau}_{\bar{x}} = \bar{s}(y), \bm{y}=y] = \E[v_{S}(\bm{\omega}) \vert \bm{\tau}_{\bar{x}} = \bar{s}(y), \bm{y}=y]$.
    Therefore, also $p^{\ast} > \E[v_{S}(\bm{\omega}) \vert \bm{\tau}_{\bar{x}} = \bar{s}(y), \bm{y}=y]$.
    Since $v_{S}$ is weakly decreasing and $s^{\ast}(y) \leq \bar{s}(y)$, we have $\E[v_{S}(\bm{\omega}) \vert \bm{\tau}_{\bar{x}} = \bar{s}(y), \bm{y}=y] \geq \E[v_{S}(\bm{\omega}) \vert \bm{\tau}_{\bar{x}} = s^{\ast}(y), \bm{y}=y]$ (this can be shown completely analogously to \Cref{lemma:strict_monotone_buyer_value_cond_trade_interval}).
    We conclude $p^{\ast} > \E[v_{S}(\bm{\omega}) \vert \bm{\tau}_{\bar{x}} = s^{\ast}(y), \bm{y}=y]$.
    This strict inequality contradicts the definition of $s^{\ast}$: in the assumed equilibrium type $y$ has a strict incentive to stay on the market at time $s^{\ast}(y)$ by offering $p^{\ast}$ for a small additional time interval rather than exiting at $s^{\ast}(y)$.

    Thus, there exists an open set of $y$ such that $s^{\ast}(y) > \bar{s}(y)$.

    \begin{step}
        If $v_{B}$ and $v_{S}$ are affine, then $s^{\ast}(y) > \bar{s}(y)$ for all $y\in [y_{0}, \bar{y}]$, where $y_{0} = \inf\lbrace y\colon \bar{s}(y) > 0\rbrace$.
    \end{step}

    Since $v_{B}$ and $v_{S}$ are affine in the state, also $v_{B}$ is affine in $v_{S}$, i.e., $v_{B} = a + bv_{S}$ for some $a\in\mathbb{R}$ and $b < 0$ (where $b < 0$ holds since $v_{B}$ and $-v_{S}$ are both strictly increasing).
    Put $h(z) = a + b z - z$ for all $z\in\mathbb{R}$.
    By linearity of expectations, it follows that for all $t$ and $y$, we can write the following posterior surplus as:
    \begin{equation*}
        \mathbb{E}[v_{B}(\bm{\omega}) - v_{S}(\bm{\omega}) \vert \bm{\tau}_{\bar{x}} = t, \bm{y} = y] = h(\mathbb{E}[v_{S}(\bm{\omega}) \vert \bm{\tau}_{\bar{x}} = t, \bm{y} = y])
        .
    \end{equation*}
    We use this expression for the posterior surplus throughout the argument.
    
    Notice that $h$ is strictly decreasing since $b< 0$.
    We also recall that the posterior surplus $t\mapsto h(\mathbb{E}[v_{S}(\bm{\omega}) \vert \bm{\tau}_{\bar{x}} = t, \bm{y} = y])$ is strictly single-crossing from above (\Cref{lemma:hatv_properties}).

    Now, by step 1 there is a type $y'$ such that $s^{\ast}(y') > \bar{s}(y')$.
    Since $s^{\ast}(y') > 0$, we get
    \begin{equation*}
        p^{\ast} = \mathbb{E}[v_{S}(\bm{\omega}) \vert \bm{\tau}_{\bar{x}} = s^{\ast}(y'), \bm{y} = y']
        .
    \end{equation*}  
    By single-crossing of the posterior surplus, and by $s^{\ast}(y') > \bar{s}(y')$, we find:
    \begin{equation*}
        0 > 
        h(\mathbb{E}[v_{S}(\bm{\omega}) \vert \bm{\tau}_{\bar{x}} = s^{\ast}(y'), \bm{y} = y'])
        .
    \end{equation*}
    In particular, $0 > h(p^{\ast})$.
    Now let $y\in [y_{0}, \bar{y}]$ be arbitrary.
    We have to show $s^{\ast}(y) > \bar{s}(y)$.    
    By definition of $y_{0}$, it is weakly efficient for $y$ to enter, i.e., the posterior surplus at zero is weakly positive,
    \begin{equation*}
        0 \leq h(\mathbb{E}[v_{S}(\bm{\omega}) \vert \bm{\tau}_{\bar{x}} = 0, \bm{y} = y])
        .
    \end{equation*}
    Since $0 > h(p^{\ast})$ and $h$ is strictly decreasing, we conclude $p^{\ast} > \mathbb{E}[v_{S}(\bm{\omega}) \vert \bm{\tau}_{\bar{x}} = 0, \bm{y} = y]$.
    By definition of $s^{\ast}$, this inequality implies $s^{\ast}(y) > 0$.
    In particular, $s^{\ast}(y)$ satisfies
    \begin{equation*}
        p^{\ast} = \mathbb{E}[v_{S}(\bm{\omega}) \vert \bm{\tau}_{\bar{x}} = s^{\ast}(y), \bm{y} = y]
        .
    \end{equation*}
    Applying $h$ to both sides and recalling $0 > h(p^{\ast})$, we infer:
    \begin{equation*}
        0 > h(\mathbb{E}[v_{S}(\bm{\omega}) \vert \bm{\tau}_{\bar{x}} = s^{\ast}(y), \bm{y} = y])
        .
    \end{equation*}
    That is, the posterior surplus given $y$ and the first $\bar{x}$-buyer arriving at $s^{\ast}(y)$ is strictly negative.
    Consequently, $s^{\ast}(y) > \bar{s}(y)$, as promised. 
    \qed

\subsection{Proof of \headercref{Proposition}{{prop:tax_equilibrium}}}

Take $\psi$ sufficiently close to $0$ satisfying
 $\int_{\Omega} (v(\omega) - \psi) g(\bar{y}\vert\omega) f(\bar{x}\vert\omega)\de\mu(\omega) > 0$,
which is possible by \Cref{assumption:trade_is_possible}.
Fixing such $\psi$, there exists an equilibrium with tax rate $\psi$ in which each type of the seller offers the good at a price $p^{\ast}_{\psi}$ up to a time $s^{\ast}_{\psi}(y)$. 
Moreover, $s^{\ast}_{\psi}$ is weakly increasing, continuous, not constantly zero, and satisfies
\begin{align*}
    p^{\ast}_{\psi} &= \E[v_{B}(\bm{\omega})\vert \bm{\tau}_{\bar{x}} \leq s^{\ast}_{\psi}(\bm{y})]
    ,
    \\
    \forall y\in Y,\quad s^{\ast}_{\psi}(y) &= \min\left\lbrace t\in\R_{+}\colon \E[v_{S}(\bm{\omega}) \vert \bm{\tau}_{\bar{x}} = t, \bm{y} = y] \geq p^{\ast}_{\psi} - \psi\right\rbrace.
\end{align*}
The argument simply applies \Cref{prop:opaque_market:eq_example} with the modified reservation value $v_{S} + \psi$.

Let $\mathcal{E}^{\ast}_{0}$ be the set of exit times $s^{\ast}_{0}$ in equilibria without tax ($\psi=0$) as in \Cref{prop:opaque_market:eq_example}.

We first argue that, fixing such $\psi > 0$ and $(p^{\ast}_{\psi}, s^{\ast}_{\psi})$, there does not exist $ s^{\ast}_{0}\in\mathcal{E}^{\ast}_{0}$ such that $s^{\ast}_{\psi}\geq s^{\ast}_{0}$.
Towards a contradiction, let $s^{\ast}_{\psi}\geq s^{\ast}_{0}$.
Let $p^{\ast}_{0}$ be the price associated with $s^{\ast}_{0}$, i.e., $p^{\ast}_{0} = \E_{\lambda}[v_{B}(\bm{\omega})\vert \bm{\tau}_{\bar{x}}\leq s^{\ast}_{0}(\bm{y})]$.
\Cref{prop:comparison_eq_vs_commitment} asserts $s^{\ast}_{0} \geq \bar{s}$.
From, e.g., the proof of \Cref{lemma:delta_maximization} we also know that, for all $y\in Y$, the map $t\mapsto \E[v(\bm{\omega})\mathbbm{1}_{\bm{\tau}_{\bar{x}} \leq t}\vert \bm{y} = y]$ is strictly quasiconcave and uniquely maximized at $\bar{s}(y)$.
Thus, $s^{\ast}_{\psi}\geq s^{\ast}_{0} \geq \bar{s}$ implies
\begin{equation*}
    \E[v(\bm{\omega})\mathbbm{1}_{\bm{\tau}_{\bar{x}} \leq s^{\ast}_{\psi}(\bm{y})}]
    \leq
    \E[v(\bm{\omega})\mathbbm{1}_{\bm{\tau}_{\bar{x}} \leq s^{\ast}_{0}(\bm{y})}].
\end{equation*}
Using $    p^{\ast}_{\psi} = \E[v_{B}(\bm{\omega})\vert \bm{\tau}_{\bar{x}}\leq s^{\ast}_{\psi}(\bm{y})]$ and $    p^{\ast}_{0} = \E[v_{B}(\bm{\omega})\vert \bm{\tau}_{\bar{x}}\leq s^{\ast}_{0}(\bm{y})]$, the previous inequality and $\psi > 0$ imply
\begin{equation*}
    \E[(p^{\ast}_{\psi} - \psi - v_{S}(\bm{\omega}))\mathbbm{1}_{\bm{\tau}_{\bar{x}} \leq s^{\ast}_{\psi}(\bm{y})}]
    <
    \E[(p^{\ast}_{0} - v_{S}(\bm{\omega}))\mathbbm{1}_{\bm{\tau}_{\bar{x}} \leq s^{\ast}_{0}(\bm{y})}].
\end{equation*}
Since for all $y\in Y$ the exit time $s^{\ast}_{\psi}(y)$ is optimal, we also know
\begin{equation*}
    \E[(p^{\ast}_{\psi} - \psi - v_{S}(\bm{\omega}))\mathbbm{1}_{\bm{\tau}_{\bar{x}} \leq s^{\ast}_{\psi}(\bm{y})}]
    \geq 
        \E[(p^{\ast}_{\psi} - \psi - v_{S}(\bm{\omega}))\mathbbm{1}_{\bm{\tau}_{\bar{x}} \leq s^{\ast}_{0}(\bm{y})}]
    .
\end{equation*}
Combining the above two inequalities, we
get
\begin{equation*}
    \E[(p^{\ast}_{\psi} - \psi - v_{S}(\bm{\omega}))\mathbbm{1}_{\bm{\tau}_{\bar{x}} \leq s^{\ast}_{0}(\bm{y})}] < \E[(p^{\ast}_{0} - v_{S}(\bm{\omega}))\mathbbm{1}_{\bm{\tau}_{\bar{x}} \leq s^{\ast}_{0}(\bm{y})}].
\end{equation*}
In particular, we conclude $p^{\ast}_{\psi} - \psi < p^{\ast}_{0}$.
(Here, we used that $s^{\ast}_{0}$ is not constantly zero.)
But this strict inequality readily implies $s^{\ast}_{\psi}(y) \leq s^{\ast}_{0}(y)$ for all $y$, with a strict inequality whenever $s^{\ast}_{0}(y) > 0$; contradiction to $s^{\ast}_{\psi} \geq s^{\ast}_{0}$.

Thus, for all $s^{\ast}_{0}\in\mathcal{E}^{\ast}_{0}$ we have $s^{\ast}_{\psi}(y) < s^{\ast}_{0}(y)$ for at least one type $y$.
By inspecting the definitions of $s^{\ast}_{0}(y)$ and $s^{\ast}_{\psi}(y)$ for this type $y$ and by using that the function $t\mapsto \E[v_{S}(\bm{\omega})\vert \bm{\tau}_{\bar{x}} = t, \bm{y}=y]$ is strictly increasing, it follows that the inequality $s^{\ast}_{\psi}(y) < s^{\ast}_{0}(y)$ requires $p_{0}^{\ast} > p^{\ast}_{\psi} - \psi$.
This inequality for the prices, in turn, implies $s^{\ast}_{\psi}(y') \leq s^{\ast}_{0}(y')$ for all $y'\in Y$, with a strict inequality whenever $s^{\ast}_{0}(y') > 0$.

Now consider a sequence $(\psi_{n})_{n\in\mathbb{N}}$ converging to $0$.
For all $n$, let $(p^{\ast}_{\psi_{n}}, s^{\ast}_{\psi_{n}})$ be as above.
Invoking Helly and possibly passing to a subsequence, let $(p^{\ast}_{\psi_{n}}, s^{\ast}_{\psi_{n}})_{n\in\mathbb{N}}$ converge pointwise to $(p^{\ast\ast}_{0}, s^{\ast\ast}_{0})$.
One may verify $s^{\ast\ast}_{0} \in \mathcal{E}^{\ast}_{0}$; i.e., $s^{\ast\ast}_{0}$ defines an equilibrium without tax as in \Cref{prop:opaque_market:eq_example}.

As a technical fact, the convergence of $(s^{\ast}_{\psi_{n}})_{n\in\mathbb{N}}$ to $s^{\ast\ast}_{0}$ is uniform.
This fact follows from, e.g., \citet[Chapter 0.1]{resnick2008extreme} by noting that $s^{\ast\ast}_{0}$ is continuous (since $s^{\ast\ast}_{0} \in \mathcal{E}^{\ast}_{0}$), and that for all $n$ the function $s^{\ast}_{\psi_{n}}$ is weakly increasing.

We complete the proof by arguing that for large enough $n$ the expected surplus in the chosen equilibrium with tax $\psi_{n}$ is strictly higher than in all equilibria without tax defined by some $s^{\ast}_{0} \in \mathcal{E}^{\ast}_{0}$.
We first describe the candidate $n$.
Since $s^{\ast\ast}_{0} \in\mathcal{E}^{\ast}_{0}$ and $\bar{s}(\ubar{y}) > 0$, \Cref{prop:comparison_eq_vs_commitment} implies $s^{\ast\ast}_{0}(y) > \bar{s}(y)$ for all $y$.
Since $s^{\ast\ast}_{0}$ and $\bar{s}$ are continuous, and since the convergence of $(s^{\ast}_{\psi_{n}})_{n\in\mathbb{N}}$ to $s^{\ast\ast}_{0}$ is uniform, for sufficiently large $n$ also $s^{\ast}_{\psi_{n}}(y) > \bar{s}(y)$ for all $y$.
Fix such $n$.
Now let $s^{\ast}_{0} \in \mathcal{E}^{\ast}_{0}$ be arbitrary.
An earlier step implies $s^{\ast}_{0}(y) > s^{\ast}_{\psi_{n}}(y)$ for all $y$.
From, e.g., the proof of \Cref{lemma:delta_maximization} we also know that, for all $y\in Y$, the map $t\mapsto \E[v(\bm{\omega})\mathbbm{1}_{\bm{\tau}_{\bar{x}} \leq t}\vert \bm{y} = y]$ is strictly quasiconcave and uniquely maximized at $\bar{s}(y)$.
Since $ s^{\ast}_{0}(y) > s^{\ast}_{\psi_{n}}(y) > \bar{s}(y)$ for all $y$, we conclude
    $
    \E[v(\bm{\omega})\mathbbm{1}_{\bm{\tau}_{\bar{x}} \leq s^{\ast}_{\psi_{n}}(\bm{y})}]
    >
    \E[v(\bm{\omega})\mathbbm{1}_{\bm{\tau}_{\bar{x}} \leq s^{\ast}_{0}(\bm{y})}]
    $.
\qed

\section{Additional results}

\subsection{On uniqueness with weak belief punishments}\label{appendix:opaque_eq:uniqueness_within_class}

In the equilibria considered here, belief punishments are weak in the sense that the seller can offer a price (along with a message) that is almost surely above the price at which trade happens in equilibrium, and this price is only accepted by $\bar{x}$-buyers. (The equilibria of \Cref{prop:opaque_market:eq_example} are examples.)
These properties are natural to the extent that, by \Cref{assumption:correlated}, signal $\bar{x}$ induces the most optimistic belief about the state and, hence, a failure of these properties would have to rely on buyers' drawing negative inferences about the state from simply being offered this price. Of course, such inferences cannot be ruled out a priori since the seller has a private type and since buyers do not observe how long the seller has been on the market. 

The next proposition shows that in all equilibria with such weak belief punishments all non-$\bar{x}$-buyers do not trade.
\begin{proposition}\label{prop:opaque_eq:uniqueness_within_class}
    Let $v_{S}$ be strictly decreasing, and let $\lambda=0$.
    Let $(\alpha, \beta, p, m)$ be an equilibrium.
    Suppose there exists an offer $(p^{\dagger}, m^{\dagger})$ such that (i) $p^{\dagger} \geq p(t, y)$ for all $(t, y)\in\R_{+}\times Y$, and (ii) $\alpha(x, p^{\dagger}, m^{\dagger}) > 0$  if $x = \bar{x}$ and (iii) $\alpha(x, p^{\dagger}, m) = 0$ for all $m$ and $x\neq\bar{x}$.
    Then, non-$\bar{x}$-buyers almost surely do not trade, i.e., 
    \begin{equation*}
        \beta(t, y)\sum_{x \neq \bar{x}} \alpha(x, p(t, y), m(t, y)) = 0
        ,
    \end{equation*}
    for almost all $(t, y)\in\R_{+}\times Y$.
\end{proposition}

\begin{proof}[Proof of \Cref{prop:opaque_eq:uniqueness_within_class}]
    Fix $y$.
    If type $y$ trades with probability $1$ in equilibrium, then type $y$ must trade at price $p^{\dagger}$ almost surely. By assumptions (ii) and (iii), only $\bar{x}$-buyers can be induced to trade at $p^{\dagger}$, and so the claim follows for such $y$.

    Henceforth, assume that there is a non-zero set of types that do not trade with probability $1$, and let $Y'$ be the set of such types. 

    If $p^{\dagger} < v_{S}(\omega)$ for all $\omega$, almost all types of the seller trade with probability $0$. Thus, in this case, the proof is also complete.
    Henceforth, let $p^{\dagger}\geq v_{S}(\omega)$ hold for at least one $\omega$.

    Let us next argue that there exists a state $\omega_{0}^{\dagger}$ such that $p^{\dagger} = v_{S}(\omega_{0}^{\dagger})$. 
    Otherwise, it holds $p^{\dagger} > v_{S}(\omega)$ for all $\omega$.
    Then almost all types of the seller must trade with probability $1$ since the on-path price is below $p^{\dagger}$ almost surely, and since the seller can guarantee to trade at $p^{\dagger}$ with probability $1$. In particular, we have a contradiction to the previous paragraph.
    Thus, there exists a state $\omega^{\dagger}_{0}$ such that $p^{\dagger} = v_{S}(\omega_{0}^{\dagger})$.

    For all $y \in Y'$, let 
    \begin{equation*}
        \xi(y, \omega) = \int_{\R_{+}} \beta(t, y) \sum_{x\in X} \alpha(x, p(t, y), m(t, y)) \frac{f(x\vert \omega)}{f(\bar{x}\vert \omega)} \de t,
    \end{equation*}
    Since $y\in Y'$ does not trade with probability $1$, this integral is finite.
    By \Cref{assumption:correlated}, the integral $\xi(y, \omega)$ is weakly decreasing in $\omega$, and strictly decreasing if there is a non-zero set of $t$ such that $\beta(t, y) \sum_{x \neq \bar{x}} \alpha(x, p(t, y), m(t, y)) > 0$.

    Now suppose a type $y$ offers $(p^{\dagger}, m^{\dagger})$ up to the time $\xi(y, \omega_{0}^{\dagger}) / \alpha(\bar{x}, p^{\dagger}, m^{\dagger})$.
    This deviation must be unprofitable for almost all $y\in Y'$.
    
    Fixing $y$, let $V_{S}(y, \sigma)$ be $y$'s on-path utility.
    Let $\mu(\cdot\vert y)$ be $y$'s posterior belief about the state (recall \cref{footnote:types_posterior_belief}).
    The utility $V_{S}(y, \sigma)$ is bounded as follows:
    \begin{align*}
        V_{S}(y, \sigma)
        = &\int_{\Omega}\int_{\R_{+}} (p(t, y) - v_{S}(\omega)) \de Q_{\sigma}(t\vert y, \omega) \de\mu(\omega\vert y)
        \\
        \leq
        &\int_{\Omega}\int_{\R_{+}} (p^{\dagger} - v_{S}(\omega)) \de Q_{\sigma}(t\vert y, \omega) \de\mu(\omega\vert y)
        \\
        =
        &\int_{\Omega}(p^{\dagger} - v_{S}(\omega)) \left(1 - e^{-  f(\bar{x}\vert \omega) \xi(y, \omega)}\right)\de\mu(\omega\vert y)
        .
    \end{align*}
    Since $v_{S}$ is strictly decreasing, the difference $p^{\dagger} - v_{S}$ strictly single-crosses $0$ from below at the state $\omega_{0}^{\dagger}$.
    Since $\xi(y, \omega)$ is weakly decreasing in $\omega$, we have
    \begin{equation*}
        V_{S}(y, \sigma) \leq
         \int_{\Omega}(p^{\dagger} - v_{S}(\omega)) \left(1 - e^{-  f(\bar{x}\vert \omega) \xi(y, \omega_{0}^{\dagger})}\right)\de\mu(\omega\vert y)
         .
    \end{equation*}
    Further, the inequality is strict if $\xi(y, \cdot)$ is strictly decreasing.
    This upper bound is precisely type $y$'s utility from offering $(p^{\dagger}, m^{\dagger})$ up to time 
    $\xi(y, \omega_{0}^{\dagger}) / \alpha(\bar{x}, p^{\dagger}, m^{\dagger})$ since this offer is rejected by all non-$\bar{x}$-buyers and accepted by $\bar{x}$-buyers with probability $\alpha(\bar{x}, p^{\dagger}, m^{\dagger})$.
    It follows that for almost all $y$, the function $\xi(y, \cdot)$ is not strictly decreasing.
    Therefore, almost all $t, y$ satisfy $\beta(t, y) \sum_{x \neq \bar{x}} \alpha(x, p(t, y), m(t, y)) = 0$.
\end{proof}

\subsection{Comparative statics}\label{appendix:comparative_statics}

This appendix presents comparative statics for the optimal exit times with respect to the seller's type, as well as comparative statics of the maximized expected surplus with respect to the seller's and the buyers' information structures.
The main text observed that the efficient exit times are increasing under conditional independence between the seller's type and buyers' signals.
Here, we will relax this assumption, generating novel comparative statics. 

\subsubsection{Setup and preliminaries.}
We relax the assumption of conditional independence between the seller's type and buyers' signals. 
Conditional on the seller's type $y$ and the state $\omega$, the buyers' signals are independently drawn according to the PMF $f(\cdot\vert y, \omega)$.
Conditional on $\omega$, the seller's type is drawn according to the density $g(\cdot\vert\omega)$, as before, while the prior distribution of the state is $\mu$.
The analogue of \Cref{assumption:correlated} is the following:
\begin{assumption}\label{assumption:correlated_correlation}
    There exists a signal $\bar{x}\in X$ satisfying:
    \begin{enumerate}
        \item for all $x\in X\setminus\lbrace\bar{x}\rbrace$ and all $y\in Y$, 
        \begin{equation*}
            \frac{f(\bar{x}\vert y, \omega)}{f(x\vert y, \omega)} \quad\mbox{is strictly increasing in $\omega$;}
        \end{equation*}
        \item the product $g(y\vert\omega)f(\bar{x}\vert y, \omega)$ is strictly log-supermodular in $(y, \omega)$.\footnote{That is, letting $h(\cdot) = g(\cdot)f(\bar{x}\vert\cdot)$, if $y > y^{\prime}$ and $\omega>\omega^{\prime}$, then $h(y, \omega) h(y^{\prime}, \omega^{\prime}) > h(y, \omega^{\prime}) h(y^{\prime}, \omega)$.}
    \end{enumerate}
\end{assumption}
The analogue of \Cref{assumption:trade_is_possible} is:
    $\int_{\Omega} v(\omega) g(\bar{y}\vert\omega) f(\bar{x}\vert \bar{y}, \omega)\de\mu(\omega)
    > 0    
    $.

As in the main text, to maximize the expected surplus, it is optimal that each type $y$ of the seller trades with the first $\bar{x}$-buyer who arrives before time $\bar{s}(y) = \min\lbrace t\in\R_{+}\colon \E[v(\bm{\omega}) \vert \bm{\tau}_{\bar{x}} = t, \bm{y} = y] \leq 0\rbrace$.

The commitment profit equals the maximized expected surplus. For example, one may verify that in the strategy profile sketched in \Cref{remark:other_commitment_imlementations}, where the seller reveals calendar time and their type, buyers are best replying even when signals and types are not conditionally independent. (By contrast, the buyers' incentives turn out to be less straightforward if the seller does not reveal calendar time and their type since the inference from the seller's type about the state is not necessarily monotonic, as explained in more detail in \Cref{sec:commitment:stopping_times}.)

Next, \Cref{sec:commitment:stopping_times} analyzes how the exit time $\bar{s}$ depends on the seller's type.
\Cref{sec:commitment_solution:cs} analyzes how the maximized surplus depends on the information structures.
The proofs are collected in \Cref{appendix:comparative_statics:proofs}.

\subsubsection{Efficient exit times.}\label{sec:commitment:stopping_times}
Without conditional independence, $\bar{s}$ need not be monotonic in the seller's type.
This section characterizes when higher types exit before or after lower types.

Type $y$'s optimal exit time is given by $\bar{s}(y) = \min\lbrace t\in\R_{+}\colon \E[v(\bm{\omega}) \vert \bm{\tau}_{\bar{x}} = t, \bm{y} = y] \leq 0\rbrace$.
The seller considers the inference from the event that (i) their private type is $y$ and an $\bar{x}$-buyer arrives exactly at time $t$, and (ii) no $\bar{x}$-buyer arrived before time $t$.
Specifically, the likelihood of state $\omega$ at time $t$ depends on
\begin{equation}\label{eq:rich_belief_dynamics}
    \underbrace{g(y\vert\omega)f(\bar{x}\vert y, \omega)}_{\substack{\mbox{(i) type is $y$ and} \\ \mbox{an $\bar{x}$-buyer arrives}}} 
    \times
    \underbrace{e^{-  f(\bar{x}\vert y, \omega) t}}_{\substack{\mbox{(ii) no $\bar{x}$-buyer} \\ \mbox{arrived before $t$}}} 
    .
\end{equation}

As $y$ increases, inference (i) makes higher states relatively more likely since $g(y\vert\omega) f(\bar{x}\vert y, \omega)$ is log-\emph{super}modular in $(y, \omega)$, by \Cref{assumption:correlated_correlation}.

The effect of $y$ on inference (ii) is not obvious and depends on how the arrival rate of $\bar{x}$-buyers $f(\bar{x}\vert y, \omega)$ varies with $y$ and $\omega$.
Recall that the state accelerates the arrival of $\bar{x}$-buyers; i.e., $f(\bar{x}\vert y, \omega)$ strictly increases in $\omega$.
If $f(\bar{x}\vert y, \omega)$ is \emph{super}modular, then higher types $y$ believe that the state $\omega$ accelerates the arrival more than lower types.
Hence, when no $\bar{x}$-buyer arrives by a fixed date $t$, higher types get more pessimistic than lower types.
That is, $e^{-  f(\bar{x}\vert y, \omega) t}$ is log-\emph{sub}modular in $(y, \omega)$.
If $f(\bar{x}\vert y, \omega)$ is instead \emph{sub}modular, the effect is reversed and $e^{-  f(\bar{x}\vert y, \omega) t}$ is log-\emph{super}modular in $(y, \omega)$.

Let us first consider the simpler case in which $f(\bar{x}\vert y, \omega)$ is submodular in $(y, \omega)$.
This case applies whenever the buyers' signals and the seller's type are independent conditional on the state ($f$ does not depend on $y$).
Under submodularity,
inferences (i) and (ii) point in the same direction. Increasing the seller's type unambiguously shifts beliefs towards higher states. Hence, one can show:
\begin{proposition}\label{prop:submodular_bars}
    If $f(\bar{x}\vert y, \omega)$ is submodular in $(y, \omega)$, then $\bar{s}(y)$ is weakly increasing in $y$, strictly so on the subinterval on which $\bar{s}$ is non-zero.
\end{proposition}

\begin{example}[Submodular arrivals]\label{example:submodf}
    Suppose each buyer's signal obtains from the following binary test.
    Types and states lie in the unit interval, $Y = \Omega \subset (0, 1)$.
    There are independent random variables, $\varepsilon_{y}$ and $\varepsilon_{\omega}$, uniformly distributed on $[0, 1]$.
    The buyer's signal tests whether at least one of $\varepsilon_{y} \leq y$ and $\varepsilon_{\omega} \leq \omega$ holds.
    Thus, the set of possible signals is $X = \lbrace 0, 1\rbrace$. The probability of the high signal ($\bar{x} = 1$) is given by
    $    f(\bar{x}\vert y, \omega) = 1 - (1 - y) (1 - \omega)$.
    Then, $f(\bar{x}\vert y, \omega)$ is strictly submodular in $(y, \omega)$.
\end{example}

Suppose now that $f(\bar{x}\vert y, \omega)$ is strictly supermodular in $(y, \omega)$.
Thus, inferences (i) and (ii) go in opposite directions.
Which inference dominates as $y$ increases depends on the strength of inference (ii) at the endogenous exit time $t = \bar{s}(y)$.
Higher exit times amplify inference (ii): if the product \eqref{eq:rich_belief_dynamics} is log-submodular in $(y, \omega)$ for a given $t$, then it is also log-submodular in $(y, \omega)$ for all larger $t$.
This observation suggests that if all types exit early, then inference (ii) tends to be weak and, hence, higher types exit after lower types; conversely, if all types exit late, then inference (ii) tends to be strong and, hence, higher types exit before lower types.

The next result confirms that the overall effect can go either way depending on the size of the exit times.
Consider a sequence $(v_{n}, \mu_{n})_{n\in\mathbb{N}}$ for the surplus and the state distribution. Both $f$ and $g$ are fixed. For all $n$, the assumptions of the model hold and $\bar{s}_{n}$ denotes the optimal exit times.
\begin{proposition}\label{prop:supermod_bars}
    Let $g$ and $f(\bar{x}\vert \cdot)$ be twice continuously differentiable, with cross-partial $\partial_{y \omega} \ln(g(y\vert\omega) f(\bar{x}\vert y, \omega))$ bounded away from zero.

    If $f(\bar{x}\vert y, \omega)$ is strictly supermodular in $(y, \omega)$ with cross-partial $\partial_{y \omega} f(\bar{x}\vert y, \omega)$ bounded away from zero, and $(\inf \bar{s}_{n})_{n\in\mathbb{N}}$ diverges, then for all but finitely many $n$ the function $\bar{s}_{n}$ is strictly decreasing.

    If $(\sup \bar{s}_{n})_{n\in\mathbb{N}}$ converges to $0$, then for all but finitely many $n$ the function $\bar{s}_{n}$ is weakly increasing, strictly so on the subinterval on which $\bar{s}_{n}$ is non-zero.
\end{proposition}

\begin{example}[Supermodular arrivals]\label{example:supermodf}
    In the setup from \Cref{example:submodf}, suppose the high signal ($\bar{x} = 1$) obtains if and only if \emph{both} $\varepsilon_{y} \leq y$ and $\varepsilon_{\omega} \leq \omega$ hold.
    That is, let 
    $f(\bar{x}\vert y, \omega) = y \omega$ so that $f(\bar{x}\vert\cdot)$ is strictly supermodular.
    \Cref{fig:informed_exit_times} depicts the optimal exit times $\bar{s}$ in such an example.
    The state is binary, $\supp\mu = \lbrace \ubar\omega, \bar\omega\rbrace$.
    All parameters are fixed except the probability $\mu(\bar{\omega})$ of the high state. 
    As $\mu(\bar{\omega})$ increases, the exit time for all types increases and diverges to $\infty$ as $\mu(\bar{\omega}) \to 1$.
    If $\mu(\bar{\omega})$ is sufficiently high, then $\bar{s}$ is strictly decreasing; if $\mu(\bar{\omega})$ is sufficiently low, then $\bar{s}$ is strictly increasing when non-zero; if $\mu(\bar{\omega})$ is intermediate, $\bar{s}$ may be non-monotonic.
    \begin{figure}[ht!]
    \centering
    \begin{subfigure}{0.45\textwidth}
        \centering
        \includegraphics[width=\textwidth]{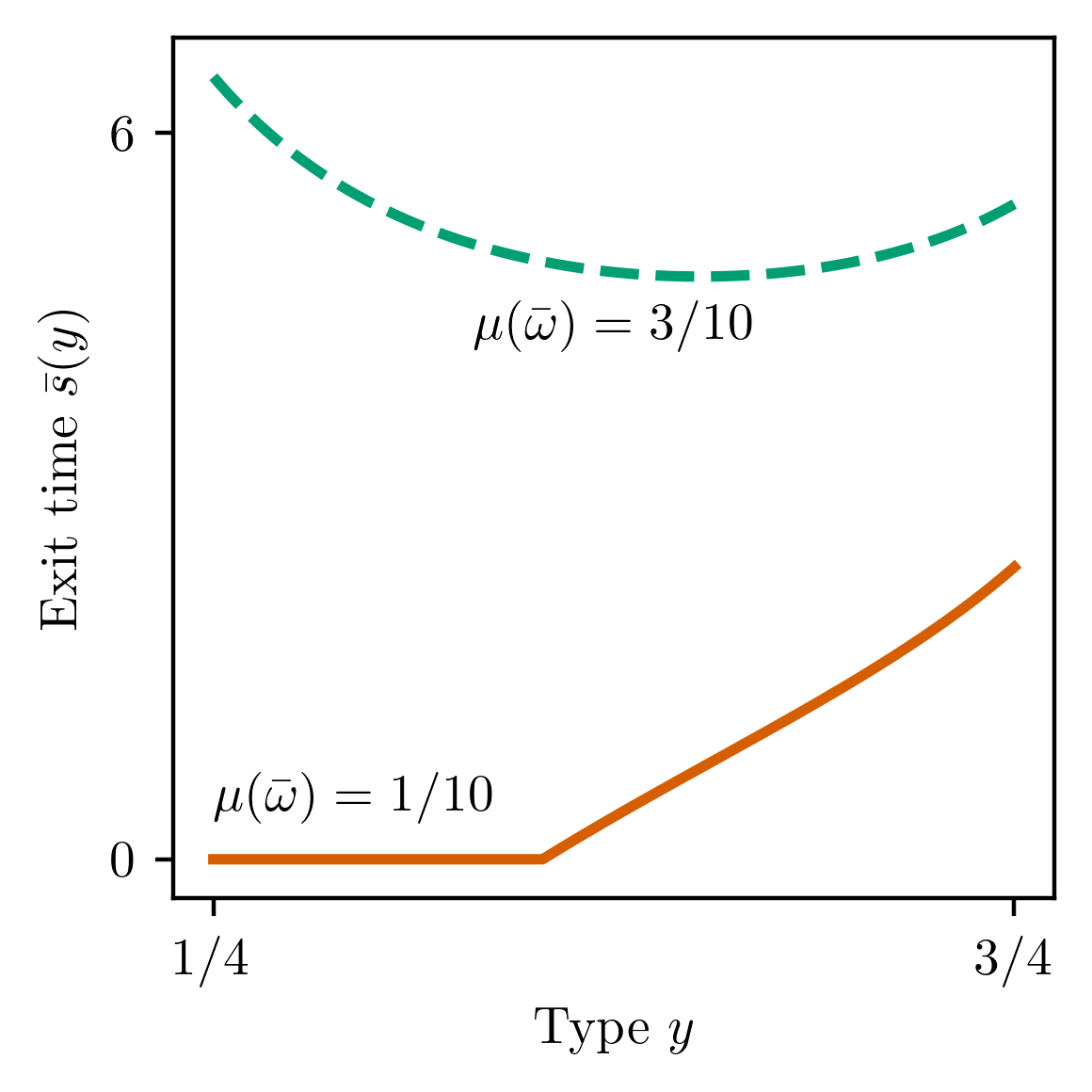}
        \caption{Priors $\mu(\bar{\omega}) \in \lbrace 1/10, 3/10\rbrace$.}
        \label{fig:informed_exit_times:small_intermediate}
    \end{subfigure}
    \hfill
    \begin{subfigure}{0.45\textwidth}
        \centering
        \includegraphics[width=\textwidth]{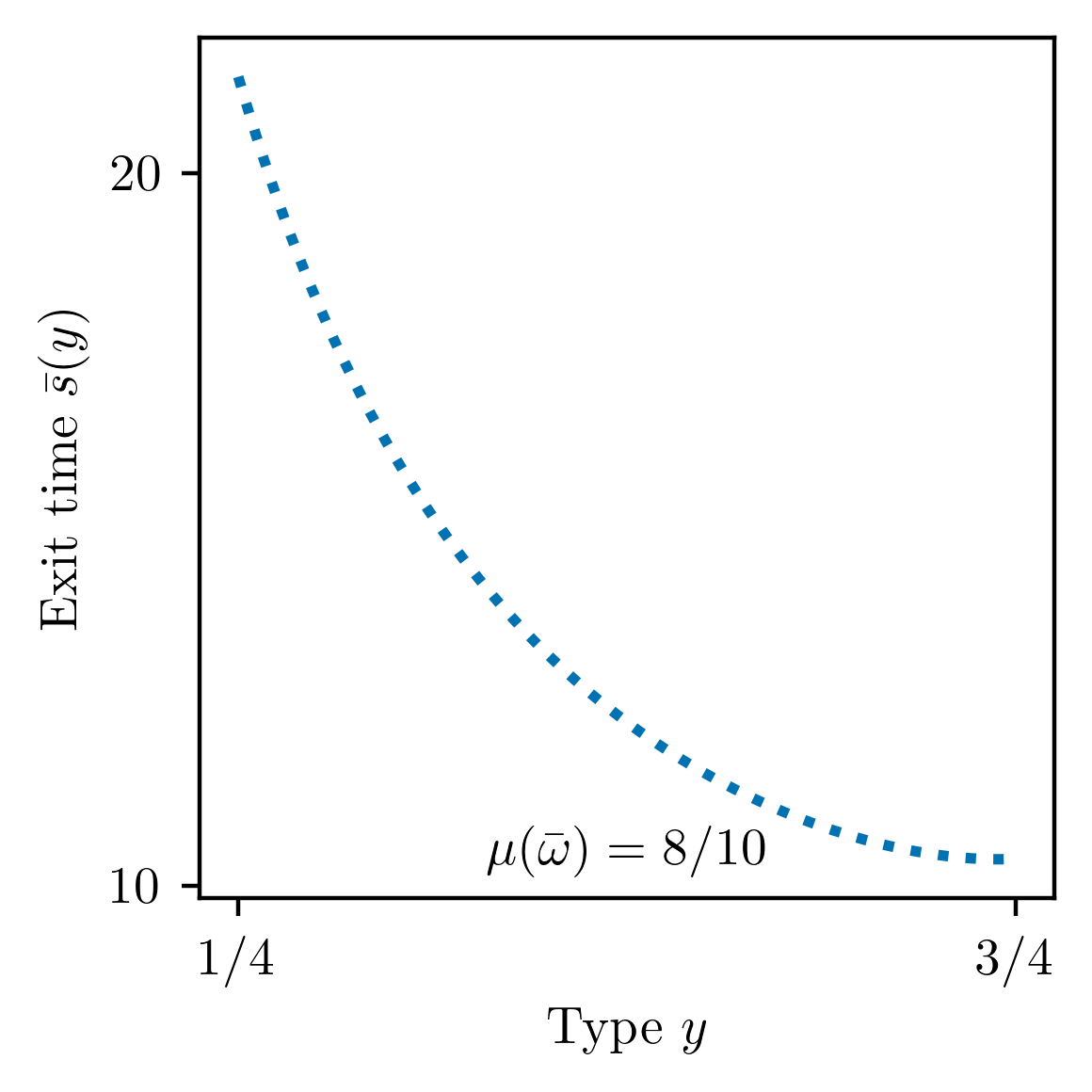}
        \caption{Prior $\mu(\bar{\omega}) = 8/10$.}
        \label{fig:informed_exit_times:big}
    \end{subfigure}
    \caption{The exit times $\bar{s}$ in \Cref{example:supermodf} for various priors. Here, $[\underline{y}, \bar{y}] = [1/4, 3/4]$ and $[\underline{\omega}, \bar{\omega}] = [1/5, 4/5]$. The surplus is given by $v(\bar{\omega}) = - v(\underline{\omega}) = 1$. The densities of the seller's type $y$ are $g(y\vert\underline{\omega}) = \underline{k} y (9/10 - y)$ and $g(y\vert\bar{\omega}) = \bar{k}y$ for normalizing constants $\underline{k}, \bar{k} > 0$.}
    \label{fig:informed_exit_times}
    \end{figure}
\end{example}

\subsubsection{Information structures.}\label{sec:commitment_solution:cs}
Without breakdowns, the commitment profit turns out to increase if $\bar{x}$ is more favorable about the state in the sense of likelihood-ratio dominance.
For the next result, let $f_{1}$ and $f_{2}$ be two signal PMFs with respective most favorable signals $\bar{x}_{1}$ and $\bar{x}_{2}$.
\begin{proposition}\label{corollary:cs_buyers_signals}
    If $\lambda = 0$ and for all $y \in Y$ the ratio 
    \begin{equation*}
        \omega\mapsto \frac{f_{2}(\bar{x}_{2}\vert y, \omega)}{f_{1}(\bar{x}_{1}\vert y, \omega)}
    \end{equation*}
    is weakly increasing, then the commitment profit is weakly higher under $f_{2}$ than under $f_{1}$.
\end{proposition}

\begin{example}[Binary testing]
    In the setting of \Cref{example:supermodf}, suppose the probability of signal $\bar{x}$ is given by $f(\bar{x}\vert y, \omega) = y \omega^{\rho}$ where the parameter $\rho$ governs the sensitivity of the test with respect to the state.
    Increasing $\rho$ raises the commitment profit.
\end{example}
With breakdowns ($\lambda > 0$), the obvious modification to \Cref{corollary:cs_buyers_signals} is to also assume $f_{2}(\bar{x}_{2}\vert y, \omega) \geq f_{1}(\bar{x}_{1}\vert y, \omega)$ for all $y\in Y$ and $\omega\in \Omega$; i.e., $\bar{x}$-buyers arrive more rapidly under $f_{2}$ than under $f_{1}$. Note that $\lambda$ should be sufficiently small so that the characterization of the commitment problem applies for both $f_{2}$ and $f_{1}$.

Specializing to the case of conditional independence, the next result shows that the commitment profit increases if the seller's type is more informative in the sense of \citet{lehmann1988comparing}.
Let $\lbrace G_{1}(\cdot\vert\omega)\rbrace_{\omega\in \Omega}$ and $\lbrace G_{2}(\cdot\vert\omega)\rbrace_{\omega\in \Omega}$ be two collections of absolutely continuous, strictly increasing CDFs whose densities satisfy the assumptions of the model.
For all $\omega$, let $G_{2}^{-1}(\cdot\vert\omega)$ be the inverse of $G_{2}(\cdot\vert\omega)$.
Say $G_{2}$ is \emph{Lehmann-more informative} than $G_{1}$ if for all $y\in \R$ the quantile $G_{2}^{-1}(G_{1}(y\vert\omega)\vert\omega)$ is weakly increasing in $\omega$.
For a canonical example, let the type be a location experiment, i.e., $y = \omega + \rho \varepsilon$ for a random noise term $\varepsilon$ and a parameter $\rho> 0$ that scales the noise; decreasing $\rho$ increases Lehmann-informativeness.

Let $\bar{\lambda}_{1}$ and $\bar{\lambda}_{2}$, respectively, be as in \Cref{lemma:delta_maximization} for $G_{1}$ and $G_{2}$, respectively.
\begin{proposition}\label{corollary:cs_sellers_type}
    Let $\lambda < \min\lbrace\bar{\lambda}_{1}, \bar{\lambda}_{2}\rbrace$, and let the seller's type and the buyers' signals be independent conditional on the state.
    If $G_{2}$ is Lehmann-more informative than $G_{1}$, then the commitment profit is weakly higher under $G_{2}$ than under $G_{1}$.
\end{proposition}
To see the role of conditional independence, let us sketch the proof in the case without breakdowns ($\lambda=0$).
Then, the commitment profit equals $\E[v(\bm{\omega})(1 - e^{-f(\bar{x}\vert\bm{\omega}) \bar{s}(\bm{y})})]$. This is roughly the expected utility of a decision-maker who learns about the state by observing $\bm{y}$, and then takes a binary action (whether to trade) for a state-dependent payoff of $v$ (trade) or $0$ (otherwise).
Unusually, the probability of the action, $1 - e^{-f(\bar{x}\vert\omega)\bar{s}(y)}$, also depends on the state through the state-dependent arrival rate $f(\bar{x}\vert\omega)$ of $\bar{x}$-buyers.
Nonetheless, since $v$ is single-crossing and $\bar{s}$ is weakly increasing (\Cref{prop:submodular_bars} and conditional independence), one can follow \citet{lehmann1988comparing} to show that a Lehmann-more informative type distribution makes this decision-maker better off. A similar argument turns out to work with breakdowns.

\Cref{corollary:cs_buyers_signals,corollary:cs_sellers_type} have a different interpretation than a result for static lemons markets due to \citet{levin2001information}.
In the market studied by Levin, the surplus is commonly known to be positive.
One side of the market has a private signal about the underlying state. 
The only possible inefficiency, stemming from asymmetric information, is that trade might not occur.
Levin shows that the informativeness of the private signal,\footnote{Levin studies Lehmann's order and the monotone information order of \citet{athey2018value}.} which determines the degree of asymmetric information, may but need not raise efficiency.
In the present setting and when the seller has commitment power, asymmetric information is not a source of inefficiency as the seller maximizes the expected surplus.
However, the realized surplus may be negative.
The remaining inefficiency is that the aggregated noisy information fails to identify the efficient allocation.
\Cref{corollary:cs_buyers_signals,corollary:cs_sellers_type} ask which information reduces this inefficiency.

\subsection{Proofs for \headercref{Appendix}{{appendix:comparative_statics}}}\label{appendix:comparative_statics:proofs}
\subsubsection{Proof of \headercref{Proposition}{{prop:submodular_bars}}.}
    Let $y\in Y$ and recall the definition $\bar{s}(y) = \min\lbrace s\in\R_{+}\colon \E [v(\bm{\omega})\vert \bm{\tau}_{\bar{x}} = s, \bm{y} = y] \leq 0\rbrace$.
    For all $t\in\R_{+}$, the sign of the conditional expectation in the definition of $\bar{s}$ equals the sign of $\hat{v}(t, y)$, where
    \begin{equation*}
        \hat{v}(t, y) = \int_{\Omega} v(\omega) g(y\vert\omega) f(\bar{x}\vert y, \omega) e^{-  f(\bar{x}\vert y, \omega) t} \de\mu(\omega).
    \end{equation*}
    Since $f(\bar{x}\vert y, \omega)$ is submodular in $(y, \omega)$, the term $e^{-f(\bar{x}\vert y, \omega) t}$ is log-supermodular in $(y, \omega)$.
    Since $v$ strictly single-crosses $0$ from below and since also $g(y\vert\omega) f(\bar{x}\vert y, \omega)$ is strictly log-supermodular, it follows that $y\mapsto \hat{v}(t, y)$ strictly single-crosses $0$ from below for fixed $t$.
    \Cref{lemma:hatv_properties} shows that $t\mapsto \hat{v}(t, y)$ is single-crossing from above (this can be shown without conditional independence). It follows that $\bar{s}(y)$ is increasing in $y$, strictly so if non-zero.
\qed

\subsubsection{Proof of \headercref{Proposition}{{prop:supermod_bars}}.}\label{appendix:prop:supermod_bars}
    Let $f(\bar{x}\vert y, \omega)$ be strictly supermodular in $(y, \omega)$, and let $(\inf \bar{s}_{n})_{n\in\mathbb{N}}$ diverge.
    Given $t\in\R_{+}$, consider the function
    \begin{equation}\label{eq:prop:supermod_bars:submod}
        (y, \omega)\mapsto
        \ln g(y\vert\omega) + \ln f(\bar{x}\vert y, \omega) -  f(\bar{x}\vert y, \omega) t.
    \end{equation}
    By assumption, $g(y\vert\omega)$ and $f(\bar{x}\vert y, \omega)$ are continuous and strictly positive for all $(y, \omega) \in Y\times \Omega$.
    Thus, the differentiability assumptions on $g$ and $f$ imply that the cross-partial derivative of $\ln g(y\vert\omega) + \ln f(\bar{x}\vert y, \omega)$ in $(y, \omega)$ is bounded across $Y\times \Omega$.
    By assumption, the cross-partial of $f(\bar{x}\vert y, \omega)$ in $(y, \omega)$ is strictly positive and bounded away from $0$ across $Y\times \Omega$.
    It follows that there exists $T\in\R_{++}$ such that \eqref{eq:prop:supermod_bars:submod} is strictly \emph{sub}modular in $(y, \omega)$ if $t \geq T$.

    Now fix $n$ such that $\inf\bar{s}_{n}\geq T$.
    We show $\bar{s}_{n}$ is strictly decreasing.
    Let $y, y^{\prime} \in Y$ be such that $y > y^{\prime}$.
    Since $T > 0$, also $\bar{s}_{n}(y) > 0$ and $\bar{s}_{n}(y^{\prime}) > 0$. Thus, $\hat{v}_{n}(\bar{s}_{n}(y), y) = \hat{v}_{n}(\bar{s}_{n}(y^{\prime}), y^{\prime}) = 0$, where $\hat{v}_{n}$ denotes the function \eqref{eq:posterior_gains_definition} in the $n$'th environment.
    By the choice of $T$, the term \eqref{eq:prop:supermod_bars:submod} is strictly submodular in $(y, \omega)$. Since also $v$ is strictly single-crossing from below, it follows that $\tilde{y}\mapsto \hat{v}_{n}(\bar{s}_{n}(y), \tilde{y})$ is strictly single-crossing from above.
    Since $t\mapsto \hat{v}_{n}(t, y^{\prime})$ is also strictly single-crossing from above, the equation $\hat{v}_{n}(\bar{s}_{n}(y), y) = \hat{v}_{n}(\bar{s}_{n}(y^{\prime}), y^{\prime}) = 0$ requires $\bar{s}_{n}(y) < \bar{s}_{n}(y^{\prime})$.

    Finally, let $\lim_{n\to\infty} \sup\bar{s}_{n} = 0$.
    Since $g(y\vert\omega) f(\bar{x}\vert y, \omega)$ is strictly log-supermodular in $(y, \omega)$, the differentiability assumptions imply that there exists $T^{\prime} \in\R_{++}$ such that the function in \eqref{eq:prop:supermod_bars:submod} is strictly \emph{super}modular in $(y, \omega)$ whenever $t \leq T^{\prime}$.
    Fixing $n$ such that $\sup\bar{s}_{n} \leq T^{\prime}$, it now follows analogously to above that $\tilde{y}\mapsto \hat{v}_{n}(t, \tilde{y})$ is strictly single-crossing from below whenever $t\leq T^{\prime}$.
    Since $t\mapsto \hat{v}_{n}(t, y)$ is also strictly single-crossing from above for all $y\in Y$, it follows that $\bar{s}_{n}(y) \geq \bar{s}_{n}(y^{\prime})$ holds whenever $y > y^{\prime}$, with a strict inequality whenever $\bar{s}_{n}(y) > 0$.
\qed

\subsubsection{Proof of \headercref{Proposition}{{corollary:cs_buyers_signals}}.}\label{appendix:proof:corollary:cs_buyers_signals}

Let $\bar{s}_{1}$ denote the optimal exit time under $f_{1}$.
Let $\omega_{0}$ denote the state at which $v$ strictly single-crosses $0$ from below.
For all $y$, let $\xi(y)$ solve 
\begin{equation*}
1 - e^{- f_{2}(\bar{x}_{2}\vert y, \omega_{0}) \xi(y)} = 1 - e^{- f_{1}(\bar{x}_{1}\vert y, \omega_{0}) \bar{s}_{1}(y)}
.
\end{equation*}
Since $f_{2}(\bar{x}_{2}\vert y, \omega) / f_{1}(\bar{x}_{1}\vert y, \omega)$ is weakly increasing in $\omega$, one may verify that 
\begin{equation*}
    \omega\mapsto
    1 - e^{- f_{2}(\bar{x}_{2}\vert y, \omega) \xi(y)} - \left(1 - e^{- f_{1}(\bar{x}_{1}\vert y, \omega) \bar{s}_{1}(y)}\right)
\end{equation*}
single-crosses $0$ from below at $\omega = \omega_{0}$.
Since $v$ single-crosses $0$ from below at $\omega_{0}$, for all $y$ and $\omega$ it holds
\begin{equation*}
    v(\omega) \left(1 - e^{- f_{2}(\bar{x}_{2}\vert y, \omega) \xi(y)}\right) \geq v(\omega) \left(1 - e^{- f_{1}(\bar{x}_{1}\vert y, \omega) \bar{s}_{1}(y)}\right)
    .
\end{equation*}
Thus, if under $f_{2}$ each type $y$ exits at $\xi(y)$ (and trade is with the first $\bar{x}_{2}$-buyer arriving before $\xi(y)$), then the expected surplus conditional on each state and type is weakly higher than under the commitment solution under $f_{1}$.
Consequently, the commitment profit under $f_{2}$ is weakly higher than under $f_{1}$.
\qed

\subsubsection{Proof of \headercref{Proposition}{{corollary:cs_sellers_type}}.}\label{appendix:proof:corollary:cs_sellers_type}

For all $y$ and $\omega$, let $\hat{z}(y, \omega) = G_{1}^{-1}(G_{2}(y\vert\omega)\vert\omega)$.
By assumption, $G_{2}^{-1}(G_{1}(y\vert\omega)\vert\omega)$ is weakly increasing in $\omega$, implying that $\hat{z}(y, \omega)$ is weakly \emph{de}creasing in $\omega$.
Let $\bar{s}_{1}$ denote the optimal exit times under $G_{1}$.
Recall that the seller's type and the buyers' signals are independent conditional on the state, and so we write $\bar{Q}(t\vert \omega) = 1 - e^{- f(\bar{x}\vert \omega) t}$ for the probability of an $\bar{x}$-buyer arriving before $t$ in state $\omega$.
By a change of variables, we can write the commitment profit $\bar{V}_{1}$ under $G_{1}$ as
\begin{align*}
    \bar{V}_{1} = & \int_{\Omega} \int_{\R_{+}} \int_{Y} v(\omega) e^{-\lambda t} \mathbbm{1}_{t\leq \bar{s}_{1}(y)} \de G_{1}(y\vert\omega) \de \bar{Q}(t\vert \omega) \de\mu(\omega) 
    \\
    = &
    \int_{\Omega} \int_{\R_{+}} \int_{Y} v(\omega) e^{-\lambda t} \mathbbm{1}_{t\leq \bar{s}_{1}(\hat{z}(y, \omega))} \de G_{2}(y\vert\omega) \de \bar{Q}(t\vert \omega) \de\mu(\omega) 
    .
\end{align*}
Recall that $\hat{z}(y, \omega)$ is decreasing in $\omega$ for fixed $y$.
Given conditional independence, also $\bar{s}_{1}$ is increasing (\Cref{prop:submodular_bars}).
In particular, for each fixed $y$ the time $\bar{s}_{1}(\hat{z}(y, \omega))$ is weakly decreasing in $\omega$.
Since $v$ strictly single-crosses $0$ from below at $\omega_{0}$, for all $\omega$ it holds
\begin{equation*}
    v(\omega) \mathbbm{1}_{t\leq \bar{s}_{1}(\hat{z}(y, \omega))}
    \leq
    v(\omega) \mathbbm{1}_{t\leq \bar{s}_{1}(\hat{z}(y, \omega_{0}))}
    .
\end{equation*}
Hence, the commitment profit under $G_{1}$ is bounded above by
\begin{equation*}
    \int_{\Omega} \int_{\R_{+}} \int_{Y} v(\omega) e^{-\lambda t} \mathbbm{1}_{t\leq \bar{s}_{1}(\hat{z}(y, \omega_{0}))} \de G_{2}(y\vert\omega) \de \bar{Q}(t\vert \omega) \de\mu(\omega) .
\end{equation*}
This upper bound equals the expected surplus under $G_{2}$ if each type $y$ of the seller trades with the first $\bar{x}$-buyer who arrives before the type-dependent exit time $\bar{s}_{1}(\hat{z}(y, \omega_{0}))$.
Importantly, this time does not depend on the unobservable state.
Thus, the upper bound is itself at most the commitment profit under $G_{2}$.
\qed

\addcontentsline{toc}{section}{References}
\newrefcontext[sorting=nyt]
\printbibliography

\end{document}